\documentclass[aip, reprint, twocolumn,jcp,english]{revtex4-1}
\usepackage[T1]{fontenc}
\usepackage{amsmath}
\usepackage{amssymb}
\usepackage{comment}
\usepackage{graphicx}
\usepackage[citecolor=blue]{hyperref}
\usepackage{enumitem}
\usepackage{bm}
\makeatletter
\usepackage{tikz}
\usetikzlibrary{calc}
\usetikzlibrary{arrows}
\usetikzlibrary{shapes.geometric}
\tikzset{
    vertex/.style={circle,draw,minimum size=1.5em},
    edge/.style={->,> = latex'}
}
\usetikzlibrary{shapes.misc, positioning}
\usetikzlibrary{decorations.markings}
\usepackage{appendix}

\usepackage{longtable}%
\usepackage{booktabs}
\usepackage{colortbl}%

\usepackage{listings}

\definecolor{codegreen}{rgb}{0,0.6,0}
\definecolor{codegray}{rgb}{0.5,0.5,0.5}
\definecolor{codepurple}{rgb}{0.58,0,0.82}
\definecolor{backcolour}{rgb}{0.95,0.95,0.92}

\lstdefinestyle{mystyle}{
    backgroundcolor=\color{backcolour},
    commentstyle=\color{codegreen},
    keywordstyle=\color{magenta},
    numberstyle=\tiny\color{codegray},
    stringstyle=\color{codepurple},
    basicstyle=\footnotesize,
    breakatwhitespace=false,
    breaklines=true,
    captionpos=b,
    keepspaces=true,
    numbers=left,
    numbersep=5pt,
    showspaces=false,
    showstringspaces=false,
    showtabs=false,
    tabsize=2
}

\setlist[itemize]{noitemsep}

\global\long\def\ket#1{\left|#1\right\rangle }%
\global\long\def\avg#1{\left\langle#1\right\rangle }%

\global\long\def\bra#1{\left\langle #1\right|}%

\global\long\def\braket#1#2{\left.\left\langle #1\right|#2\right\rangle }%
\makeatother
\usepackage{babel}

\begin{document}

\title{pyTTN: An Open Source Toolbox for Open and Closed System Quantum Dynamics Simulations Using Tree Tensor Networks}
\begin{abstract}
We present the Python Tree Tensor Network package (pyTTN) for the evaluation of dynamical properties of closed and open quantum systems that makes use of Tree Tensor Network (TTN), or equivalently the multi-layer multiconfiguration time-dependent Hartree (ML-MCTDH), based representations of wavefunctions. This package includes several features allowing for easy setup of zero- and finite-temperature calculations for general Hamiltonians using single and multi-set TTN ans\"atze with an adaptive bond dimension through the use of subspace expansion techniques. All core features are implemented in C\texttt{++} with Python bindings provided to simplify the use of this package. In addition to these core features, pyTTN provides several tools for setting up efficient simulation of open quantum system dynamics, including the use of the TTN ansatz to represent the auxiliary density operator space for the simulation of the Hierarchical Equation of Motion (HEOM) method and generalised pseudomode methods; furthermore we demonstrate that the two approaches are equivalent up to a non-unitary normal mode transformation acting on the pseudomode degrees of freedom. We present a set of applications of the package, starting with the widely used benchmark case of the photo-excitation dynamics of 24 mode pyrazine, following which we consider a more challenging model describing the exciton dynamics at the interface of a $n$-oligothiophene donor-C$_{60}$ fullerene acceptor system.  Finally, we consider applications to open quantum systems, including the spin-boson model, a set of extended dissipative spin models, and an Anderson impurity model. By combining ease of use, an efficient implementation, as well as an extendable design allowing for the addition of future extensions, pyTTN can be integrated in a wide range of computational modelling software.
\end{abstract}
\author{Lachlan P Lindoy}
\email{lachlan.lindoy@npl.co.uk}
\affiliation{National  Physical  Laboratory,  Teddington,  TW11  0LW,  United  Kingdom}
\author{Daniel Rodrigo-Albert}
\affiliation{National  Physical  Laboratory,  Teddington,  TW11  0LW,  United  Kingdom}
\author{Yannic Rath}
\affiliation{National  Physical  Laboratory,  Teddington,  TW11  0LW,  United  Kingdom}
\author{Ivan Rungger}
\email{ivan.rungger@npl.co.uk}
\affiliation{National  Physical  Laboratory,  Teddington,  TW11  0LW,  United  Kingdom}
\affiliation{Department of Computer Science, Royal Holloway, University of London, Egham, TW20 0EX, United Kingdom}
\maketitle

\section{Introduction}
In recent years, tensor network methods have emerged as some of the most successful for calculating ground state and dynamical properties of quantum systems. For molecular systems, the multiconfiguration time-dependent Hartree (MCTDH)~\citep{MEYER199073,MANTHE1992,BECK20001}
and its multi-layer extension (ML-MCTDH)~\citep{WANG2003,MANTHE2008,VENDRELL2011,WANG2015,lindoy_mctdh_1}, which uses a tree tensor network (TTN) ansatz for the wavefunction \footnote{throughout this work we use the terms ML-MCTDH ansatz and TTN interchangeably}
have found widespread success in obtaining systematically
convergable dynamics beyond that obtainable with exact diagonalisation. These approaches have been successfully applied to
study the dynamics of small isolated systems~\citep{MANTHE1992,MANTHE199212,MANTHE19937,HAMMERICH1994,RAAB1999,VENDRELL20071,VENDRELL20072,VENDRELL20073,VENDRELL20074,WORTH2008,MENG2013} and to larger condensed-phase systems, using both unitary~\citep{CRAIG2007,WANG2007,KONDOV2007,CRAIG2011,BORRELLI2012,RABANI2015,SCHULZE2016,SHIBL2017,MENDIVE2018,10.1063/5.0218773} and more recently non-unitary methods for simulating the dynamics~\cite{10.1063/5.0050720,lindoy_thesis,LindoyMandalReichman,10.1063/5.0153870}.

Within these methods, a multidimensional wavefunction is expanded in terms of a linear combination of a set of hierarchically defined, time-dependent and variationally optimised basis functions.  These approaches can help to alleviate the so-called curse of dimensionality, that is the exponential scaling of cost of quantum dynamics calculations with increasing numbers of dimensions and, in principle, can be converged to give the exact quantum dynamics of large-scale systems by systematically increasing the number of basis functions included in the representation.  As a result, they are often used to generate benchmark results for other approximate approaches~\citep{Wang2010,Miller2018,Richardson2020}. However, whether or not this convergence can be obtained in practice depends strongly on the extent of entanglement observed within the system of interest and how well this entanglement is captured by the topology of the network used.

Due to the success of these methods in handling the quantum dynamics of large scale quantum systems, several software packages have been developed over the years for applying these approaches to the unitary dynamics of quantum systems~\cite{Heidelberg,WORTH2020107040,10.1063/5.0180233,10.1063/5.0218773,  milbradt2024pytreenetpythonlibraryeasy, lacroix2024mpsdynamicsjltensornetworksimulations,renormalizer}, each with varying features and support for wavefunction ans\"atze.  In this manuscript, we present pyTTN, an efficient, open source, Python library with core features implemented in C\texttt{++} for the simulation of ground-state properties and quantum dynamics using tree tensor network state-based ans\"atze. In the current release, available at \url{https://gitlab.npl.co.uk/qsm/pyttn}, pyTTN supports:
\begin{figure*}[tp]
\includegraphics[width=\textwidth]{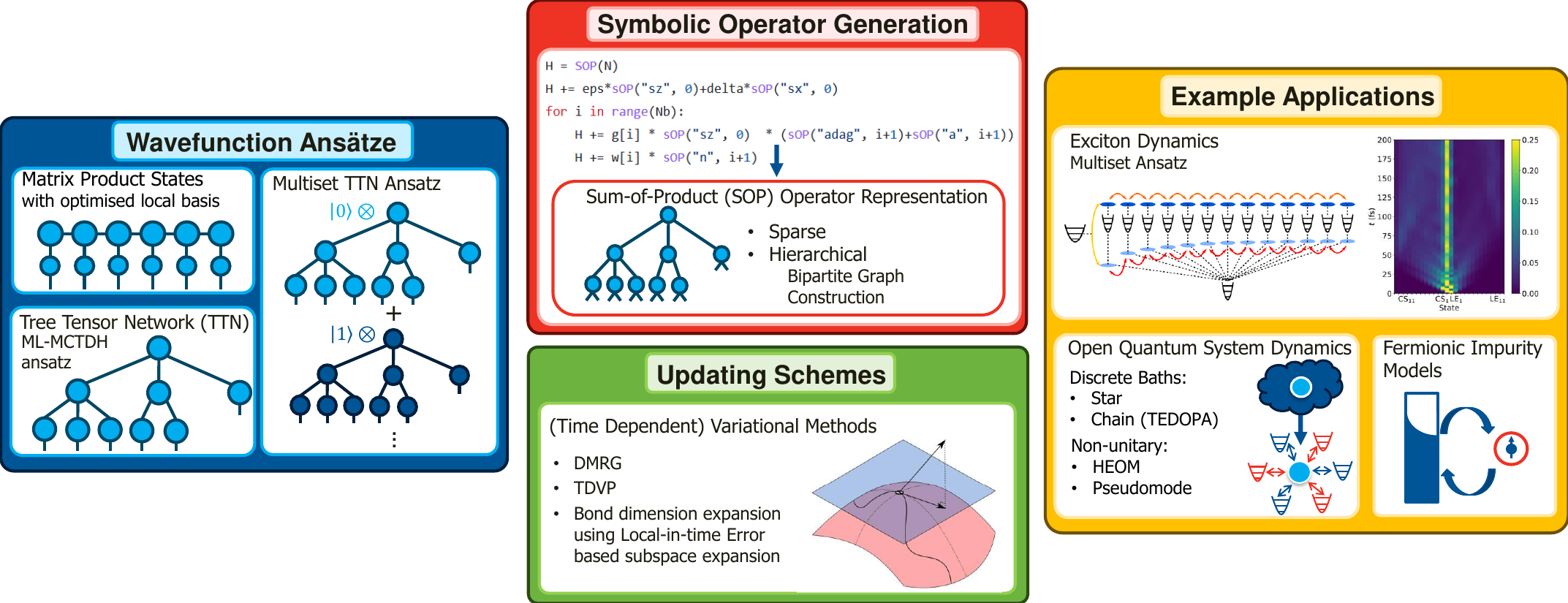}%
\caption{ \label{fig:pyttn_schematic}  Schematic outlining the key features and applications of pyTTN described in this manuscript.  The pyTTN repository can be found at \url{https://gitlab.npl.co.uk/qsm/pyttn}. The key wavefunction ans\"atze supported within pyTTN, which are discussed in Sec.~\ref{sec:wavefunction_ansatz}, are illustrated in the blue square.  The Hamiltonian generation tools (Sec.~
\ref{sec:op_gen}) and updating schemes (Sec.~\ref{sec:updating} and~\ref{sec:subspace}) that enable calculations using these ans\"atze are illustrated in red and green rectangles, respectively.  Finally, the orange rectangle illustrates a subset of applications of pyTTN that are are discussed in Sec.~\ref{sec:unitary_dynamics} and~\ref{sec:oqs_dynamics}.}
\end{figure*}
\begin{itemize}[leftmargin=*, topsep=0pt]
    \item Time evolution through the use of the time-dependent variational principle (TDVP) and the one-site projector splitting integrator~\cite{Lubich2014a,Kloss2017a, Bauernfeind2020,Kloss2020,Ceruti2021,lindoy_mctdh_2}.
    \item Ground state optimisation through the use of the one-site alternating least squares based approach~\cite{Larsson2019}; i.e. Density Matrix Renormalisation Group (DMRG) on trees.
    \item Adaptive control of bond dimension through subspace expansion methods that can use either natural orbital population based metrics and/or a generalisation of the local-in-time error metric considered in Ref.~\onlinecite{MendiveTapia2020} to the multi-layer tree case. 
    \item Symbolic operator construction with automatic construction of either sum-of-product operator~\cite{RAAB1999} or hierarchical sum-of-product operator~\cite{10.1063/1.4856135, OTTO2018116} representations.  
    \item Multi-set tree tensor network states, including all fixed-bond dimension algorithms and support for symbolic operator generation.
    \item Features for supporting open quantum system dynamics including:
    \begin{itemize}[topsep=0pt]
        \item Methods for discretising continuous baths to obtain discrete Hamiltonians approximating the dynamics of these open systems.
        \item Inclusion of finite temperature effects through the use of thermal spectral density~\cite{PhysRevLett.123.090402}.
        \item Unitary bath transformations to transform the open quantum system Hamiltonian to other topologies, such as the chain representation used by the Time Evolving Density operator with Orthogonal Polynomials Algorithm (TEDOPA) method~\cite{PhysRevLett.105.050404, 10.1063/1.4866769, 10.1063/1.3490188,Rosenbach_2016}.
        \item Non-unitary methods, namely the Hierarchical Equations of Motion (HEOM)~\cite{TanimuraJPSJ1989,IshizakiJPSJ2005,TanimuraJCP2020} and the generalised pseudomode~\cite{park2024quasilindbladpseudomodetheoryopen} methods, that make use of more compact exponential decompositions of bath correlation functions.
    \end{itemize}
\end{itemize}
Throughout this article we present examples highlighting the use of these features within pyTTN. A summary, of these features and applications are provided in Fig.\ref{fig:pyttn_schematic}.

In an effort to facilitate use of the library for problems beyond those considered in this manuscript, the current version of the pyTTN repository additionally includes a set of example Python scripts (found in \protect\path{pyttn/examples}) that demonstrate applications of the library for the simulation of the quantum dynamics of several open and closed quantum systems, including all applications considered in this manuscript. In addition, the repository includes a set of tutorials presented as Jupyter notebooks outlining the use of the core features of the library:
\begin{itemize}[leftmargin=*,topsep=0pt]
    \item The construction of tree tensor networks with various structures (\protect\path{pyttn/tutorials/tree_topologies.ipynb}).
    \item The generation of arbitrary operators using both predefined (supporting a standard set of spin, bosonic, and fermionic operators) and user-defined single-site operators, with support for both sparse and dense formats for matrix representations of operators (\protect\path{pyttn/tutorials/operator_generation.ipynb}).
    \item The use of some additional features of the pyTTN package that enable efficient simulation of open quantum system dynamics (including both bosonic and fermionic baths) with approaches based on the wavepacket dynamics and the pseudomode/hierarchical equations of motion (HEOM) methods (found in \protect\path{pyttn/tutorials/oqs/}).
\end{itemize}

This manuscript is structured as follows.  In Sec.~\ref{sec:features}, we outline the key features of the pyTTN software package.  This section begins with a review of the single and multi-set ML-MCTDH/tree tensor network ans\"atze supported by pyTTN, before describing the Hamiltonian representations used within pyTTN, an adaptive bond dimension scheme used to allow for change of bond dimension throughout evolution that generalises existing approaches for MCTDH wavefunction to the multi-layer case, and a series of tools used for simplifying the simulation of open quantum system dynamics.  In Sec.~\ref{sec:numerical_performance}, we discuss the runtime scaling of the time evolution scheme.  In Sec.~\ref{sec:unitary_dynamics} we present example applications to the simulation of the unitary dynamics of the photoexcitation of a 24-mode pyrazine model and the exciton dynamics in a vibronic coupling model for the interface of an $n$-oligothiophene donor-C$_{60}$ fullerene acceptor system, before illustrating the use of open quantum system dynamics methods in Sec.~\ref{sec:oqs_dynamics}, where we consider
the dynamics of the spin-boson model, a set of dissipative spin models, and a set of single impurity Anderson models.  Finally, we conclude and provide an outlook for future developments of this software package in Sec.~\ref{sec:conclusions}. 

\section{Methods \label{sec:features}}

We start this section by briefly reviewing the ML-MCTDH/TTN ansatz and highlighting some of the notation required to explain key features of the pyTTN software package. More detailed overviews of the ansatz can be found in Ref.~\onlinecite{WANG2003,MANTHE2008,VENDRELL2011, doi:10.1080/00268976.2024.2306881, lindoy_mctdh_1}.
\subsection{Wavefunction Ansatz \label{sec:wavefunction_ansatz}}
The ML-MCTDH approach makes use of a hierarchical decomposition of high-dimensional time-dependent wavefunctions, $\left|\Psi(t)\right\rangle $ in terms of
a basis constructed as a direct product of $d^{(1)}$ sets of single-particle
functions (SPFs), $\left|\phi_{i}^{(1,k)}(t)\right\rangle $, as~\citep{WANG2003,MANTHE2008,VENDRELL2011}
\begin{equation}
\left|\Psi(t)\right\rangle =\sum_{i_{1}}^{n_{1}^{(1)}}\dots\sum_{i_{d^{(1)}}}^{n_{d^{(1)}}^{(1)}}A_{i_{1}\dots i_{d^{(1)}}}^{(1)}(t)\bigotimes_{k=1}^{d^{(1)}}\left|\phi_{i_{k}}^{(1,k)}(t)\right\rangle,\label{eq:wavefunction_expansion_spf}
\end{equation}
In Eq. \ref{eq:wavefunction_expansion_spf}, $n_k^1$ for $k\in\{1, 2,\dots, d^(1)\}$ is the number of single particle functions in the $k$-th set that is the number of distinct $\left|\phi_{i}^{(1,k)}(t)\right\rangle $ used in the expansion, and we consider general complex coefficients in the coefficient tensor, that is $A_{i_{1}\dots i_{d^{(1)}}}^{(1)}(t) \in \mathbb{C}$.
Each of these SPFs can be expanded in terms of direct product basis of $d^{(1,k)}$ sets of SPF acting on further-reduced subsets of the physical degrees of freedom, giving rise to new sets of SPFs~\citep{WANG2003,MANTHE2008,VENDRELL2011}
\begin{equation}
\left|\phi_{i}^{(1,k)}(t)\right\rangle =\!\sum_{i_{1}}^{n_{1}^{(1,k)}}\!\dots\!\sum_{i_{d^{(1,k)}}}^{n_{d^{(1,k)}}^{(1,k)}}\!A_{i_{1}\dots i_{d^{(1,k)}}i}^{(1,k)}(t)\bigotimes_{l=1}^{d^{(1,k)}}\left|\phi_{i_{l}}^{(1,k,l)}(t)\right\rangle, \label{eq:spf_recursion}
\end{equation}

Continuing this process of expanding each set of SPFs in terms of direct product basis of sets of SPF acting on further-reduced subsets of the physical degrees of freedom, we obtain a general expression of the form
\begin{equation}
\left|\phi_{i}^{z}(t)\right\rangle =\!\sum_{i_{1}}^{n_{1}^{z}}\!\dots\!\sum_{i_{d^{z}}}^{n_{d^{z}}^{z}}\!A_{i_{1}\dots i_{d^{z}}i}^{z}(t)\bigotimes_{k=1}^{d^{z}}\left|\phi_{i_{k}}^{(z,k)}(t)\right\rangle, \label{eq:spf_recursion}
\end{equation}
where the label $z=(b_{1}\dots b_{l})$, contains the
indices $b_{i}$ that denote the path required to reach this SPF starting
from Eq.\,\ref{eq:wavefunction_expansion_spf}, and $d^{z}$ is
the number of sets of SPFs that form the direct product basis. This recursive process is terminated by expanding lower level SPFs
in terms of a primitive set of basis functions for each of the physical degrees
of freedom of the system
\begin{equation}
\left|\phi_{i_{k}}^{(z,k)}(t)\right\rangle =\Big|\chi_{\alpha}\Big\rangle.
\end{equation}

\begin{figure}
\centering
\includegraphics{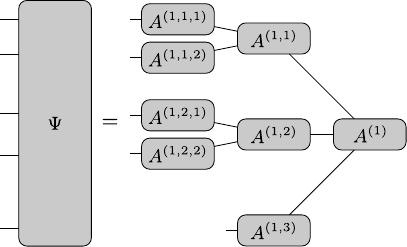}
\caption{\label{fig:mctdh_wavefunction}A diagram representing a ML-MCTDH ansatz for
a five-mode wavefunction, here we consider a tree diagram of maximum degree three, and have $d^{(1)}=3$, $d^{(1,1)}=2$, and $d^{(1, 2)}=2$. Each node of the tree corresponds
to a coefficient tensor, and the lines connecting nodes represent
the contraction over indices common to the two nodes. }
\end{figure}

This hierarchical expansion of the wavefunction gives rise to a tree tensor network (TTN) as illustrated in Fig.\,\ref{fig:mctdh_wavefunction}.  Here the wavefunction coefficient tensor is expressed in terms of a
series of contractions (represented by lines linking nodes in the tensor network) between the coefficient tensors (the nodes in the tensor network), $A_{i_{1}\dots i_{d^{z}}}^{z}(t)$.  The quantity, $n^z_k$,  controls the expressiveness of the ansatz, and is the bond dimension, $\chi$, connecting node $z$ to node $(z,k)$. Throughout this work we will use both the terms ``bond dimension'' and ``number of single-particle functions''.

Working with individual coefficient tensors and SPFs
becomes cumbersome for large tree structures. In order to simplify
expressions it is useful to define the ``single-hole functions'' (SHFs),
$\Psi_{i}^{z}$, associated with each node, $z$, by
    \begin{equation}
    \begin{aligned}[b]\left|\Psi(t)\right\rangle  & \!=\!\sum_{i_{1}}^{n_{1}^{z}}\!\dots\!\sum_{i_{d^{z}}}^{n_{d^{z}}^{z}}\!\sum_{i}^{N^{z}}\!A_{i_{1}\dots i_{d^{z}}i}^{z}(t)\left|\Psi_{i}^{z}(t)\right\rangle\!\otimes\!\left(\!\bigotimes_{k=1}^{d^{z}}\!\left|\phi_{i_{k}}^{(z,k)}(t)\!\right\rangle \!\right)\\
     & =\!\!\!\!\!\sum_{I^{z}}^{n_{1}^{z}\dots n_{d^{z}}^{z}}\!\sum_{i}^{N^{z}}A_{I^{z}i}^{z}(t)\left|\Psi_{i}^{z}(t)\right\rangle\!\otimes\!\left|\Phi_{I^{z}}^{z}(t)\right\rangle ,
    \end{aligned}
    \label{eq:wavefunction_expansion_mctdh-states}
    \end{equation}
where $I^{z}=(i_{1},\dots, i_{d^{z}})$ is a combined
index for all indices involved in the contractions with the children
of node $z$, and $\left|\Phi_{I^{z}}^{z}(t)\right\rangle $
is the direct product of the SPFs associated with the children of
node $z$. Here we have $N^{z}$ SHFs, where $N^{z}$ corresponds to the bond dimension associated with the bond connecting node $z$ to its parent.  The SHFs associated with a node $z$ account for all nodes in the tree tensor network that are not part of the subtree with root $z$. If we append an additional index $i = 1$ to the coefficient tensor of the root node such that $A^1_{1_1\dots i_{d^1} i} \equiv A^1_{1_1\dots i_{d^1}}$, and define the number of SHFs at the root nodes as $N^1 = 1$, from the above expressions we can obtain a recursive definition for the SHFs of a node $(z, k)$ in terms of the coefficient tensor and
SHFs of its parent node, $z$, and the SPFs of the sibling nodes
of node $(z,k)$ (that is all nodes $(z,j)$ with $j\neq k$),
by
\begin{equation}
\begin{aligned}\Big|\Psi_{n}^{(z,k)}(t) &\Big\rangle =  \!\sum_{i_{1}}^{n_{1}^{z}}\!\dots\!\sum_{i_{k-1}}^{n_{k-1}^{z}}\!\sum_{i_{k+1}}^{n_{k+1}^{z}}\!\dots\!\sum_{i_{d^{z}}}^{n_{d^{z}}^{z}}\sum_{m=1}^{N^{z}}\left|\Psi_{m}^{z}(t) \right\rangle \cdot \\
 & A_{i_{1}\dots i_{k-1}ni_{k+1}\dots i_{d^{z}}m}^{z}\left(\bigotimes_{\substack{j=1\\
{j\neq k}
}
}^{d^{z}}\left|\phi_{i_{j}}^{(z,j)}(t)
\right\rangle \right)\\
&=  \!\!\sum_{I^{z\setminus k}}\!\sum_{i}^{n^{z}}A_{I^{z\setminus k;n}i}^{z}(t)\left|\Psi_{i}^{z}(t)\right\rangle\!\otimes\!\left|\Phi_{I^{z\setminus k}}^{z\setminus k}(t)\right\rangle ,
\end{aligned}
\label{eq:shf_recursive_definition}
\end{equation}
where $I^{z \setminus k }=(i_{1}, \dots, i_{k-1}, i_{k+1}\dots, i_{d^{z}})$ is the set of indices connecting the node $(z)$ to all of its children except for node $(z, k)$, $I^{z \setminus k; n}=(i_{1}, \dots, i_{k-1}, n, i_{k+1}\dots, i_{d^{z}})$, and the sum over $I^{z \setminus k }$ corresponds to the sum over all values of the indices in this multi-index.  The choice of coefficient tensors for a given wavefunction is not unique, and it is necessary to impose a gauge choice to fix the values of the tensor network. In this work, and within the pyTTN library in general, we will typically employ a  canonical gauge~\cite{Kloss2020} representation of the network when updating site tensors.

These SHFs and SPFs are useful for recursively defining Hamiltonian matrix elements that are required in the evolution schemes used for time evolution of TTNs.  Additionally, this notation will be used  below when discussing subspace expansion schemes for adaptively expanding the number of single-particle functions and single hole functions (or alternatively the bond dimension of the tensor network).

In addition to allowing for the use of the ML-MCTDH ansatz, pyTTN also supports the use of the multi-set ansatz
\begin{equation}
    \ket{\Psi(t)} = \sum_{\alpha} \ket{\alpha} \ket{\Psi^{\alpha}(t)},
\end{equation}
for an arbitrary (but same for all $\alpha$) tree structure for the $\ket{\Psi^{\alpha}(t)}$.  Within this representation, the wavefunction is partitioned into system, $\alpha$, and environment degrees of freedom.  For each system state, a TTN with a different set of coefficient tensors is employed for the environment.  This representation has found success in treating problems in nonadiabatic dynamics~\cite{RAAB1999} and exciton dynamics~\cite{PhysRevLett.123.126601, 10.1063/1.5051498} where different electronic configurations may have vastly different environment dynamics.

\subsection{Symbolic Operator Generation \label{sec:op_gen}}
A key aim of the pyTTN package is to provide an easy-to-use interface for defining operators acting on high-dimensional spaces.  Within pyTTN, we have provided a set of tools for symbolic operator definition as well as tools for converting this representation to formats necessary for efficient implementation of algorithms on tree tensor networks.  In this section we describe the underlying Hamiltonian representations, and briefly outline the approach used to convert from the symbolic representation to these formats.  Details on the Hamiltonian generation procedure can be found in the tutorial \protect\path{pyttn/tutorials/operator_generation.ipynb}.
\subsubsection{Sum-of-Product Operator Form}
In order to efficiently implement the ML-MCTDH equations of motion, it is necessary to have a representation of the Hamiltonian that is compatible with the tree tensor network structure of the Hamiltonian.  The standard representation that is employed in the vast majority of ML-MCTDH calculations is the sum-of-product (SOP) operator representation.  In this representation, the Hamiltonian is given as a sum over $N_{\mathrm{prod}}$ operators each of which is expressed as a direct product of operators ($\hat{h}_{\alpha,r}$) that each act on a single physical degree of freedom
\begin{equation}
\hat{{H}}=\sum_{r=1}^{N_{\mathrm{prod}}}c_r \bigotimes_{\alpha=1}^{D}\hat{h}_{\alpha,r}.\label{eq:sop_ham}
\end{equation}
where $D$ is the total number of physical degrees of freedom in the problem and each coefficient $c_r\in\mathbb{C}$.
At each node $z$, we can make use of this representation
to write the Hamiltonian operator as a sum over a set of $N_{\mathrm{prod}}$
operators, each of which is expressed as a direct product of operators
acting on the physical degrees of SPFs and SHFs of this node
\begin{equation}
\hat{{H}}=\sum_{r=1}^{N_{\mathrm{prod}}}c_r\hat{h}_{r}^{z} \hat{H}_{r}^{z}.\label{eq:sop_node_naive}
\end{equation}
Here $\hat{h}_{r}^{z}$
and $\hat{H}_{r}^{z}$ are operators acting on the physical degrees
of freedom associated with the SPFs and SHFs of node $z$, respectively.
We refer to these operators as the single-particle operators (SPOs) and mean-field operators (MFOs), respectively.
These operators are simply direct products of the operators $\hat{h}_{\alpha,r}$,
and can be constructed recursively.  For the SPOs of a given node, $z$, we have
\begin{equation}
\hat{h}_{r}^{z}=\bigotimes_{k}\hat{h}_{r}^{(z,k)},\label{eq:spf_hamiltonian_op}
\end{equation}
where $\hat{h}_{r}^{(z,k)}$ is a SPF Hamiltonian operator
associated with the $k$-th child of node $z$.  The MFOs for node $z$ can be constructed recursively from the mean field operators of its parent node and the single-particle operators of its sibling nodes,
\begin{equation}
\hat{H}_{r}^{(z,k)}=\hat{H}_{r}^{z}\bigotimes_{j\neq k}\hat{h}_{r}^{(z,j)}.\label{eq:shf_hamiltonian_op}
\end{equation}

\subsubsection{Hierarchical Sum-of-Product Operator Form}
If two (or more) of the operators $\hat{h}_{r}^{z}$ are identical or if any
two (or more) operators $\hat{H}_{r}^{z}$ are identical, then it
is possible to simplify Eq.~\ref{eq:sop_node_naive}. In order to do this we start by introducing the notation $\hat{h}_{c;i}^{z}$
($\hat{H}_{c;i}^{z}$) to denote an operator acting on the SPF
(SHF) degrees of freedom that is common to the set of
terms with indices $\boldsymbol{r}_{i}^{h;z}$($\boldsymbol{r}_{i}^{H;z})$ in the sum in Eq.~\ref{eq:sop_node_naive}.  We suppose that we have $I^{h;z}$ and $I^{H;z}$ common SPF and SHF operators, respectively.
We also introduce the notation $\boldsymbol{r}^{ind;z}$ to denote
the indices of operators that do not contain any of these common operators.
Using this notation, we may rewrite Eq.~\ref{eq:sop_node_naive} as
\begin{equation}
\begin{split}
\hat{H}=&\sum_{r\in\boldsymbol{r}^{ind;z}}c_r\hat{h}_{r}^{z}\hat{H}_{r}^{z}+\sum_{i=1}^{I^{h;z}}\hat{h}_{c;i}^{z}\sum_{r\in\boldsymbol{r}_{i}^{h;z}}c_r\hat{H}_{r}^{z} \\ &+\sum_{i=1}^{I^{H;z}}\hat{H}_{c;i}^{z}\sum_{r\in\boldsymbol{r}_{i}^{H;z}}c_r\hat{h}_{r}^{z}.\label{eq:sop_node_naive-1}
\end{split}
\end{equation}
 We next introduce the notation
\begin{equation}
\hat{h}_{s;i}^{z}=\sum_{r\in\boldsymbol{r}_{i}^{H;z}}c_r \hat{h}_{r}^{z}\ \ \ \ \ \text{and}\ \ \ \ \ \hat{H}_{s;i}^{z}=\sum_{r\in\boldsymbol{r}_{i}^{h;z}}c_r\hat{H}_{r}^{z}
\end{equation}
 for the sums of operators that act on the physical degrees
of freedom associated with the SPFs and SHFs, respectively. Now Eq.~\ref{eq:sop_node_naive-1}
becomes
\begin{equation}
\hat{H}=\underbrace{\sum_{r\in\boldsymbol{r}^{ind;z}}c_r \hat{h}_{r}^{z}\hat{H}_{r}^{z}}_{\text{(A)}}+\underbrace{\sum_{i=1}^{I^{h;z}}\hat{h}_{c;i}^{z}\hat{H}_{s;i}^{z}}_{\text{(B)}}+\underbrace{\sum_{i=1}^{I^{H;z}}\hat{h}_{s;i}^{z}\hat{H}_{c;i}^{z}}_{\text{(C)}}.\label{eq:sop_node}
\end{equation}
 This expression is still in the form of a sum-of-product operator representation, however now with a reduced number of terms due to the compression of sums of operators into a single object. The terms in (A) are simply standard SOP terms, the terms in (B) involve a compressed set of mean-field operators, and the terms in (C) involve a compressed set of single-particle operators. This hierarchical Sum-of-Product (hSOP) operator form, is the form used by the multilayer potfit method~\cite{10.1063/1.4856135, OTTO2018116} for representing general potential energy surface.

 In contrast to the SOP representation where $\sim N_{prod}$ SPOs and MFOs need to be stored at each node in the ML-MCTDH wavefunction (with smaller numbers being possible by exploiting optimisations involving the presence of identity SPOs and MFOs in the Hamiltonian), at each node the hSOP representation requires storage of a single SPO and MFO for each term in the sums in Eq.~\ref{eq:sop_node}.  Depending on the extent of compression, the use of the hSOP representation can significantly reduce computer time and memory requirements compared to standard SOP representations. We note, however, that while the memory requirements scale as the number of terms in Eq.~\ref{eq:sop_node}, the number of operations scales as the total number of terms in Eq.~\ref{eq:sop_node_naive-1}.  This representation corresponds to a sparse tensor version of the Tree Tensor Network Operator (TTNO)~\cite{10.1063/5.0218773}, in which additional information is stored to facilitate efficient sparse tensor contractions both up and down the tree structure.  Whether this provides performance advantages over the dense operator TTNO form likely depends on the details of the operator considered and is beyond the scope of this work.

 For efficient ML-MCTDH calculations it is necessary to be able to recursively construct each SPF and SHF Hamiltonian operators recursively from nearby nodes in the tree (e.g. analogous to Eqs. \ref{eq:spf_hamiltonian_op} and \ref{eq:shf_hamiltonian_op}).  Such a construction has been presented in Supplementary Information Sec. IV of reference \onlinecite{lindoy_mctdh_2}.  This approach allows for the recursive definition of the hSOP representation given a user-specified partitioning of modes. In recent years several schemes have been proposed for automatic partitioning of terms in the Hamiltonian suitable for such tree representations~\cite{milbradt2024pytreenetpythonlibraryeasy,10.1063/5.0218773}. pyTTN supports the use of a bipartite graph based implementation for the automatic construction~\cite{10.1063/5.0018149, 10.1063/5.0218773} of the hierarchical sum-of-product operator form for both the single and multi-set ML-MCTDH ansatz.  In order to construct the sparsity structure of operators acting down the tree present in the hSOP representation, the implementation provided in pyTTN adds an additional step to the approach; a pre-order traversal of the tree structure to recursively define compressed mean-field operators. Making use of the hierarchical SOP representation enables several additional optimisations to be performed compared to the dense TTNO representation.  Namely, this allows for identification of any SPOs and MFOs that are exactly the identity operator, allowing for such terms to be treated using a copy operation rather than tensor contractions.  The extent to which this optimisation impacts the performance of the approach depends strongly on the number of terms in the sum.  Recent work, has demonstrated that the use of symbolic Gaussian elimination throughout the operator compression step can improve upon the compression obtained using the Bipartite graph approach for operator rational function valued coefficients~\cite{cakır2025optimalsymbolicconstructionmatrix}. The implementation of this approach within pyTTN is left as future work.

Within the pyTTN package, the automatic construction of SOP and hSOP representations of operators is supported through the use of a symbolic sum-of-product operator representation.  The package supports a set of standard spin, boson, fermion, and Pauli operators.  The use of general operators is supported through optional user-specified operator dictionaries.  Additionally, pyTTN supports the use of time-dependent Hamiltonian coefficients. Further details of these features can be found in the tutorial \protect\path{pyttn/tutorials/operation_generation.ipynb}. In addition, we support symbolic generation of multi-set SOP and hSOP operators.

\subsection{Variational Updating Schemes \label{sec:updating}}
Within pyTTN we make use of (time-dependent) variational approaches for updating the coefficient tensors within the TTNs in both the single and multi-set forms.  We use the projector splitting integrator~\cite{Lubich2014a,Kloss2017a, Bauernfeind2020,Kloss2020,Ceruti2021, lindoy_thesis, lindoy_mctdh_1, lindoy_mctdh_2} (PSI) for time evolution of the coefficient tensors within the network.

Within this scheme, the problem of evolving all coefficient tensors in the network is reduced to a series of local coefficient tensor updates, where at each coefficient tensor we solve the (time-dependent) Schr\"odinger equation projected into the basis defined by the rest of the tensor network.  In order to do this, we make use of the sum-of-product (and can readily generalise to the hierarchical sum-of-product) form of the Hamiltonian associated with a node $z$ of the form
\begin{equation}
\hat{{H}}=\sum_{r=1}^{N_{\mathrm{prod}}}c_r\hat{h}_{r}^{z} \hat{H}_{r}^{z}.
\end{equation}

Projecting the operators $\hat{h}_r^z$ and $\hat{H}_r^z$ onto the  single-particle functions and single-hole functions of the node, respectively, we obtain a set of single-particle operator matrices ($\boldsymbol{{h}}_{r}^{z}(t)$) and mean-field operator matrices ($\boldsymbol{{H}}_{r}^{z}(t)$) defined by
\begin{align}
\left[\boldsymbol{{h}}_{r}^{z}(t)\right]_{IJ} & =\bra{\Phi_{I}^{z}(t)}\hat{h}_{r}^{z}\ket{\Phi_{J}^{z}(t)},\\
\left[\boldsymbol{{H}}_{r}^{z}(t)\right]_{ij} & =\bra{\Psi_{i}^{z}(t)}\hat{H}_{r}^{z}\ket{\Psi_{j}^{z}(t)}.
\end{align}

The (time-dependent) Schr\"odinger equation projected into the basis defined by the rest of the tensor network, is then given by
\begin{equation}
\begin{aligned}\dot{\boldsymbol{A}}^{z}(t)= & -\frac{{i}}{\hbar}\sum_{r} c_r \boldsymbol{{h}}_{r}^{z}(t){{\boldsymbol{{A}}}}^{z}(t){{\boldsymbol{{H}}}}_{r}^{z}(t),\end{aligned}
\label{eq:psi_eom_1}
\end{equation}
where $\boldsymbol{A}^z$ is a matrisation of the tensor $A^z$ obtained when flattening the composite index $I^z$ into a single composite index.

As this scheme involves updates of individual site tensors in the network, it is an example of a single-site update scheme.  pyTTN also supports analogous algorithms  for the evaluation of ground state properties, that is the density matrix renormalisation group (DMRG) method.  For further details of the TDVP and DMRG approaches for TTNs see Refs.~\onlinecite{Bauernfeind2020, Larsson2019,lindoy_mctdh_2}.

\subsection{Subspace Expansion using Two-Site Energy Variance \label{sec:subspace}}
The standard projector splitting integrator for the (ML-)MCTDH ansatz is a single-site integration scheme, defining a scheme for updating individual site tensors while keeping their bond dimensions constant.  Within the context of MPSs it is standard to use two-site algorithms, in which two sites of the tensor network are contracted together and the resulting two-site object is updated, before being decomposed to reform two one-site objects with a potentially altered bond dimension. Such approaches can readily be extended to the ML-MCTDH ansatz~\cite{Bauernfeind2020}; however, in contrast to the MPS case where the scaling of computational cost with respect to bond dimension is the same for one and two-site algorithms ($\mathcal{O}(\chi^3)$), two-site algorithms are considerably more expensive for general tree structures. If we consider a tree where each node has $D$ children (and constant bond dimension on all bonds), the cost of a single-site update scales as $\mathcal{O}(\chi^{D+2})$, whereas a two-site update scales as $\mathcal{O}(\chi^{2D+1})$.  As such, these schemes are only practical for sufficiently small bond dimension.

In recent years several subspace expansion based techniques have been proposed that aim to allow for adaptation of bond dimension without requiring direct evolution of two-site objects~\citep{PhysRevB.104.214302,doi:10.1021/jacsau.1c00474,PhysRevB.97.045145,MENDIVETAPIA2017113,Larsson2017,Ceruti2022, https://doi.org/10.48550/arxiv.2201.10291, MendiveTapia2020,PhysRevLett.133.026401}. These approaches differ in the choice of metric used for determining when to expand the bond dimension and the choice of new vectors to include when doing so.  Of particular interest here is the approach developed by Mendive-Tapia and Meyer for the spawning of SPFs in MCTDH~\citep{MendiveTapia2020}. This approach makes use of a local-in-time error estimate~\cite{PhysRevLett.124.150601} to determine optimal basis functions to add to the ML-MCTDH wavefunction to minimise the error in evolution.  Central to this approach is the evaluation of eigenvalues of the two-site energy variance object that are not captured by the current one-site bases.  While this approach provides a controllable approach for expanding bond dimension in the MCTDH case, the explicit formation of the two-site energy variance leads to this approach scaling as a two-site algorithm.  In what follows, we present a generalisation of the approach of Mendive-Tapia and Meyer to enable the spawning of SPFs and SHFs, or equivalently the expansion of bond dimensions, in general TTN wavefunctions.

Consider two wavefunctions $\ket{\Psi_e(t)}$ and $\ket{\Psi_{M}(t)}$ corresponding to the exact solution of the Schr\"odinger equation, and the approximate ML-MCTDH solution obtained with a given set of bond dimensions throughout the tensor.
\begin{equation}
    \begin{aligned}[b]\ket{\Psi_{M}(t_0)}  &=\!\!\!\!\!\sum_{I^{z}}^{n_{1}^{z}\dots n_{d^{z}}^{z}}\!\sum_{i}^{n^{z}}A_{I^{z}i}^{z}(t_0)\left|\Psi_{i}^{z}(t_0)\right\rangle\!\otimes\!\left|\Phi_{I^{z}}^{z}(t_0)\right\rangle ,
    \end{aligned}
    \end{equation}

We further assume that at $t=t_0$, the two wavefunctions represent the same wavefunction
\begin{equation}
    \ket{\Psi_e(t_0)} = \ket{\Psi_{M}(t_0)}.
\end{equation}
Within the framework of the Dirac-Frenkel Variational principle, the ML-MCTDH equations of motion are obtained by finding the time evolution of the coefficient tensors of $\ket{\Psi_{M}(t_0)}$ that minimize the difference in the time-derivative of the exact and ML-MCTDH wavefunction
\begin{equation}
E = \left\|\ket{\dot{\Psi}_e(t)} - \ket{\dot{\Psi}_{M}(t)}\right\|^2.
\end{equation}
We next consider an augment wavefunction $\ket{\Psi_{A}(t_0)}$ obtained from $\ket{\Psi_{M}(t_0)}$ by expanding the bond connecting two nodes $(z)$ and $(z, k)$, while leaving the value of the wavefunction unchanged, that is $\ket{\Psi_{A}(t_0)} = \ket{\Psi_{M}(t_0)}$.  This can be done by augmenting the coefficient tensor associated with node $(z)$ and defining a new set of single-particle functions $\ket{\tilde{\phi}_i^{(z, k)}}$ associated with node $(z, k)$ that are mutually orthogonal to the existing single-particle functions $\ket{\phi^{(z, k)}}$ associated with node $(z, k)$.  This orthogonality constraint may be written as
\begin{equation}
    P^{(z, k)} \ket{\tilde{\phi}^{(z, k)}} =0,
\end{equation}
where
\begin{equation}
    P^{(z, k)} = \sum_i \ket{\phi_i^{(z, k)}}\bra{\phi_i^{(z, k)}}
\end{equation}
is the projector onto the orthonormal set of single-particle function states of node $(z, k)$. The augmented wavefunction can now be written as
\begin{equation}
    \begin{aligned}[b]&\ket{\Psi_{A}(t_0)}  =\sum_{I^{z}, i}A_{I^{z}i}^{z}(t_0)\left|\Psi_{i}^{z}(t_0)\right\rangle\!\otimes\!\left|\Phi_{I^{z}}^{z}(t_0)\right\rangle \\
    &+ \sum_{m, I^{z \setminus k}, i}\tilde{A}_{I^{z\setminus k; m}i}^{z}(t_0)\left|\Psi_{i}^{z}(t_0)\right\rangle\!\otimes\!\left|\Phi_{I^{z\setminus k}}^{z\setminus k}(t_0)\right\rangle \!\otimes\! \left|\tilde{\phi}_{m}^{(z, k)}(t_0)\right\rangle \label{eq:augmented_mctdh},
    \end{aligned}
    \end{equation}
where when working in the mixed-canonical gauge with orthogonality centre at $(z)$, the set of states
\begin{equation}
    \ket{\Theta_{iI^{z\setminus k}}^{(z, k)}(t_0)} = \left|\Psi_{i}^{z}(t_0)\right\rangle\!\otimes\!\left|\Phi_{I^{z\setminus k}}^{z\setminus k}(t_0)\right\rangle,
\end{equation}
represent an orthonormal basis for the degrees of freedom that are not descendents of the node $(z, k)$.  Expressing the augmented wavefunction in terms of these functions, we have
\begin{equation}
    \begin{aligned}[b]\ket{\Psi_{A}(t_0)}&  =
    \sum_{I^{z \setminus k}, i}\bigg[
    \sum_{i_k}A_{I^{z}i}^{z} (t_0)\ket{\Theta_{iI^{z\setminus k}}^{(z, k)}(t_0)}\otimes \left|\phi_{i_k}^{(z, k)}(t_0)\right\rangle\\
    &+ \sum_m \tilde{A}_{I^{z\setminus k; m}i}^{z}(t_0)\ket{\Theta_{iI^{z\setminus k}}^{(z, k)}(t_0)}\otimes \left|\tilde{\phi}_{m}^{(z, k)}(t_0)\right\rangle \bigg] \label{eq:augmented_mctdh},
    \end{aligned}
    \end{equation}
where in order to ensure equality of the augmented and original wavefunctions we required the augmenting coefficients $\tilde{A}_{I^{z\setminus k; m}i}^{z}(t_0)$ to initially be zero.

Generalising the discussion in Ref.~\onlinecite{MendiveTapia2020}, it is straightforward to show that the error in the time-derivatives of this augmented ML-MCTDH wavefunction can be written as
\begin{equation}
\begin{split}
\tilde{E} &= \big\|\ket{\dot{\Psi}_e(t_0)} - \ket{\dot{\Psi}_{A}(t_0)}\big\|^2 \\
          &= \big\|\ket{\dot{\Psi}_e} - \ket{\dot{\Psi}_{M}}\big\|^2 \\&- \!\!\sum_{i, m, I^{z \setminus k}} \!\!\!\!\!\braket{\Theta_{iI^{z\setminus k}}^{(z, k)}\tilde{\phi}_{m}^{(z, k)}}{\dot{\Psi}_e\!-\!\dot{\Psi}_{M}\!} \!\!\braket{\!\dot{\Psi}_e\!-\!\dot{\Psi}_{M}}{\Theta_{iI^{z\setminus k}}^{(z, k)}\tilde{\phi}_{m}^{(z, k)}},
\end{split}
\end{equation}
that is, the error in the time-derivatives of the original ML-MCTDH wavefunction minus the contribution to the error that lies in the space spanned by these augmented basis functions. Enforcing the orthogonality of the new single-particle functions with respect to the original single-particle functions, the difference in error between the original wavefunction and the augmented wavefunction may be obtained as
\begin{equation}
    \Delta_d^{ (z, k)} = E-\tilde{E} = \sum_m \bra{\tilde{\phi}_{m}^{(z, k)}} \hat{\Delta}_d^{(z, k)}\ket{\tilde{\phi}_{m}^{(z, k)}},
\end{equation}
where we have introduced the operator
\begin{equation}
\begin{split}
\hat{\Delta}^{(z, k)} = \big(1-& P^{(z, k)} \big)  \!\!\sum_{i, I^{z \setminus k}} \!\!\!\braket{\Theta_{iI^{z\setminus k}}^{(z, k)}}{\dot{\Psi}_e\!-\!\dot{\Psi}_{M}\!} \\&\braket{\!\dot{\Psi}_e\!-\!\dot{\Psi}_{M}}{\Theta_{iI^{z\setminus k}}^{(z, k)}}  \big(1- P^{(z, k)} \big).
\end{split}
\end{equation}

This operator provides a convenient means for choosin optimal single-particle functions to add to the wavefunction that best capture the remaining error.  Namely, the eigenfunctions of this operator corresponding to the largest eigenvalues define a set of new unoccupied SPFs that have the most significant impact on reducing the error in evolution.  By adding all eigenvectors with eigenvalue satisfying some spawning criteria we can improve our wavefunction representation in a controllable way.  Within pyTTN, several spawning criteria are supported, but by default the adaptive subspace approaches will introduce new SPFs (and as discussed below SHFs) corresponding to eigenvectors of the matrix $\hat{\Delta}^{(z, k)}$, whenever the corresponding eigenvalue, $\lambda_s$, satisfies the condition
\begin{equation}
    c\sqrt{\lambda_s} \geq \epsilon_S,
\end{equation}
where $\epsilon_S$ is some user defined spawning tolerance and $c$ depends on the type of algorithm being considered ($c=1$ for DMRG calculations or $c=\delta t$, the integration timestep, for TDVP calculations).

In practice, there may be situations in which this metric fails to capture important search directions, and in such cases more global metrics may be preferable~\cite{MANTHE2018279,Yang2020}, although typically come at increased computation cost. In particular, care must be taken when considering small initial bond dimensions, in which case local (single and two-site) metrics may not be able to accurately capture the optimal expansion directions as they restrict the choice of directions to those captured by the two-site tangent space.  To partially alleviate these issues, the use of larger bond dimensions than those strictly required to represent the state (by including unoccupied single-particle functions) enables a larger search space to potentially identify the optimal expansion direction.  So in all calculations here, even when starting with a product state considered an initial TTN with bond dimension throughout the network of $\mathcal{O}(10)$ is used.

Due to the symmetry between SPFs and SHFs when using the projector splitting integrator gauge choice, an analogous expression and set of operators can be obtained for finding optimal SHFs.  Within pyTTN, the default subspace expansion approach employs both expansions of the wavefunction at node $(z, k)$.  Evaluation of these objects can be performed using a generalisation of the scheme described in Ref.~\onlinecite{MendiveTapia2020}.  However, a direct generalisation to the multilayer case requires the formation of two-site tensor objects, which for high degree trees has high memory and computational time overhead.  As we are only interested in the evaluation of the dominant eigenvalues, iterative linear algebra-based approaches are a potentially useful alternative approach.  Additionally, due to the structure of these operators, we can obtain the action of these operators onto a set of SPFs, at considerably reduced cost (comparable to single-site algorithms).  Discussion of the numerical cost associated with this approach is presented in Sec.~\ref{Sec:subspace_performance}.

Within pyTTN, we have implemented a Krylov subspace expansion-based scheme for the evaluation of the dominant eigenvalues and eigenvectors of these energy error objects.  This subspace expansion scheme is easily integrated within the standard projector splitting integrator (see Ref.~\onlinecite{lindoy_mctdh_2} for implementation details of the PSI approach), by adding a subspace expansion step for each bond whenever the orthogonality centre is shifted across this bond.  Given the symmetry between SPFs and SHFs when using the mixed-canonical gauge employed in the PSI~\cite{WEIKE2021,lindoy_mctdh_1}, an analogous metric can be obtained for spawning SHFs by considering an expansion of the form given in Eq.~\ref{eq:augmented_mctdh}, except at node $(z, k)$ for the case where it is the orthogonality centre. In doing this we obtain an approach for spawning ``optimal'' SHFs when traversing up the tree, in addition to the scheme outlined above for spawning SPFs when traversing down the tree.  Within the subspace expansion scheme implemented within pyTTN, each bond is traversed twice throughout an integration step, once up and once down, at which stage spawning of SPFs and SHFs is attempted depending on the direction of traversal.

In the current release, the two-site energy variance-based bond dimension expansion algorithm has been implemented for single-set ML-MCTDH wavefunctions.  An extension of the two-site energy variance-based subspace expansion algorithm to the multi-set ansatz is planned for future work.

\subsection{Open Quantum System Simulation Tools}
In addition to the tools described above, pyTTN implements several features to help facilitate simulations of open quantum system dynamics, in which a quantum system (with interactions described by a system Hamiltonian $\hat{H}_S$) interacts with a continuum bath.  Specifically, we consider the case in which the system is linearly coupled (through a system operator of the form $\hat{S}$) to a continuous set of bath operators providing an additional Hamiltonian contribution of the form
\begin{equation}
    \hat{H}_{SB}= \int_0^\infty\!\!\left(  g(\varepsilon)\hat{S} \left(\hat{a}^\dagger(\varepsilon) + \hat{a}(\varepsilon) \right)  + \omega(\varepsilon) \hat{a}^\dagger(\varepsilon) \hat{a}(\varepsilon) \right) \mathrm{d} \varepsilon, \label{eq:oqs_continuum}
\end{equation}
where $\hat{a}(\varepsilon)$ is the annihilation operator for a bosonic bath mode at energy $\varepsilon$, $\hat{a}^{\dagger}(\varepsilon)$ the creation operator for a bath mode at energy $\varepsilon$, $g(\varepsilon)$ the coupling distribution, and $\omega(\varepsilon)$ the bath frequency distribution.

For the case of a Gaussian initial configuration of the bath, the influence of the bath on the system is entirely characterised by its spectral density, $J(\omega)$, and the initial bath state.  Several approaches have been presented for handling the dynamics of such models.  These include unitary-dynamics based approaches that consider a discretised representation of the bath in various geometries~\cite{PhysRevA.105.032406,WANG2003,10.1063/1.4866769,PhysRevLett.105.050404,Rosenbach_2016,PhysRevLett.123.090402,lacroix2024mpsdynamicsjltensornetworksimulations,lindoy_mctdh_2,10.1063/5.0218773}; Time-Evolving Matrix Product Operators (TEMPO)~\cite{Strathearn2018} and related methodologies~\cite{Jorgensen2019, Gribben2022, FowlerWright2022, PhysRevResearch.5.033078,Cygorek2024} for handling the continuum bath limit; and Hierarchical Equation of Motion (HEOM)~\cite{TanimuraJPSJ1989,IshizakiJPSJ2005,TanimuraJCP2020,10.1063/1.5026753, 10.1063/5.0050720,10.1063/5.0153870,10.1063/1.5099416, 10.1063/5.0200410, PhysRevLett.129.230601,Mangaud2023} and pseudomode methods~\cite{Lambert2019,PhysRevResearch.2.043058,park2024quasilindbladpseudomodetheoryopen} that employ decompositions of the bath in terms of a set of dissipative bath modes.

In this section we discuss the tools included in pyTTN to simplify the setup of tensor network based simulations of unitary dynamics, and pseudomode/HEOM based approaches for the simulation of open quantum systems with bosonic baths.  Details on the use of these tools can be found in the tutorials on open quantum systems \protect\path{pyttn/tutorials/oqs/}.

\subsubsection{Unitary Dynamics: Bath Discretisation \label{sec:bath_discretisation}}
In order to apply unitary dynamics methods for the simulation of open quantum systems, it is first necessary to approximate the influence of the continuous bath on the system, by considering a discretised bath containing $N$ bosonic modes.  In doing this, we are interested in obtaining an approximate discretised system-bath coupling Hamiltonian of the form
\begin{equation}
\hat{H}_{SB}=\hat{S}\sum_{k=1}^Ng_{k}(\hat{{a}}_{k}^{\dagger}+\hat{{a}}_{k})+\sum_{k=1}^N\omega_{k}\hat{{a}}_{k}^{\dagger}\hat{{a}}_k, \label{eq:oqs_discretised_hamiltonian}
\end{equation}
where the discretised form of coupling constants and bath frequencies are related to the spectral density by~\citep{Leggett1987,Weiss2012}
\begin{equation}
\begin{aligned}J(\omega) = \pi \sum_{k=1}^Ng_{k}^{2}\delta(\omega-\omega_{k})\end{aligned}. \label{eq:discrete_spec}
\end{equation}

Two discretisation strategies are provided by the pyTTN package, namely a density of frequencies based discretisation that is typically used within the ML-MCTDH community~\cite{10.1063/1.1385561}, and orthogonal polynomial based discretisation schemes~\cite{deVega2015}, that are commonly used in schemes such as TEDOPA.

The first is based on the use of a (user specified) density of frequencies, $\rho(\omega)$, in which the frequencies of the $N$ bosons chosen to discretise the bath are determined using the relation~\cite{wang2001c}
\begin{equation}
    \int_{\omega_{\mathrm{min}}}^{\omega_k} \rho(\omega)\mathrm{d} \omega = k.
\end{equation}
Once the frequencies of the bath have been obtained, we can obtain the coupling constants as
\begin{equation}
 g_k^2 = \frac{1}{\pi} \frac{J(\omega_k)}{\rho(\omega_k)}.
\end{equation}
The accuracy of this discretisation strategy depends significantly on the choice of the density of frequencies.  A common choice employed within the ML-MCTDH community (and the default within pyTTN) is~\cite{WANG2003,10.1063/1.1385561,Wang2019}
\begin{equation}
    \rho(\omega) = \frac{J(\omega)}{\omega},
\end{equation}
which gives rise to a discretisation in which each mode accounts for the same contribution to the total bath reorganisation energy.

The second is an orthogonal polynomial based discretisation scheme, in which the central object for constructing the discretisation are the orthonormal polynomials, $\pi_j(\omega)$, $j=0, 1, \dots, N-1$, that satisfy the orthogonality constraint
\begin{equation}
    \int_{\omega_{\mathrm{min}}}^{\omega_\mathrm{max}} J(\omega) \pi_i(\omega) \pi_j(\omega) \mathrm{d}\omega = \delta_{ij}.
\end{equation}
The coefficients for the discretised Hamiltonian can be obtained by constructing the gaussian-quadrature rule associated with these polynomials, which within pyTTN is performed using the modified Chebyshev method~\citep{doi:10.1137/0903018,10.1145/174603.174605}.
Here $\omega_{\mathrm{min}}$ and $\omega_{\mathrm{max}}$ are the minimum and maximum bath frequencies to be included in the problem, and at $T=0$ $\omega_{\mathrm{min}}=0$. From this we construct the $N$-point Gaussian quadrature rule
\begin{equation}
    \int_{\omega_{\mathrm{min}}}^{\omega_\mathrm{max}} F(\omega) J(\omega) \mathrm{d}\omega \approx \sum_{i=1}^{N} \alpha_i F(\omega_i),
\end{equation}
which is exact whenever $F(\omega)$ is a polynomial of degree less than $2N-1$, using the Golub-Walsch algorithm~\citep{10.2307/2004418}. This procedure provides a discretised bath with spectral density given by
\begin{equation}
    J(\omega) \approx \sum_{i=1}^{N} \alpha_i \delta(\omega - \omega_i),
\end{equation}
giving us a discrete set of bath frequencies, $\omega_i$, and coupling constants
\begin{equation}
    g_i = \sqrt{\frac{\alpha_i}{\pi}},
\end{equation}
that exactly captures integrals of the form
\begin{equation}
    \int_{\omega_{\mathrm{min}}}^{\omega_\mathrm{max}} P_n(\omega) J(\omega) \mathrm{d}\omega ,
\end{equation}
for polynomials $P_n$ of order $n \leq 2N-1$, and consequently accurately captures bath correlation function up to the time at which a $2N-1$-th order polynomial approximation of the trigonometric functions is accurate.

Finite-temperature bath initial conditions can be included within this scheme through the use of a temperature-dependent spectral density~\cite{PhysRevLett.123.090402}
\begin{equation}
    J_{\beta}(\omega) = \frac{1}{2}J(\omega)\left(1+\coth\left(\frac{\beta\omega}{2}\right)\right). \label{eq:effective_spec_dens}
\end{equation}
Discretisation of this spectral density gives a discrete Hamiltonian containing negative frequency bath modes, where the dynamics of a bath initially at zero temperature reproduces the
finite-temperature dynamics of the original model.

So far we have considered a star-topology representation of the system-bath Hamiltonian, in which each of the (non-interacting) bosonic modes couples directly to the system degrees of freedom only.  The specific coupling topology is arbitrary and can be altered by applying global unitary transformations of the bath modes that lead to alternative, unitarily equivalent representations.  While these approaches are equivalent when solved in exact arithmetic, the accuracy and efficiency of numerical schemes for performing dynamics can depend significantly on the choice of coupling topology.
In particular, the entanglement structure---and in turn the numerical effort associated with tensor network simulations of the model---can depend significantly on the specific form of the Hamiltonian considered. pyTTN provides direct support for two additional Hamiltonian coupling topologies, namely the chain-topology central to the TEDOPA method~\cite{10.1063/1.4866769,PhysRevLett.105.050404,Rosenbach_2016,PhysRevLett.123.090402,lacroix2024mpsdynamicsjltensornetworksimulations}, and a more recent interaction-picture chain representation form that gives rise to time-dependent coupling constants~\cite{PhysRevA.105.032406}.
In principle, the discussion presented in this section can be generalised to the case of Fermionic baths, an example of this is shown in Sec.\ref{sec:siam}.
\subsubsection{Unitary Dynamics: Bath Transformations}
The system-bath coupling Hamiltonian given in Eq.~\ref{eq:oqs_discretised_hamiltonian} has a star-topology. The use of unitary transformations on the bosonic modes enables transformations of the Hamiltonian into different forms. In particular, tridiagonalisation of the coupling matrix leads to a chain form of the system-bath coupling Hamiltonian~\cite{10.1063/1.4866769, 10.1063/1.3490188,Huh_2014} (Eq.~\ref{eq:oqs_discretised_hamiltonian})that is the central object employed within TEDOPA~\cite{PhysRevLett.105.050404, 10.1063/1.4866769, 10.1063/1.3490188,Rosenbach_2016,lacroix2024mpsdynamicsjltensornetworksimulations}:
\begin{equation}
\begin{split}
\hat{{H}}_{SB}^c=& \kappa_{0}\hat{S} (\hat{{a}}_{1}^{\dagger} +\hat{{a}}_{1}) +\sum_{k=1}^{N-1}t_k(\hat{{a}}_{k}^{\dagger}\hat{a}_{k+1}+\hat{a}_{k+1}^{\dagger}\hat{{a}}_{k})\\ & +\sum_{k=1}^N\epsilon_{k}\hat{{a}}_{k}^{\dagger}\hat{{a}}_k, \label{eq:sbm_hamiltonian_chain}
\end{split}
\end{equation}
where $\epsilon_k$ is the onsite energy of each bosonic site of the chain and $t_k$ the hopping element between chain sites $k$ and $k+1$.

More recently, the use of time-dependent transformations has been suggested as a means for reducing the amount of entanglement that needs to be captured within a tensor network simulation~\cite{interaction_picture_haobin}.  Here we consider the use of a transformation to the interaction picture of the chain Hamiltonian, giving rise to a Hamiltonian of the form
\begin{equation}
\begin{split}
\hat{H}_{SB}^{ic}(t)&=e^{-i\hat{H}_B t} \hat{H}^{c}_{SB} e^{i\hat{H}_B t} \\&= \hat{S}\sum_{k}(g_{k}(t)\hat{{a}}_{k}^{\dagger}+g_{k}^*(t)\hat{{a}}_{k}), \label{eq:sbm_hamiltonian_ipchain}
\end{split}
\end{equation}
in which the time-dependent coefficient $g_{k}(t)$ can be expressed in terms of the star-bath on-site energy terms and the transformation matrix that takes the coupling matrix from the chain-topology to the star-topology~\cite{PhysRevA.105.032406}.

While all of these Hamiltonians are equivalent up to a unitary transformation, the entanglement structure---and in turn the numerical effort associated with tensor network simulations of the model---can depend significantly on the specific form of the Hamiltonian considered.  The script \protect\path{pyttn/examples/spin_boson/sbm_unitary.py} provides an example showing the use of the different bath topologies within the pyTTN package.

\subsubsection{Non-unitary Dynamics: the Hierarchical Equations of Motion and Pseudomode Methods \label{sec:non-unitary_dynamics}}
An alternative strategy for treating the dynamics of open-quantum systems is provided by the Hierarchical Equations of Motion (HEOM) method~\cite{TanimuraJPSJ1989,IshizakiJPSJ2005,TanimuraJCP2020}. The HEOM method makes use of the Vernon-Feynman influence functional representation of the reduced system dynamics in terms of the bath correlation functions.  By employing a discretisation of the bath correlation function as a sum of decaying exponential
\begin{equation}
    C(t) = \sum_{k=1}^K c_k e^{-\alpha t}, \label{eq:corr_func_expansion}
\end{equation}
it becomes possible to construct a linear set of equations of motion for the reduced system density operator and an infinite set of auxiliary
density operator (ADO), that fully captures the dynamics of the system:
\begin{equation}
\begin{split}
    &\frac{\partial}{\partial t}\hat{\rho}_{\boldsymbol{m},\boldsymbol{n}}(t) = - \left(i\mathcal{L}_s(t) + \sum_{k=1}^K (\nu_k m_k + \nu_k^* n_k) \right)\hat{\rho}_{\boldsymbol{m}, \boldsymbol{n}}(t)\\&- i\sum_{k=1}^N \left[\sqrt{\frac{m_k}{\|\alpha_k\|}} \alpha_k \hat{S} \hat{\rho}_{\boldsymbol{m}_k^-, \boldsymbol{n}}(t)  - \sqrt{\frac{n_k}{\|\alpha_k\|}} \alpha_k^*\hat{\rho}_{\boldsymbol{m}, \boldsymbol{n}_k^-}(t)\hat{S}\right] \\ & - i\sum_{k=1}^N \left[\sqrt{(m_k +1)\|\alpha_k\|} \left[\hat{S}, \hat{\rho}_{\boldsymbol{m}_k^+, \boldsymbol{n}}(t)\right] \right. \\& \ \ \ \ \ \ \ \ \left. + \sqrt{(n_k +1)\|\alpha_k\|} \left[\hat{S}, \hat{\rho}_{\boldsymbol{m}, \boldsymbol{n}_k^+}(t)\right]\right],
\end{split}\label{eq:fpheom}
\end{equation}
where $\boldsymbol{m}_k^\pm$ ($\boldsymbol{n}_k^\pm$) corresponds to the set of indices $\boldsymbol{m}$ ($\boldsymbol{n}$) but with the $k$-th element incremented (+) or decremented (-) by one.  Here there are a total of $2K$ bath modes for the $K$ terms in the bath correlation function.  Truncation of this equation at some fixed depth of the hierarchy provides an efficient scheme for generating open-quantum system dynamics.

For rational functional spectral densities at high temperatures, expansion of the form given in Eq.~\ref{eq:corr_func_expansion} can be constructed efficiently with very few terms.  For more general, spectral densities, there have been significant developments in algorithms for constructing such decompositions; however, in general these decompositions can contain more terms than what is feasible using exact treatments of the hierarchy of ADOs.  In such cases, the use of tensor network based ans\"atze for the hierarchy of ADOs has proven successful in obtaining accurate dynamics with the HEOM approach~\cite{lindoy_thesis, 10.1063/1.5026753, 10.1063/5.0027962, 10.1063/5.0050720,10.1063/5.0153870,10.1063/1.5099416, LindoyMandalReichman,10.1063/5.0200410, PhysRevLett.129.230601,Mangaud2023,ivander2024unifiedframeworkopenquantum, doi:10.1021/acs.jctc.4c00711, 10.1063/5.0226214,10.1063/5.0202312}.
The Hierarchical Equations of Motion can be rewritten in terms of two sets of bosonic operators $\hat{a}_k$ and $\hat{\tilde{a}}_k$ acting on the degrees of freedom associated with the $m$ and $n$ indexed ADOs, respectively, as~\cite{xu2023universalframeworkquantumdissipationminimally}
\begin{equation}
\begin{split}
    \frac{\partial}{\partial t}&\left|\hat{\rho}(t)\right) = -i\left(\mathcal{L}_s(t) -i\sum_{k=1}^K (\nu_k \hat{n}_k + \nu_k^*  \hat{\tilde{n}}_k) \right)\left|\hat{\rho}(t)\right)\\
    &- i\sum_{k=1}^N \left[\frac{ \alpha_k}{\sqrt{\|\alpha_k\|}} \hat{S}^L \hat{a}_k^{\dagger}  - \frac{ \alpha_k^*}{\sqrt{\|\alpha_k\|}} \hat{S}^R \hat{\tilde{a}}_k^{\dagger}\right]\left|\hat{\rho}(t)\right) \\
    &
    - i\sum_{k=1}^N \sqrt{\|\alpha_k\|}\hat{S}^- \left[\hat{a}_k + \hat{\tilde{a}}_k\right]\left|\hat{\rho}(t)\right),
\end{split}\label{eq:fpheom_op}
\end{equation}
 The generator of the HEOM dynamics has the same structure as a star-topology open quantum system Hamiltonian with non-hermitian couplings and (generally) complex frequencies, and as such can readily be represented with an efficient hSOP form.  At this stage it is worth noting that a similarity transformation of the bath modes can be applied to transform the HEOMs into the standard pseudomode form~\cite{xu2023universalframeworkquantumdissipationminimally}. As shown in Sec.~\ref{sec:supplementary_fpheom} of the Supplementary Information, analogous non-unitary normal-mode transformations can be performed to the generalised pseudomode representation~\cite{park2024quasilindbladpseudomodetheoryopen} to recover the HEOMs. Consequently, both methods are equivalent when solved with exact arithmetic, although they may have different numerical properties when treated using floating-point numbers and approximate simulation techniques (such as the tensor network strategies implemented within pyTTN).
In the script \path{pyttn/examples/spin_boson_model/sbm_nonunitary.py} we demonstrate how the Free Pole HEOM~\cite{PhysRevLett.129.230601} and generalised quasi-Lindblad pseudomode equations of motion can be used within the pyTTN pacakage for the simulation of the spin-boson model. The generalisation to multiple bosonic baths follows trivially. pyTTN supports the use of the adaptive Antoulas-Anderson (AAA)~\cite{doi:10.1137/16M1106122} and the estimation of signal parameters via rotational invariant techniques (ESPRIT)~\cite{1457851, 9000636,10.1063/5.0209348,park2024quasilindbladpseudomodetheoryopen} methods to construct approximate representations of the bath correlation function.
Further details of these decomposition methods, as well as of the bath discretisation schemes used for unitary dynamics simulations that were discussed in Sec.~\ref{sec:bath_discretisation} are provided in the pyTTN tutorials (\protect\path{pyttn/tutorials/oqs/}).

\section{Numerical Performance \label{sec:numerical_performance}}
We consider the scaling of the pyTTN implementation of the TDVP algorithm for the evolution of tree tensor network states.  Throughout this section we consider the spin-boson model using the star-topology form of the Hamiltonian
\begin{equation}
\hat{{H}}=\frac{\varepsilon}{2}\hat{{\sigma}}_{z}+\frac{\Delta}{2}\hat{{\sigma}}_{x}+\hat{{\sigma}}_{z}\sum_{k}g_{k}(\hat{{a}}_{k}^{\dagger}+\hat{{a}}_{k})+\sum_{k}\omega_{k}\hat{{a}}_{k}^{\dagger}\hat{{a}}. \label{eq:sbm_hamiltonian}
\end{equation}
Here the coupling constants and frequencies for the discrete bath are related to the bath spectral density according to Eq.~\ref{eq:discrete_spec}.
 \begin{figure}[tp]
\includegraphics[width=0.5\columnwidth]{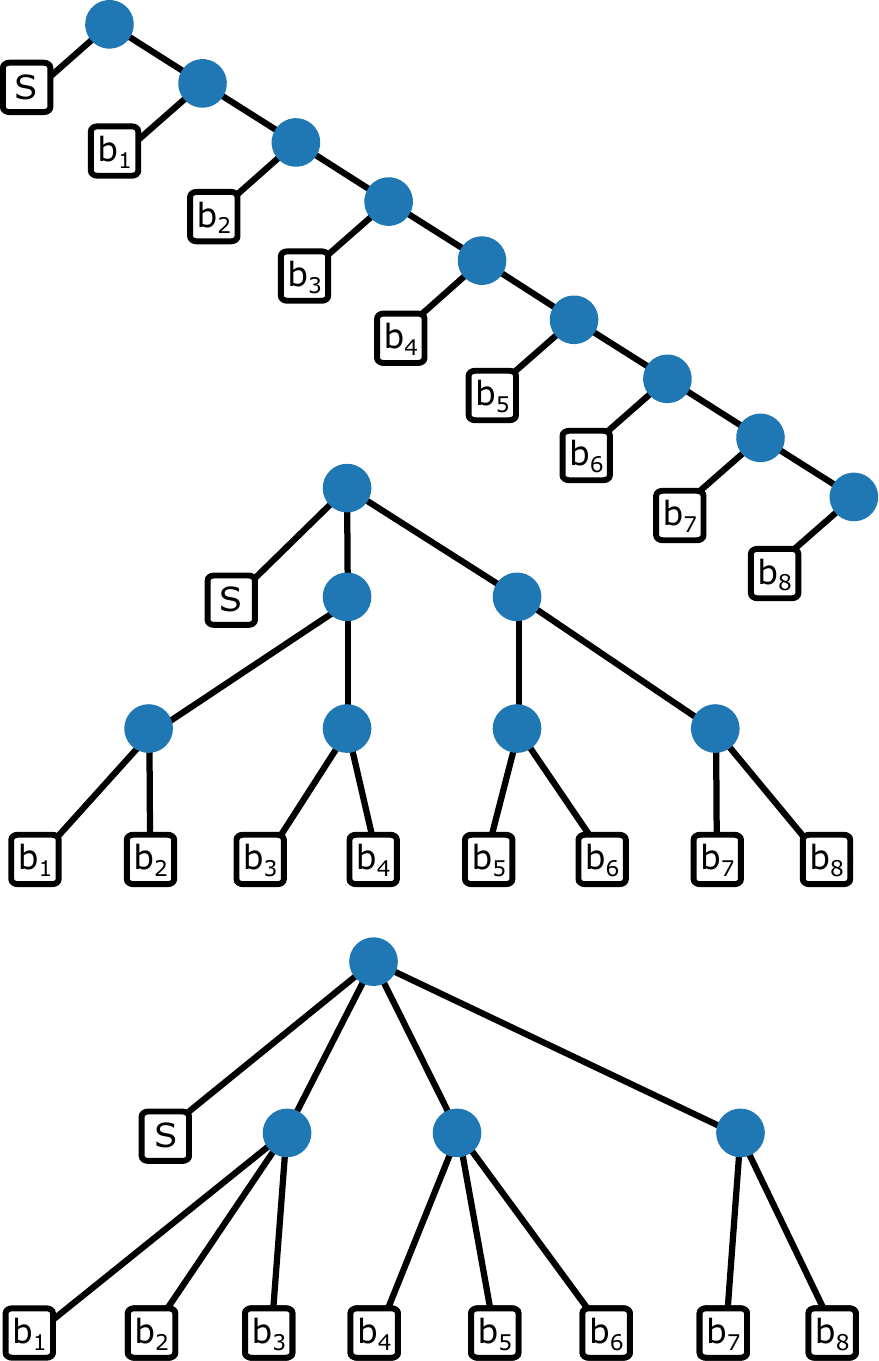} \vspace{-1em}
\caption{\label{fig:topologies} Tree tensor network topologies corresponding to an MPS treatment of the environmental bosonic degrees of freedom (top), a balanced binary tree tensor network (middle) and balanced ternary tree representation (bottom).   Here the node labelled S corresponds to the system degree of freedom and the nodes label b$_i$ correspond to the $i$-th bath degree of freedom.}
\end{figure}
We consider a continuous Ohmic spectral density with an exponential cutoff
\begin{equation}
J(\omega)= \frac{\pi}{2} \, \alpha \,\omega \, e^{-\omega/\omega_{c}},
\end{equation}
 where $\alpha$ is the dimensionless Kondo parameter, and $\omega_c$ is the bath cutoff frequency.  In all calculations presented in this section we use the orthogonal polynomial-based discretisation strategy with an additional hard cutoff of $\omega_{\mathrm{max}} = 10\,\omega_c$, such that when scaling the number of bath modes we capture the same underlying bath dynamics but with a more accurate description of the influence of the continuum bath on the system.

 Throughout this section we consider several different topologies for the tree tensor network representation of the bath degrees of freedom, in order to explore the scaling properties of TTN-based approaches as a function of the number of bonds in the tensors in the network.  Namely, we consider a matrix product state topology with the use of an optimised boson basis, a balanced binary tree, and a balanced ternary tree. The resultant tensor network topologies are shown in Fig.~\ref{fig:topologies}, where the node labelled S corresponds to the system degree of freedom and the nodes labelled b$_i$ correspond to the $i$-th bath degree of freedom.  Mode combination is not used throughout this section, but due to the relatively minor dependence of the cost of ML-MCTDH calculations on the local Hilbert space dimension, particularly when handling systems with large bond dimension, this approach can often considerably reduce the overall cost of calculations.  All timings presented in this section were obtained using a single core of a 2.6 GHz Intel Xeon Gold 6142 processor.

\subsection{Comparison of hSOP and SOP Hamiltonian Representation}

\begin{figure}[tp]
\includegraphics[width=\columnwidth]{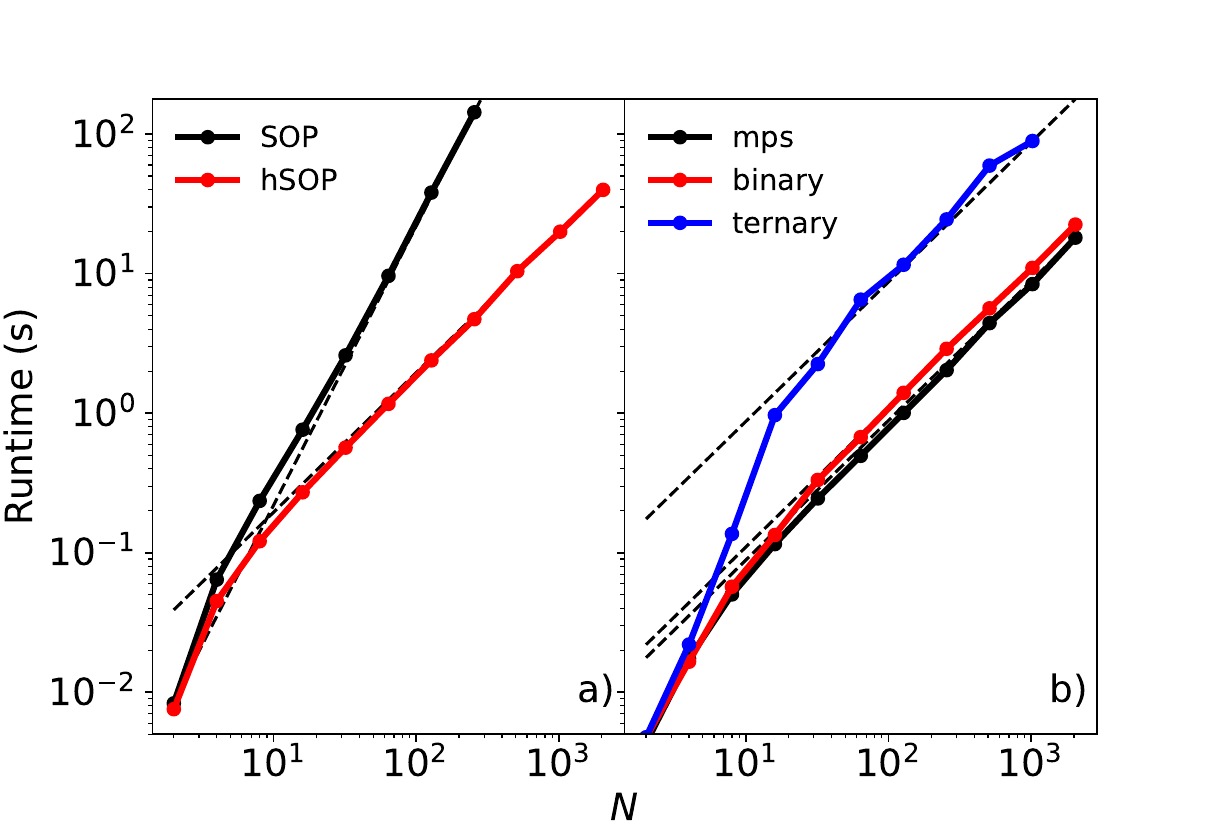} \vspace{-1em}
\caption{\label{fig:sbm_sop_vs_hsop} The average walltime cost (over 10 steps) of a single one-site TDVP step as a function of the number of bosonic modes for an ohmic spin-boson model with $\alpha=1.0$ and $\omega_c=5$ at $T=0$.  a) comparison of the SOP and hSOP Hamiltonian representations for a binary tree topology. b) comparison of the runtimes obtained for different tree geometries. The dotted lines represent asymptotic scalings corresponding to $N^2$ and $N$ for the SOP and hSOP representations, respectively.   Here we used a local Hilbert space dimension of $D=10$ for each bosonic mode, and a constant bond dimension of $\chi=20$.
The results presented here were generated using the script \protect\path{pyttn/examples/scaling/spin_boson_model/sbm_SOP_vs_hSOP.py} and \protect\path{pyttn/examples/scaling/spin_boson_model/sbm_Nb_scaling.py}.  }
\end{figure}

 In Fig.~\ref{fig:sbm_sop_vs_hsop}a) we provide a comparison between performance of the SOP and hSOP representations for a spin-boson model. We observe a quadratic scaling of the computational cost when using the SOP representation.
 The hSOP representation provides considerable improvements, observing a linear scaling with the number of bath modes, $N$.  This result is readily explained by the fact that in the SOP representation we are evolving a Hamiltonian with $\mathcal{O}(N)$ terms, and need to update $\mathcal{O}(N)$ modes throughout a single integration step. In contrast, using the hSOP representation we perform a sum over $\mathcal{O}(1)$ terms to represent the Hamiltonian locally at each of the $\mathcal{O}(N)$ nodes, giving the required scaling. Additionally, this representation significantly reduces memory requirements at this stage, as it is only necessary to store a total of $\mathcal{O}(1)$  single-particle function and mean-field Hamiltonian matrices at each node, in contrast to the $\mathcal{O}(N)$ matrices required by the SOP representation.

It is worth noting that this speed-up depends on the specific form of the Hamiltonian considered, and the performance improvements in moving from the SOP to hSOP representation are considerably less dramatic for problems involving denser Hamiltonians, such as the Molecular Hamiltonian that we consider in Sec.~\ref{sec:si_hsop_vs_sop} of the Supplementary Information.  For the case of the molecular Hamiltonian, which contains quartic interaction terms between modes, the dominant numerical cost arises from Hamiltonian evaluations across the bond in the tensor network partitioning the system into two approximately equal halves.  The cost of this scales quartically with the number of orbitals, regardless of whether the SOP or hSOP representation is used, although the use of the hSOP representation does reduce the memory required to store the matrices necessary for this evaluation. The use of approximate tensor network compression techniques may enable more efficient simulations. Exploring how such an approach compares to recently proposed canonical polyadic decomposition-based compression schemes~\cite{10.1063/5.0192012} is a potential avenue for future work. For the types of Hamiltonian considered in this manuscript, the use of the hSOP representation leads to dramatic performance improvements, particularly for large baths, and as such in all calculations presented throughout the remainder of the manuscript we make use of the hSOP representation.

In Fig.~\ref{fig:sbm_sop_vs_hsop}b), we compare runtimes for a single step of the one-site TDVP algorithm obtained using the hSOP representation and three different tree topologies.  Here we observe the expected $\mathcal{O}(N)$ scaling for all topologies, however observe considerably larger overhead when using a ternary tree representation.  This is unsurprising for the case considered here, in which we have considered $\chi=20$ for all tree structures.  For the ternary tree representation typical tensors will contain $\mathcal{O}(\chi^4)$ elements compared to the $\mathcal{O}(\chi^3)$ used in the binary tree representation and $\mathcal{O}(\chi^2 D)$ used with the MPS representation, and consequently ternary tree operations are considerably more expensive for the same bond dimension.
\subsection{Performance of Adaptive Subspace Expansion \label{Sec:subspace_performance}}
\begin{figure}[!b]
\includegraphics[width=0.95\columnwidth]{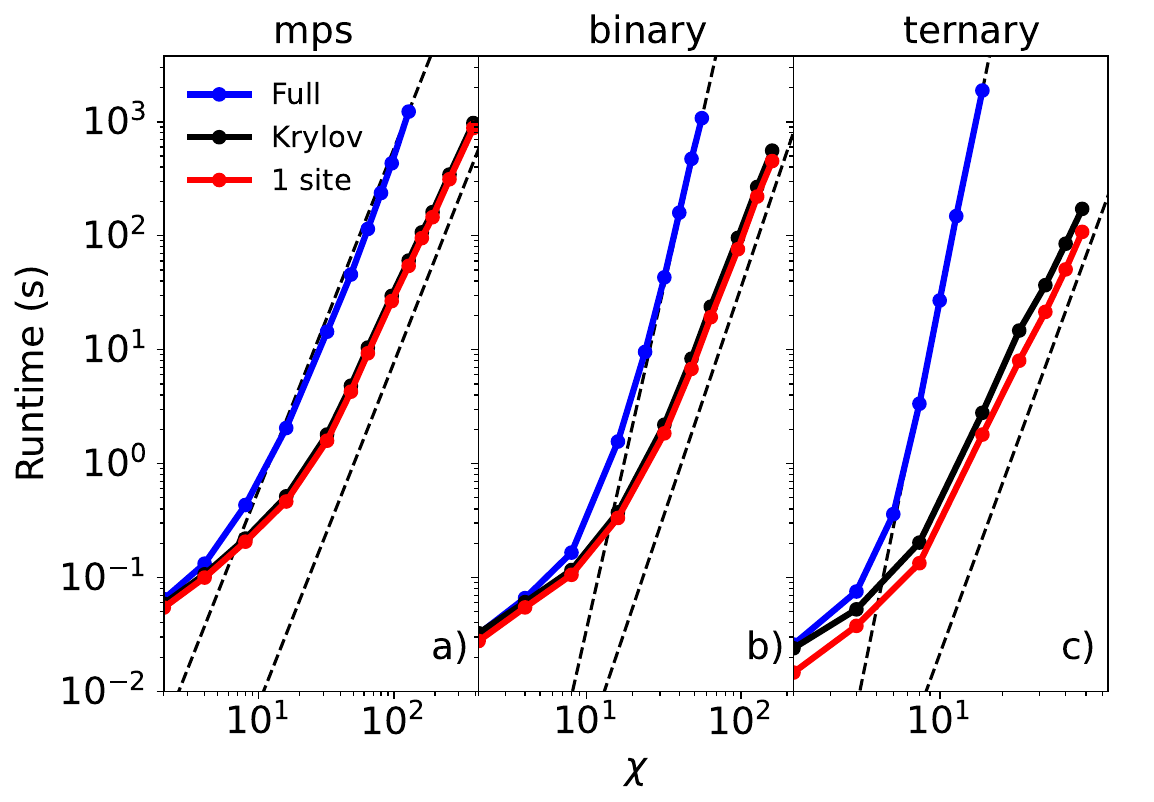} \vspace{-1em}

\caption{\label{fig:sbm_subspace_expansion_scaling} The average walltime cost (over 10 steps) of a single step of the TDVP algorithm using the full evaluation of the two-site energy variance (blue), Krylov subspace methods to evaluate a few dominant eigenvalues relevant to the bond dimension expansion (black), and the standard non-adaptive single-site algorithm (red). Each panel corresponds to a different tree topology ((a) MPS , (b) binary tree and (c) ternary tree). Dotted lines represent asymptotic scalings associated with a single-site algorithm and two-site algorithm (full approach) corresponding to $\mathcal{O}(\chi^3)$ and $\mathcal{O}(\chi^3)$ (left panel), $\mathcal{O}(\chi^4)$ and $\mathcal{O}(\chi^6)$ (middle panel), and $\mathcal{O}(\chi^5)$ and $\mathcal{O}(\chi^9$ (right panel).  The single-site results presented here were generated using the script \protect\path{pyttn/examples/scaling/spin_boson_model/sbm_chi_scaling.py}. For all calculations we used a local Hilbert space dimension of $d=10$, and for the MPS and binary tree calculations $N=32$ bath modes were used to discretise the bath while for the ternary tree case $N=27$ bath modes were used. }
\end{figure}
In Fig.~\ref{fig:sbm_subspace_expansion_scaling}, we compare timings for the evaluation of a single TDVP step obtained using the adaptive subspace expansion employing a dense matrix-based evaluation of the singular vectors of the two-site energy object and through the use of iterative linear solvers that make use of the action of the two-site energy variance on a one-site object to obtain the dominant singular vectors of the two-site energy object.  Here we observe that, when using the full two-site algorithm, the formation and decomposition of the two-site variance is the dominant contribution to the runtime cost.  This is demonstrated by runtimes that scale with bond dimension as $\mathcal{O}(\chi^{3X})$ with $X=1,2$, and $3$ for MPS (a), binary tree (b), and ternary tree (c) bath representations, respectively.  For the case of an MPS wavefunction, this corresponds to the same asymptotic cost as expected for a one-site algorithm. However, we observe a significant increase in runtime costs when using the two-site approach compared to the one-site approach, which can be attributed to an increased dependence of runtime on the size of the local Hilbert space dimension when using a two-site algorithm compared to a one-site approach.  When moving to high degree trees, the two-site algorithm becomes even less favourable with considerably worse scaling of runtime cost.  In contrast, the use of iterative linear algebra techniques, which avoid the need to explicitly form or compute the full singular value decomposition of the two-site algorithm, leads to a considerable improvement in performance of the subspace expansion approach.  This is highlighted in Fig.~\ref{fig:sbm_subspace_expansion_scaling}, where we observe comparable runtime costs with both the single-site and the Krylov subspace based implementation of the subspace expansion method, and importantly observe the same asymptotic runtime scaling with bond dimension. As such, when working with generic tree structures or large local Hilbert space dimensions, the use of iterative linear algebra based approaches provides significant improvements in the performance of subspace expansion based approaches for expanding the bond dimension of the tensor networks.

\begin{figure}[tp]
\includegraphics[width=0.95\columnwidth]{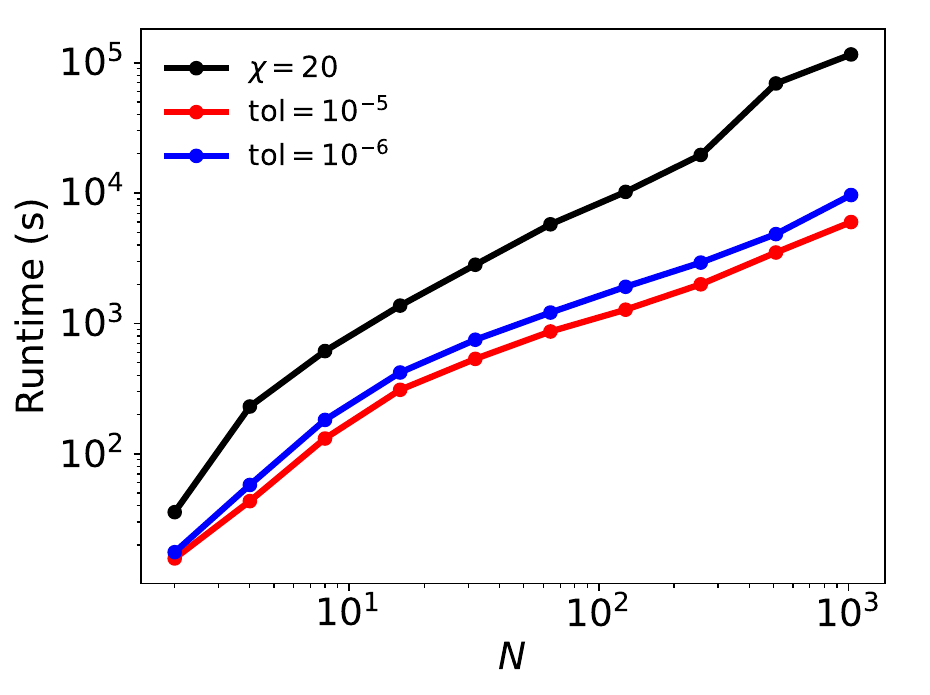}
\vspace{-1em}
\caption{\label{fig:sbm_subspace_expansion_timings} The runtime required to obtain the zero temperature polarisation dynamics of the Ohmic spin-boson model with an exponential cutoff with $\omega_c=5$ and $\alpha=1$, using the single-site algorithm with a fixed bond dimension $\chi=20$ (black), and the adaptive subspace expansion scheme using a tolerance of $10^{-5}$ (red) and $10^{-6}$ (blue). The timings presented here were obtained from simulations using the script \protect\path{pyttn/examples/oqs/spin_boson/sbm_unitary.py}}
\end{figure}

In Fig.~\ref{fig:sbm_subspace_expansion_timings}, we compare the total computer time required for the dynamics of a model spin-boson model using the single-site algorithm with a bond dimension of $\chi=20$ and the subspace expansion algorithms with a spawning tolerance of $10^{-5}$ and $10^{-6}$.  Here we have considered a symmetric spin-boson model using the star geometry Hamiltonian with $\Delta=1$, and an Ohmic bath with exponential cutoff using a cutoff frequency of $\omega_c=5 \Delta$ and Kondo parameter of $\alpha=1.0$.  We have considered dynamics up to a time of $t \Delta=30$ using a time step of $\delta t \Delta=0.005$. We observe significant reductions in computer time ($>\!\!10\times$ for $N\!=\!1024$) when using the subspace expansion approach compared to the TDVP single-site algorithm. This reduction in computer time can be attributed to time-dependent reductions in the bond dimension used throughout the tensor network compared to the use of a constant bond dimension in the single-site algorithm.  In principle, careful tuning of the bond dimensions within the network can significantly reduce the computational cost associated with the single-site algorithm without loss of accuracy, but this considerably complicates the process of setting up new calculations.  The adaptive subspace-based approach provides a scheme for automatically adjusting the bond dimension used throughout the network, and additionally allows for the use of a reduced bond dimension at short time to reduce the cost of the initial steps, without the need for careful tuning of bond dimensions through the network.  As such, unless otherwise specified, we use the adaptive subspace expansion integrator in the following calculations.

\section{Illustrative Applications: Unitary Dynamics \label{sec:unitary_dynamics}}
In this section we present a series of example applications of pyTTN to the simulation of unitary quantum dynamics of vibronic coupling models relevant to nonadiabatic and exciton dynamics of molecular systems. Here we consider two systems, the well studied 24-mode Pyrazine model~\cite{VENDRELL2011} and a more recent model for exciton dynamics at the interface of a $n$-oligothiophene donor-C$_{60}$ fullerene acceptor system (p3ht:pcbm heterojunction)~\cite{doi:10.1021/acs.jpclett.2c01928,doi:10.1021/ja4093874,doi:10.1021/acs.jctc.4c00751, doi:10.1021/acs.jpclett.5b00336}. All scripts for performing the simulations presented in this paper are included in the \path{pyttn/examples} folder of the Gitlab repository.  These scripts demonstrate the general workflow involved in setting up a unitary dynamics calculation within pyTTN, namely definition and construction of the the Hamiltonian of the system, setting up the tree structure for the wavefunction ansatz, preparing the initial condition and evolving the initial state.

\subsection{24-mode Pyrazine}
\begin{figure}[tp]
\includegraphics[width=\columnwidth]{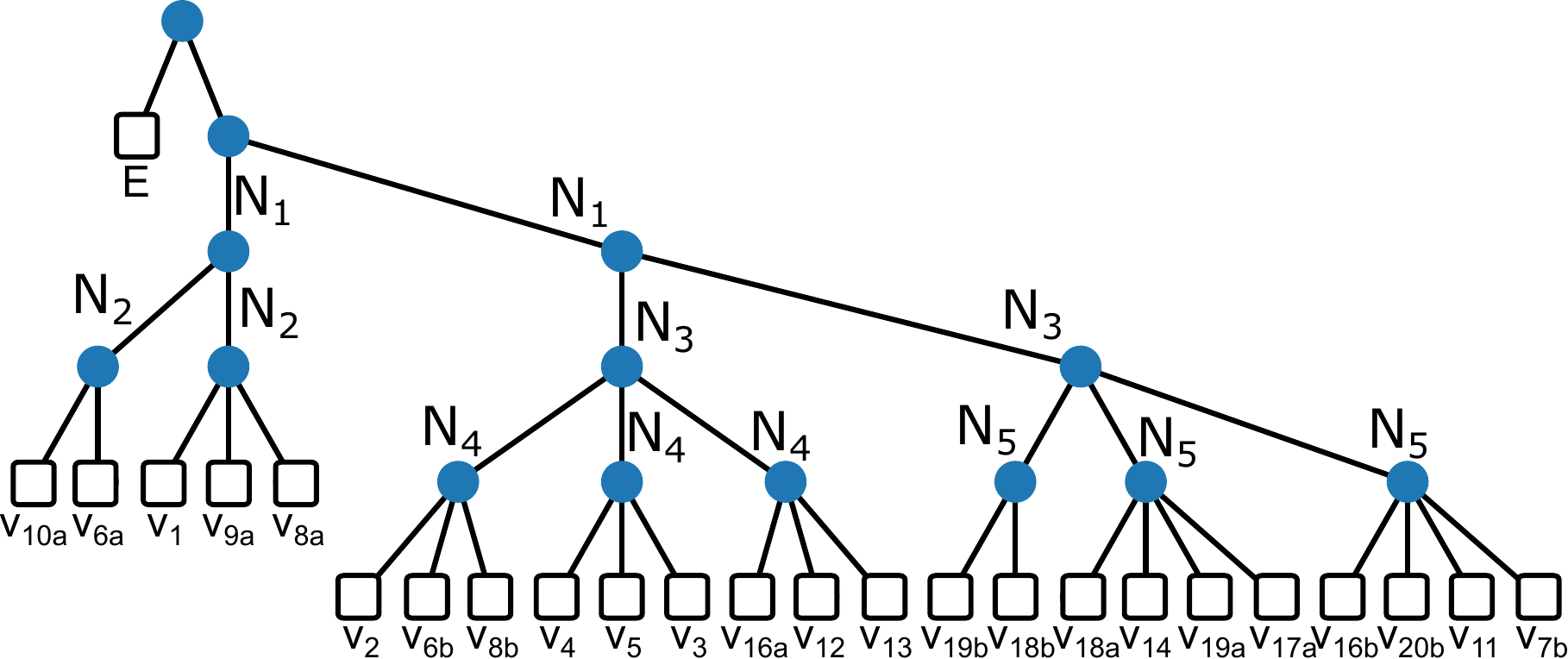}
\caption{\label{fig:pyrazine_tree} Tree topology used for all pyrazine calculations. Square nodes correspond to physical degrees of freedom.  Here the labels $\{N_1, N_2, N_3, N_4, N_5\}$ indicate the number of single-particle functions associated with each tensor in the TTN. Alternatively, the values $\{N_1, N_2, N_3, N_4, N_5\}$ correspond to the bond dimension of the bond connecting a node to its parent node.  }
\end{figure}

As a first illustration of the use of the pyTTN software package, we consider the dynamics of the 24-mode pyrazine model following excitation to the $S_2$ electronic state, which is a widely used benchmark system for methods for simulating quantum dynamics of molecular systems.~\cite{10.1063/1.481668,10.1063/1.1776111,10.1063/1.2356477,10.1063/1.2996349,RAAB1999,10.1063/1.4993219,10.1063/1.5125945,10.1063/5.0226214,doi:10.1021/jacsau.1c00474}  Within this model, we have sufficiently many degrees of freedom such that exact diagonalisation approaches become infeasible, additionally, the presence of a conical intersection between the $S_1$ and $S_2$ (that is the first and second excited triplet states) leads to non-trivial dynamics for the model.  Here we make use of the quadratic vibronic Hamiltonian presented in Ref.~\onlinecite{RAAB1999},
\begin{equation}
\begin{split}
    \hat{H} &= \sum_{i=1}^{24} \omega_i \left( - \frac{\partial^2 }{\partial \hat{Q}_i^2} + \hat{Q}_i^2\right) - \Delta \hat{\sigma}_z \\
        &+ \sum_{i\in G_1} \left( a_i \ket{0}\!\!\bra{0} + b_i \ket{1}\!\! \bra{1} \right) \hat{Q}_i  + \sum_{i\in G_3} c_i \hat{\sigma}_x \hat{Q}_i \\
    &+\!\!\!\!\!\sum_{(i,j)\in G_2}\!\!\!\!\!\left( a_{ij} \ket{0}\!\!\bra{0} + b_{ij} \ket{1}\!\!\bra{1} \right) \hat{Q}_i \hat{Q}_j\! +\!\!\!\!\!\!\sum_{(i,j)\in G_4}\!\!\!\!\!c_{ij} \hat{\sigma}_x \hat{Q}_i \hat{Q}_j,
\end{split}
\end{equation}
where
\begin{equation}
\hat{{\sigma}}_{x}=\left|0\right\rangle \left\langle 1\right|+\left|1\right\rangle \left\langle 0\right| \text{ and }
\hat{{\sigma}}_{z}=\left|0\right\rangle \left\langle 0\right|-\left|1\right\rangle \left\langle 1\right|
\end{equation}
are Pauli operators, and $\hat{Q}_i$ is the position operator associated with vibrational mode, $i$.
This Hamiltonian describes the vibrational motion of the pyrazine molecule on the $S_1$ and $S_2$ potential energy surfaces, and involves summations over various groups of vibrational modes $G_k$ that have different symmetry properties.  The parameters used in this Hamiltonian are taken from Ref.~\onlinecite{RAAB1999}.  The script \protect\path{pyttn/examples/pyrazine/pyrazine_hamiltonian.py} provides a definition of this Hamiltonian for use within the pyTTN package.

In the calculations presented in this section, we make use of the tree structure that has been employed in previous ML-MCTDH calculations for this system~\cite{VENDRELL2011}, which we present for completeness in Fig.~\ref{fig:pyrazine_tree}.  Here the set of indices $\{N_1, N_2, N_3, N_4, N_5\}$ defines the maximum bond dimensions (or equivalently, numbers of single-particle functions) allowed at each bond within the tree, and we perform simulations with varying numbers of maximum allowed bond dimensions (see Tab.~\ref{tab:pyrazine}). We have used mode combination to combine all vibrational modes connected to the a given interior node into a single composite degree of freedom.  For each primitive mode, we use a harmonic oscillator number operator, containing the same number of primitive basis functions as was used in Ref.~\onlinecite{10.1063/1.4993219}.  All calculations are performed with the adaptive one-site TDVP scheme and a spawning tolerance of $10^{-6}$, and varying numbers of single-particle functions (bond dimensions) as shown in Tab.~\ref{tab:pyrazine}.  We note that in all simulations the largest bond dimensions allowed is reached by at least one bond in the tensor network throughout evolution.  All other simulation parameters are provided in the example script \protect\path{pyttn/examples/pyrazine/pyrazine.py}.

We consider dynamics of the pyrazine molecular following photoexcitation from the ground state (both electronic and vibrational) to the $S_2$ electronic state.  We consider the dynamics constrained to the $S_1$, $S_2$ manifold, and as such, we consider an initial wavefunction of the form
\begin{equation}
\ket{\Psi(0)} = \ket{S_2} \bigotimes_{k=1}^{24} \ket{0}_k,
\end{equation}
where $\ket{0}_k$ corresponds to the vacuum state of the $k$-th vibrational degree of freedom.
While this state corresponds to a product state, and can therefore be accurately represented with a wavefunction containing a single SPF at each node, in all simulations reported here we include unoccupied initial single-particle functions in the initial state, so that initial we have $N_i = \mathrm{min}(16, N_i)$.

\begin{figure}[tp]
\includegraphics[width=\columnwidth]{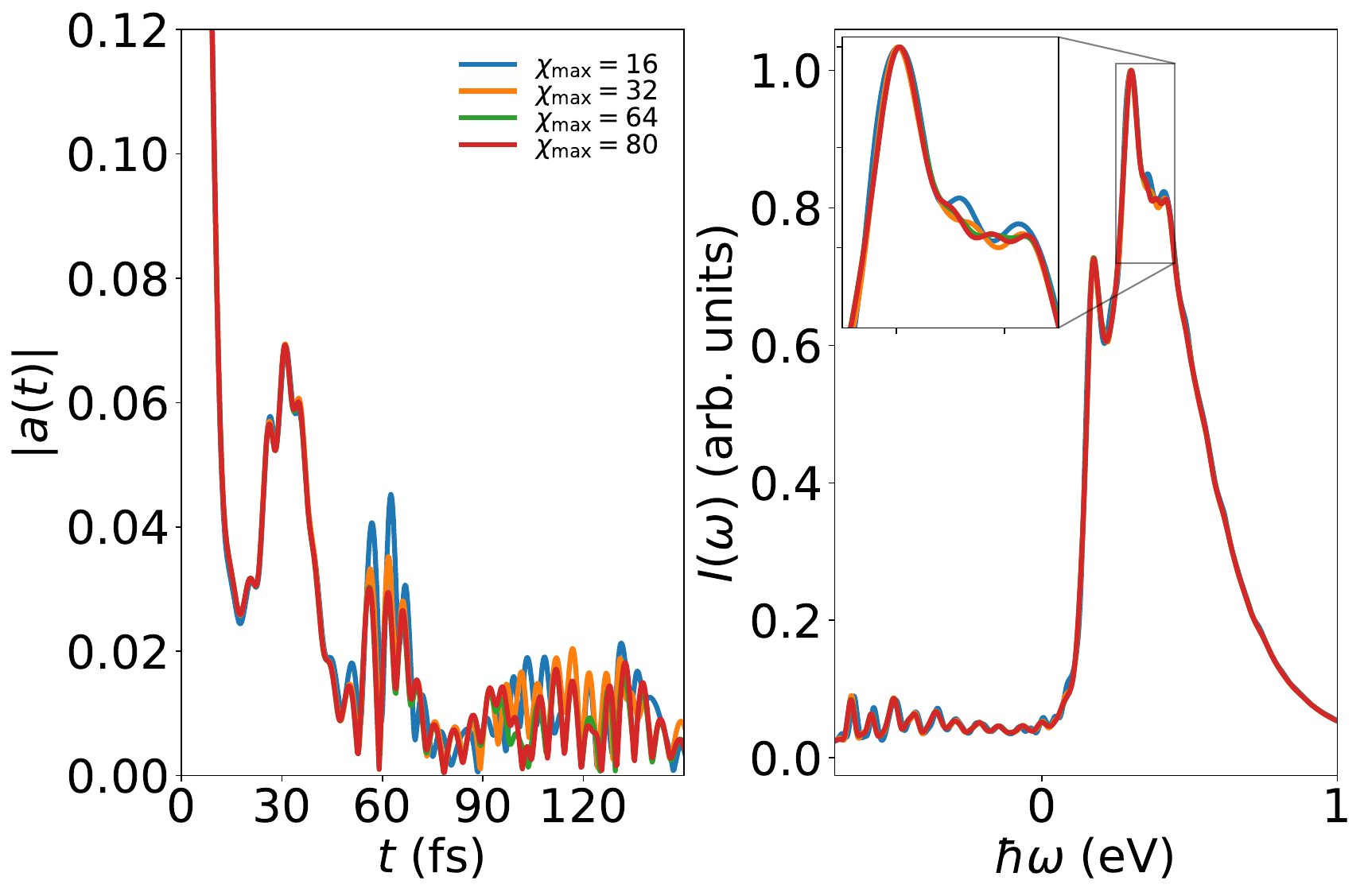} \vspace{-1em}
\caption{\label{fig:pyrazine} Left panel: auto-correlation function $a(t) = \braket{\Psi(t/2)^*}{\Psi(t/2)}$ for the pyrazine model as a function of time for varying numbers of single-particle functions (bond dimensions), $\chi_{\mathrm{max}}$.  Right panel: shifted and scaled spectrum of pyrazine obtained from the auto-correlation functions according to Eq.~\ref{eq:shifted_scaled_spectrum}.  The inset shows a zoomed-in region of the peak in the spectrum.  These results were generated using the script \protect\path{pyttn/examples/pyrazine/pyrazine.py}.}
\end{figure}

Fig.~\ref{fig:pyrazine} shows the auto-correlation function
\begin{equation}
    a(t) = \braket{\Psi(0)}{\Psi(t)},
\end{equation}
and corresponding spectrum~\cite{VENDRELL2011, 10.1063/1.4993219}
\begin{equation}
    I(\omega) \propto \int_0^{t_{\mathrm{max}}} \mathrm{Re}\left[a(t) e^{-\frac{t}{\tau}} \cos\left(\frac{\pi t}{2t_{\mathrm{max}}}\right) e^{i\omega t} \right]\mathrm{d} t,  \label{eq:shifted_scaled_spectrum}
\end{equation}
where $\tau=150$ fs. For the real-valued initial wavefunction, $\ket{\Psi(0)}$, considered in this example, the evaluation of the autocorrelation may be simplified by splitting the time-evolution operator to give
\begin{equation}
    a(t) = \braket{\Psi(-t/2)}{\Psi(t/2)} = \braket{\Psi(t/2)^*}{\Psi(t/2)},
\end{equation}
where in the final step we have used that $\ket{\Psi(-t/2)} = \ket{\Psi(t/2)^*}$.

Results are obtained from dynamics up to a maximum time of $t_{\mathrm{max}}=150$ fs using a timestep of $\mathrm{d} t = 0.5$ fs.
Here we observe systematic improvement in the auto-correlation function function upon increasing the value of $\chi_{\mathrm{max}}$.  For $\chi_{\mathrm{max}}=4$, a simulation which took $\sim 1$ minute to perform and corresponds to a similar structure to the ML-1 tree in Ref.~\onlinecite{VENDRELL2011}, we observe significant deviations in the absolute value of the auto-correlation function compared to results obtained with larger numbers of single-particle functions.  These deviations are evident in the resultant spectrum, as significant additional oscillation not present in the other results.

\begin{table}[h]
\begin{tabular}{c | c c c c c c c c }\hline\hline
$\chi_{\mathrm{max}}$ & $N_1$ &  $N_2$ & $N_3$ & $N_4$ & $N_5$ & Runtime & $\Delta I$ & $\Delta a$  \\\hline
4 & $4$ & $4$& $3$& $2$& $3$ & 0:00:51& $1.5\!\!\times\!\!10^{-1}$ & $2.2\!\!\times\!\!10^{-2}$ \\
16 & $16$ & $16$& $10$ &$8$& $12$ & 0:04:35 & $4.5\!\!\times\!\!10^{-2}$ &$9.6\!\!\times\!\!10^{-3}$\\
24 & $24$ &$24$& $12$& $8$& $12$ & 0:05:31 & $2.5\!\!\times\!\!10^{-2}$ & $5.4\!\!\times\!\!10^{-3}$\\
32 & $32$ &$32$ &$20$& $12$& $16$ & 0:11:27 & $1.6\!\!\times\!\!10^{-3}$ & $3.7\!\!\times\!\!10^{-3}$\\
48 & $48$ &$48$ &$30$& $18$&$24$ & 0:43:33 & $6.5\!\!\times\!\!10^{-3}$ & $1.5\!\!\times\!\!10^{-3}$ \\
64 & $64$ &$64$ &$30$& $24$& $32$ & 1:36:55& $4.5\!\!\times\!\!10^{-3}$ & $1.3\!\!\times\!\!10^{-3}$ \\
80 & $80$ &$80$ &$50$& $30$& $40$ & 4:19:00 & 0  & 0 (reference) \\\hline
\end{tabular}
\caption{\label{tab:pyrazine} Values for the tree topology parameters  (see Fig.~\ref{fig:pyrazine_tree}) for the pyrazine calculations shown in Fig.~\ref{fig:pyrazine}. The reported runtimes are given are presented in the format hours:minutes:seconds, and were obtained from single core simulations performed on 2.0 GHz Intel Xeon Gold 6338 CPUs. For each simulation we report the error relative to the reference calculation ($\chi_{\mathrm{max}}=80$) in the spectrum, $\Delta I = \frac{1}{\omega_{\mathrm{max}}-\omega_{\mathrm{min}}}\int_{\omega_{\mathrm{min}}}^{\omega_{\mathrm{max}}} |I(\omega) - I_{\mathrm{ref}}(\omega)| \mathrm{d}\omega$, and the error in the auto-correlation, $\Delta a = \frac{1}{t_{\mathrm{max}}}\int_{0}^{t_{\mathrm{max}}} |a(t) - a_{\mathrm{ref}}(t)| \mathrm{d}t$. }
\end{table}

Increasing the maximum number of single-particle functions to $32$ the simulation requires $\sim 11$ minutes. It corresponds to a wavefunction between the ML-7 and ML-8 wavefunctions considered in Ref.~\onlinecite{VENDRELL2011} in complexity. We find considerably better agreement of the results with those obtained with the $\chi_{\mathrm{max}}=80$ ($\sim 4$ hours 20 minutes), and these results accurately capture the structure observed in the $\|a(t)\|$ at $t=30$ fs. At longer times, small deviations in the auto-correlation function become evident, but we observe a significant decrease in these longer time deviations upon further increasing the number of single-particle functions to $\chi_{\mathrm{max}}=64$.  The key features in the spectra are well converged for all $\chi_{\mathrm{max}} > 32$, with only small deviation being observed in the peak regions, as shown in the inset of the right panel of Fig.~\ref{fig:pyrazine}.

Further details of this convergence, as well as timings for the simulations, are shown in Tab.~\ref{tab:pyrazine}.  Here we have presented the mean absolute errors in the spectrum $\Delta I$ and auto-correlation function $\Delta a$ defined by averaging the absolute deviation of the results at a given $\chi_{\mathrm{max}}$ from the reference simulations obtained with $\chi_{\mathrm{max}}=80$, over the domains shown in Fig.~\ref{fig:pyrazine}. These results show more clearly the systematic convergence of the error in the auto-correlation and spectra with increasing $\chi_{\mathrm{max}}$.

\subsection{Exciton Dissociation at a P3HT:PCBM Heterojunction}
We next consider a larger-scale linear vibronic coupling model for exciton dynamics at the interface of a $n$-oligothiophene donor-C$_{60}$ fullerene acceptor system~\cite{doi:10.1021/acs.jpclett.2c01928,doi:10.1021/ja4093874,doi:10.1021/acs.jctc.4c00751, doi:10.1021/acs.jpclett.5b00336}, here considering a model consisting of an array of 13 oligothiophene molecules.  For each oliogothiophene molecule, $n$, we consider two distinct electron states: the local exciton states $\ket{\mathrm{LE}_{n}}$, and the charge-separated states $\ket{\mathrm{CS}_{n}}$~\cite{doi:10.1021/acs.jctc.4c00751}. In total, this model includes $26$ electronic states and $113$ vibrational modes.

\begin{figure}[b!]
\includegraphics[width=\columnwidth]{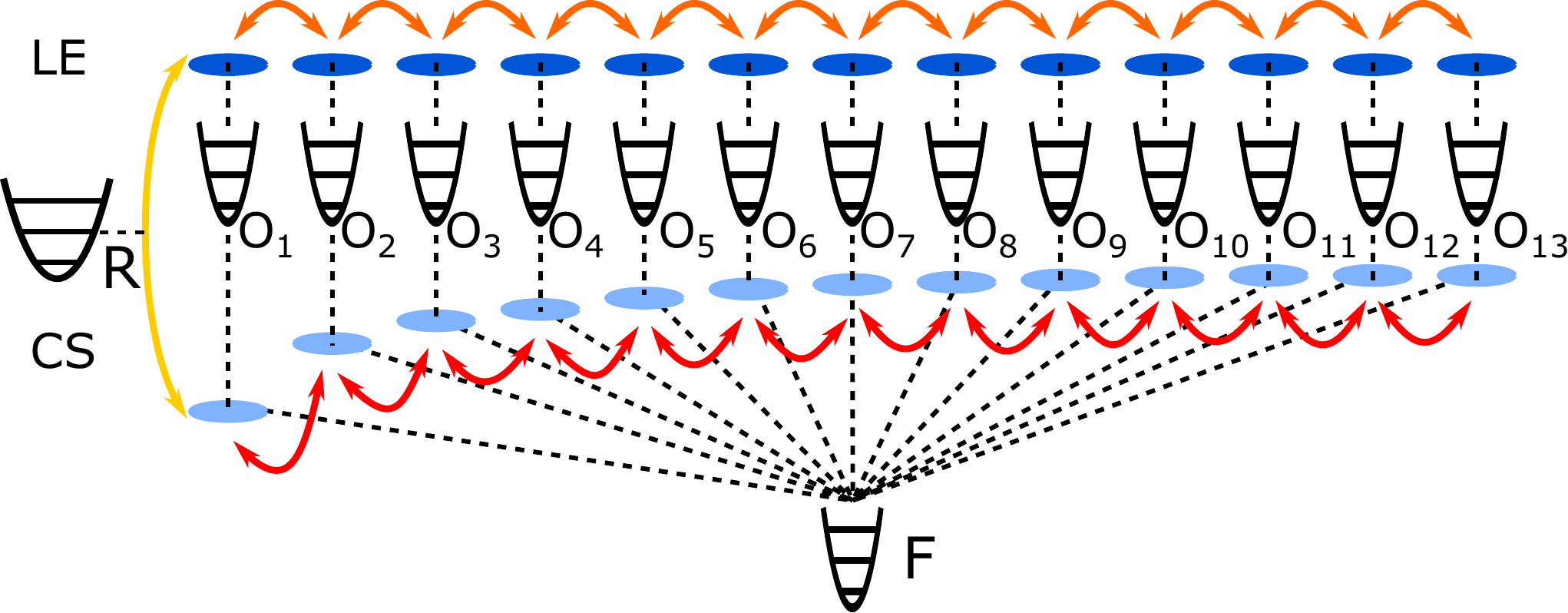} \vspace{-1em}
\caption{ \label{fig:p3ht_pcbm_schematic} Illustration of the the interactions present in the P3HT:PCBM heterojunction model, where ellipses denote electronic states, parabolas represent vibrational modes, arrows depict electronic state couplings, and dotted lines vibronic couplings.  This model contains 26 total electronic states, 13 local exciton states and 13 charge-separated states, associated with the 13 oligothiophene molecules considered. For each oligiothiophene molecule we consider 7 vibrational states ($O$ modes) that couple to both local exciton and charge separated states associated to this molecule.  Additionally, a set of 7 vibrational states associated with the fullerene acceptor system ($F$ modes) couple to all charge separated states.  Finally, interactions between the charge separated and local exciton states of  the oligiothiophene at the junction are mediated via molecular vibrations of the $R$ mode.}
\end{figure}

Recently, a comparative study of the performance of MPS and ML-MCTDH based approaches was performed, in which significant deviations were observed between the results obtained using MPS and ML-MCTDH based methods~\cite{doi:10.1021/acs.jctc.4c00751}. These deviations were attributed to different descriptions of entanglement between the two tensor network ans\"atze.  Here we consider this problem using pyTTN to simulate the dynamics of this model.  We present converged results obtained using multi-set MPS and ML-MCTDH (TTN) based wavefunctions, and using these results consider a systematic study of the convergence of the single-set MPS and ML-MCTDH (TTN) approaches comparable to Ref.~\onlinecite{doi:10.1021/acs.jctc.4c00751}, but with the use of the same integration scheme across both network topologies. As in previous treatments of this model, we split the vibrational modes into four distinct sets, the R-mode that mediates coupling between the local exciton and charge-separated states on the first oligothiophene molecules ($\{$R$\}$), $7$ vibrational modes corresponding to fullerene vibrations ($\{$F$\}$) , and the low ($\{$O$_\mathrm{l}\}$) and high frequency ($\{$O$_\mathrm{h}\}$) modes associated with the $13$ oligothiophene molecules.

\begin{figure}[tp]
\includegraphics[width=0.75\columnwidth]{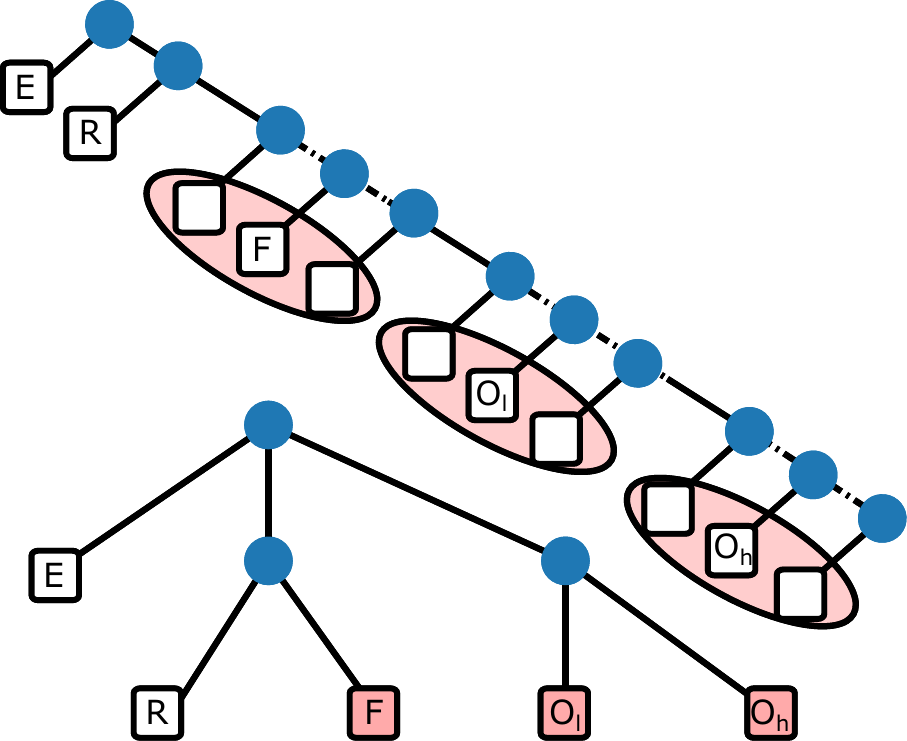}
\caption{ \label{fig:p3ht_pcbm_trees} Tensor network topologies used for the P3HT:PCBM heterojunction model.  Top: The matrix product state structure demonstrating ordering of modes.  The different sets of vibrational modes ($\{$F$\}$, $\{$O$_\mathrm{l}\}$, and $\{$O$_\mathrm{h}\}$) have been indicated with red shaded circles. Bottom: The ML-MCTDH multilayer tree.  In the ML-MCTDH tree structure, the red nodes corresponding to the different sets of vibrational modes $\{$F$\}$, $\{$O$_\mathrm{l}\}$, and $\{$O$_\mathrm{h}\}$ are expanded using balanced binary trees.}
\end{figure}

For both the single-set MPS and ML-MCTDH wavefunctions we have used a similar topology to that considered in Ref.~\onlinecite{doi:10.1021/acs.jctc.4c00751}.  Further details of the specific tensor network topologies used are shown in Fig.~\ref{fig:p3ht_pcbm_trees}.  For the multi-set calculations we treat the electronic degrees of freedom as the system, $\alpha$, and use the same the same tree topologies for representing vibrational configurations as in the single-set cases.  Full specifications of the tensor network topologies used for this model can be found in the example script \protect\path{pyttn/examples/p3ht_pcbm_heterojunction/p3ht_pcbm.py} for both the single- and multi-set calculations.

\begin{figure*}[tp]
\includegraphics[width=0.335\textwidth]{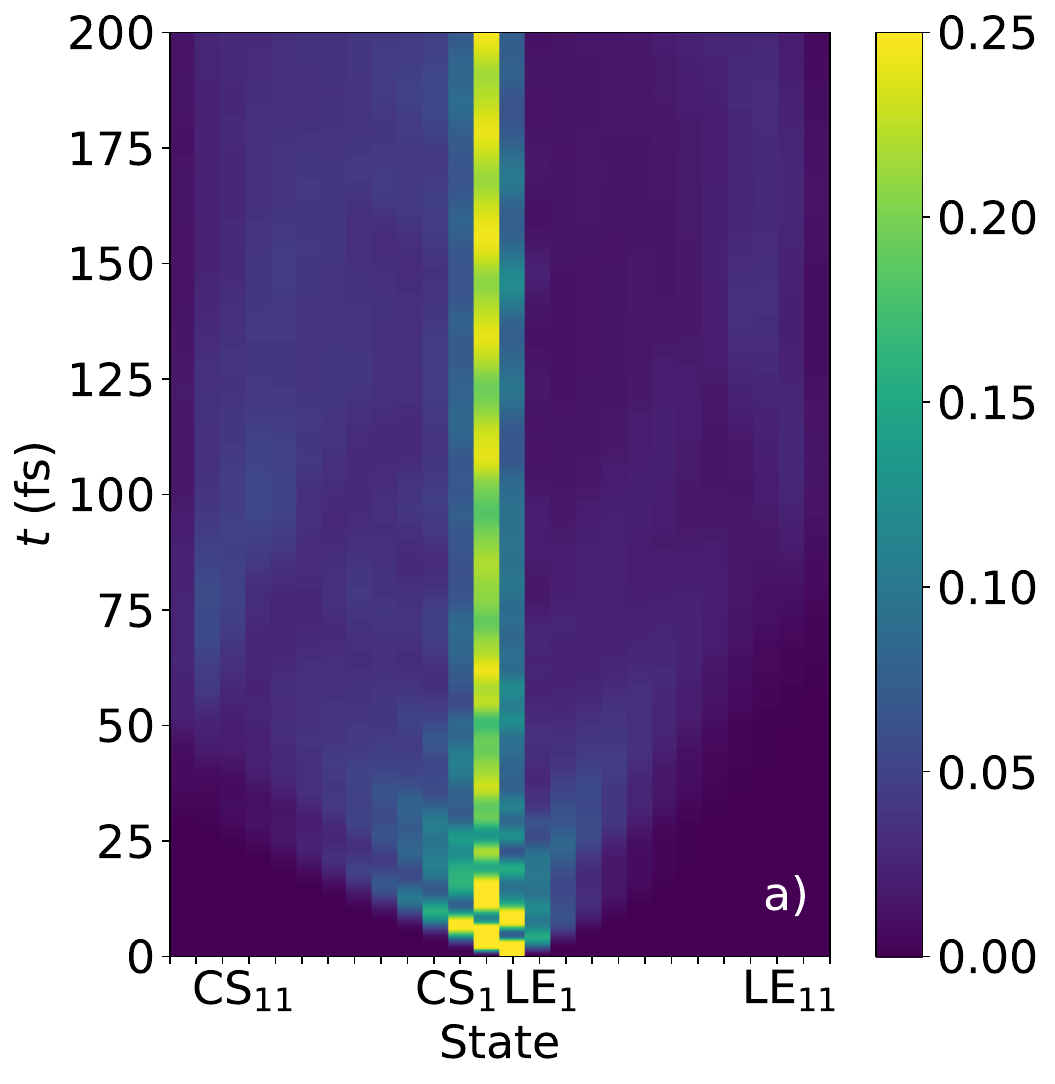}%
\includegraphics[width=0.665\textwidth]{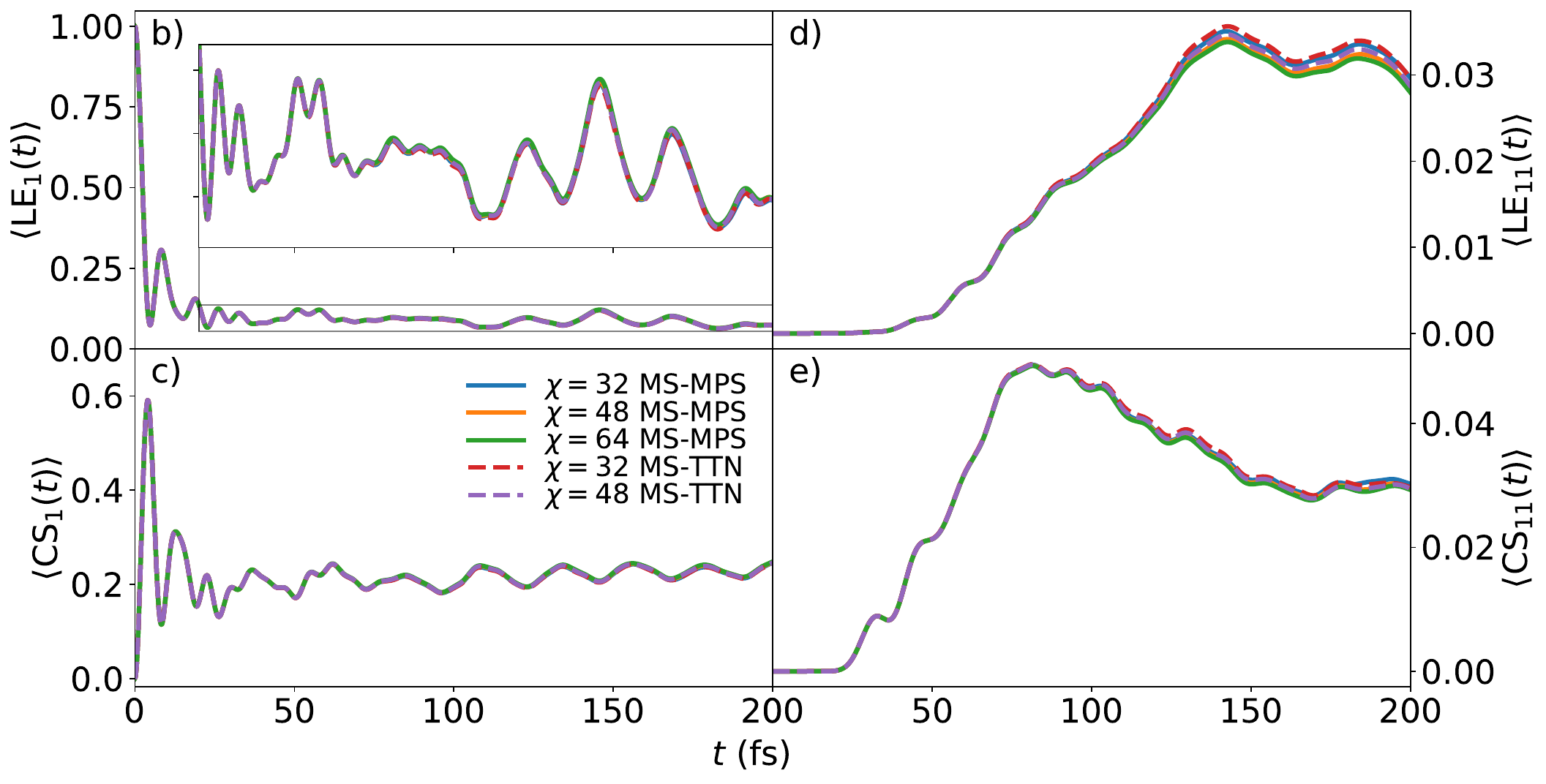}  \vspace{-2em}
\caption{ \label{fig:p3ht_multiset_results}  a) Electronic state population dynamics of the P3HT:PCBM heterojunction model obtained using the multi-set MPS approach with a bond dimension of $\chi=64$.  b-e): Convergence of the electronic state dynamics obtained using both multi-set calculations for the local exciton (b) and charge-separated state population dynamics (c) for oligiothiophene $1$,  the local exciton (d) and charge-separated state population dynamics (e) for oligiothiophene $11$. All results were generated using the example script \protect\path{pyttn/examples/p3ht_pcbm_heterojunction/p3ht_pcbm.py}. }
\end{figure*}

To be consistent with Ref.~\onlinecite{doi:10.1021/acs.jctc.4c00751}, we consider the electronic state dynamics of the model following an initial Franck-Condon excitation of the oligothiophene molecule at the junction. The initial state corresponds to one in which the electron and vibrational degrees of freedom are uncorrelated, with
\begin{equation}
    \ket{\Psi(0)} = \ket{\mathrm{LE}_1} \bigotimes_{k=1}^{113} \ket{0}_k.
\end{equation}
Here $\ket{0}_k$ corresponds to the vacuum state of the $k$-th vibrational degree of freedom, and $\ket{\mathrm{LE}_1}$ is the state where the initial exciton is localised on oligothiophene $1$.

For all multi-set simulations we use the fixed bond dimension single-site TDVP approach. For all single-set simulations use the one-site TDVP algorithm with subspace expansion based bond dimension adaptation for both single-set tensor network topologies. In each case we use a combination of two-site energy variance based subspace expansion with a cutoff tolerance of $10^{-6}$, in addition to a natural population based subspace expansion in which additional randomly selected SPFs and SHFs are added whenever all natural populations of a mode are found to be above a cutoff tolerance of $10^{-6}$.  The bond dimension is allowed to expand to a maximum value of $\chi_{\mathrm{max}}$. We find that in all single-set calculations the bond dimension bounds used are rapidly saturated within the first $<20$ fs, for both MPS and ML-MCTDH ansatz.


In  Fig.~\ref{fig:p3ht_multiset_results} we present the electronic state population dynamics obtained using the multi-set MPS and multi-set ML-MCTDH ans\"atze.  Fig.~\ref{fig:p3ht_multiset_results}a) presents the population dynamics of all electronic states obtained using the multi-set MPS approach with a bond dimension of $\chi=64$.  Initially, rapid transfer between the local exciton and charge-separated states of oligothiophene $1$ is observed.  The excitation is then observed to propagate through both the charge-separated and local exciton state manifolds with differing propagation speeds, until at time $t\sim 60$ fs we observe reflections in the dynamics of the charge separated state manifold due to the excitation reaching the final oligiothiophene in the system.  Similar reflections are observed at a later time $t\sim 100$ fs in the local excitation state manifold.  At long times significant population is observed in both the $\ket{\mathrm{CS}_1}$ and $\ket{\mathrm{LE}_1}$ states, with clear signature of coherent population transfer between these two states continuing to times $t\sim 200$ fs.  These results presented are qualitatively consistent with the results presented in Ref.~\onlinecite{doi:10.1021/acs.jctc.4c00751}; however, quantitative differences in the long time populations are observed compared to both their MPS and ML-MCTDH calculations, with our resolution showing more delocalisation of the excitation over the electronic state manifold than their MPS results but less than their ML-MCTDH results.

In Fig.~\ref{fig:p3ht_multiset_results}b-e) we present the dynamics of individual electronic state populations, the local exciton and charge-separated state populations obtained at oligothiophenes $1$ and $11$, obtained with the two multi-set wavefunction ans\"atze at varying bond dimensions.  Excellent agreement is observed between the dynamics obtained with the two multi-set approaches, with only very minor deviation of about $\sim 10^{-3}$ observed in in the long time dynamics ($t>100$ fs) obtained with the multi-set MPS and ML-MCTDH results with these bond dimensions.  Upon increasing the bond dimensions used with each wavefunction topology, no significant changes are observed in the population dynamics of the $\ket{\mathrm{LE}_1}$ and $\ket{\mathrm{CS}_1}$ states.  However, a minor systematic reduction in the long time populations in the $\ket{\mathrm{LE}_{11}}$ and $\ket{\mathrm{CS}_{11}}$ is observed when increasing from $\chi=36$ to $\chi=48$, while only very minor changes are observed between the multi-set MPS results obtained with $\chi=48$ and $\chi=64$, indicating that we are reaching convergence with respect to bond dimension in this case.  This behaviour is consistent between both the multi-set MPS and multi-set ML-MCTDH wavefunctions.

\begin{figure*}[htp]
\includegraphics[width=0.695\textwidth]{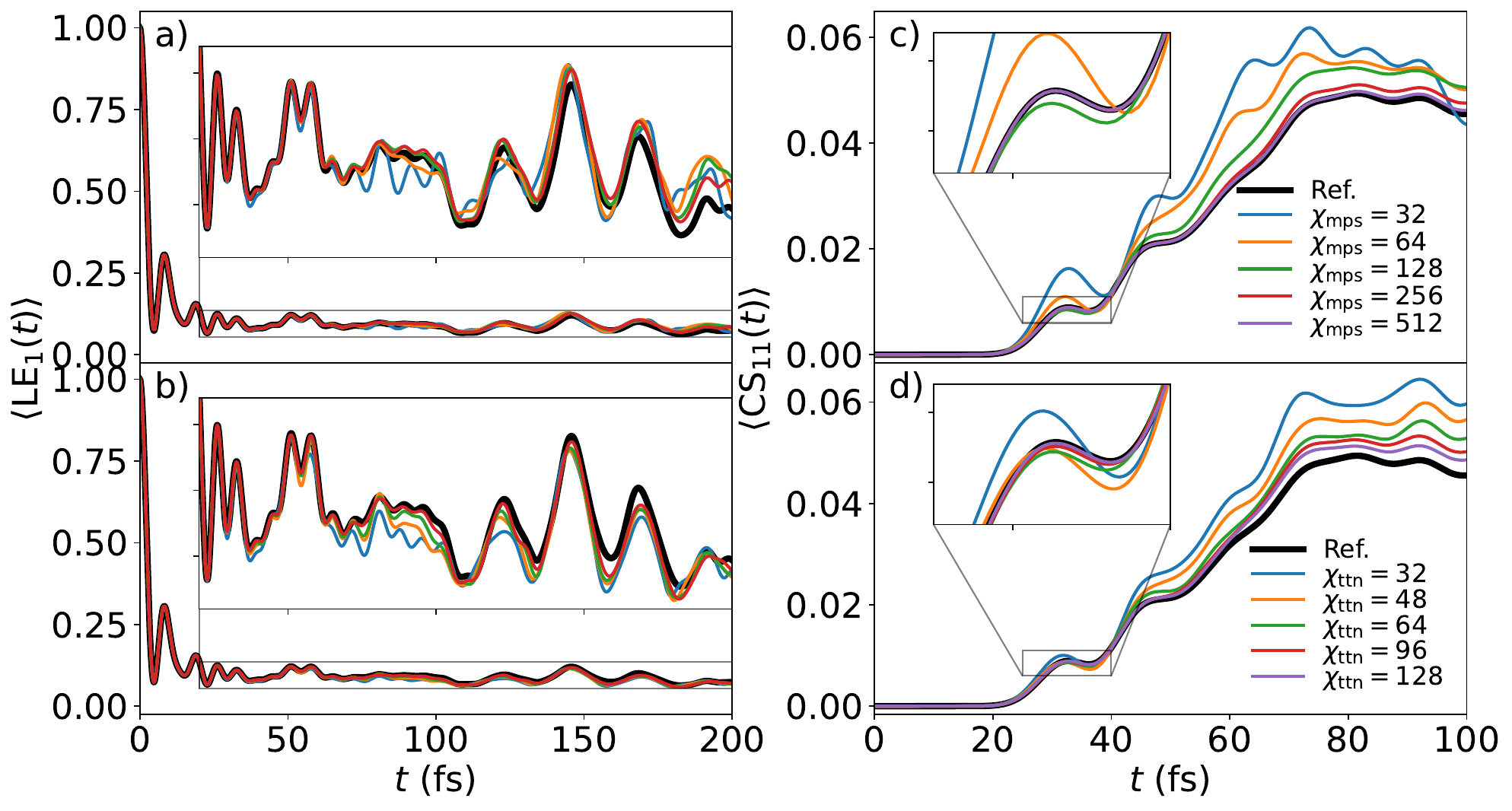}%
\includegraphics[width=0.305\textwidth]{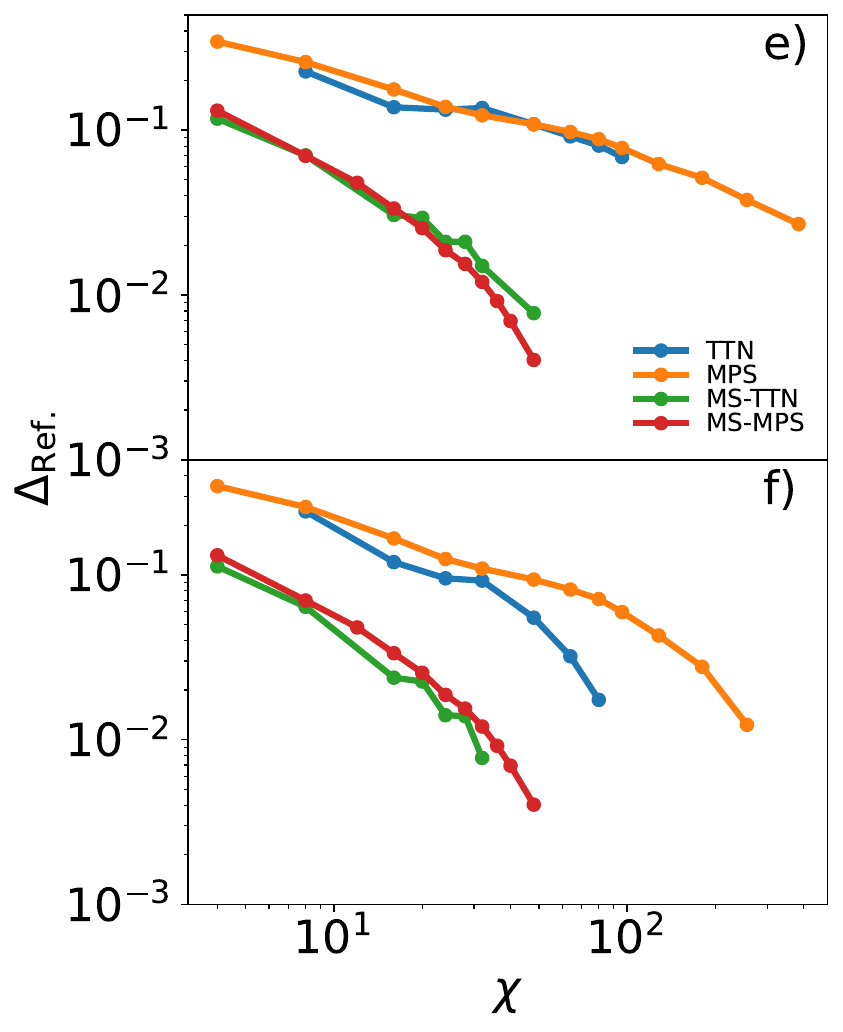} \vspace{-2em}
\caption{ \label{fig:p3ht_multiset_conv} Convergence of the electronic state population dynamics for state $\ket{\mathrm{LE}_1}$, obtained using MPS (a) and TTN (b), and $\ket{\mathrm{CS}_{11}}$, obtained using MPS (c) and TTN (d), for varying maximum bond dimensions.  e,f) Convergence of the time-averaged absolute error in electron state populations summed over all electronic states.  In the panel e), errors are measured with respect to the multi-set MPS calculation with $\chi=64$.  In panel f), a different reference is used for each method, corresponding to the calculation obtained using the largest maximum bond dimension considered for that method ($\chi = 96$, $384$, $64$, and $48$ for TTN (ML-MCTDH), MPS, multi-set MPS, and multi-set TTN, respectively). All results were generated using the example script \protect\path{pyttn/examples/p3ht_pcbm_heterojunction/p3ht_pcbm.py}.  }
\end{figure*}

The convergence of the multi-set calculations is further illustrated in the right most panels of Fig.~\ref{fig:p3ht_multiset_conv}, in which we plot the deviation from a reference calculation, defined as
\begin{equation}
    \Delta_{\mathrm{Ref}.} = \sum_{\alpha} \frac{1}{T} \int_0^T | \langle \alpha(t) \rangle - \langle \alpha_{\mathrm{Ref.}}(t) \rangle| \mathrm{d}t,
\end{equation}
where we are summing over each electronic state $\alpha$, and $\alpha_{\mathrm{Ref.}}(t)$ is the population dynamics obtained for state $\alpha$ from a reference calculation.

In Fig.~\ref{fig:p3ht_multiset_conv}e) the multi-set MPS calculation with $\chi=64$ was used as the reference, whereas in the bottom right panel a different reference is used for each method corresponding to the calculation obtained using the largest maximum bond dimension considered for that method; as an example, for the MPS results the MPS calculation with $\chi=512$ is used as reference.
By considering the red and green lines in each panel, we see clear convergence with respect to bond dimension of both the multi-set MPS and multi-set ML-MCTDH ans\"atze to both their internal reference (as shown in Fig.~\ref{fig:p3ht_multiset_conv}f)) and the multi-set MPS results (as shown in Fig.~\ref{fig:p3ht_multiset_conv}e)).  This figure also presents the convergence of the single-set ansatz.  As in the case of the multi-set ansatz, we observe that the deviation obtained in each case are of the same order of magnitude, providing indication that the deviation between the different approaches can at least partially be attributed to insufficient bond dimension to reach tighter convergence. Furthermore, for the single-set approaches, we observe considerably slower convergence than in the multi-set case.  In particular, we find a single-set simulation with a bond dimension of $\chi\sim 48$ to have comparable error to the multi-set calculation with the lowest bond dimension $\chi=4$ considered here. Further increasing the bond dimension, we observe a systematic decrease in the deviation from the reference reaching a value of $2.7\times 10^{-2}$ by $\chi=512$, a value obtained with $\chi=20$ when using the multi-set ansatz.

To further highlight the convergence behaviour of the single-set ansatz, we compare individual electron state populations, obtained using these approaches to the reference result obtained with the multi-set case, in Fig.~\ref{fig:p3ht_multiset_conv}a-d).  We observe rather rapid qualitative agreement in the local exciton state populations on oligiothiophene $1$, that is the oligiothiophene closest to the junction, seeing qualitatively consistent populations with both methods by $\chi=32$.  Increasing the bond dimension, we observe improved agreement to longer times, as visible in the inset. However, even for the largest bond dimension presented here, $\chi=512$ for MPS and $\chi=128$ for the ML-MCTDH ansatz, small deviations are observed in the long time dynamics, with the MPS calculations overestimating the long time population and the ML-MCTDH calculations underestimating this population.  These results are  consistent with those presented in Ref.~\onlinecite{doi:10.1021/acs.jctc.4c00751}, although we note that the extent of deviations is considerably smaller in the results presented here, suggesting that the deviations between MPS and ML-MCTDH observed in this prior work can likely be attributed to other origins than the manner in which the two tensor network ans\"atze capture entanglement between degrees of freedom in the system.

The deviations in the dynamics obtained using the single-set methods presented here can likely be entirely attributed to insufficient bond dimensions in the calculations.  This is more clearly visible in the central panels, in which we consider the dynamics of the charge separated state of oligiothiophene $11$.  Here we observe that both the MPS and ML-MCTDH wavefunctions overestimate the initial population dynamics, predicting a significantly larger initial population transfer observed in the peak near $t\sim 30$ fs.  Increasing the bond dimension systematically improves the results, with the first deviation appearing at later times as the bond dimension increases.  Importantly, we observe substantial differences when increasing the bond dimension of the ML-MCTDH calculations from $\chi=48$, comparable to that considered in Ref.~\onlinecite{doi:10.1021/acs.jctc.4c00751}, to $\chi=128$.  These results suggest that more stringent convergence criteria should be considered when performing ML-MCTDH calculations of these large scale vibronic coupling models.

Finally, it is worth noting that a multi-set ansatz with $N_{\mathrm{set}}$ set variables and bond dimension $\chi$ can be represented by a single-set ansatz with bond dimension $
\chi \leq \chi'\leq N_{\mathrm{set}} \chi$ (smaller bond dimensions are possible if some additional low rank structure is found).  For higher-accuracy simulations, we observe that the bond dimensions required approach the upper limit, suggesting significant differences in the vibrational state wavefunctions obtained for the different electronic states.  As the cost of single-set calculations scales as $\mathcal{O}(\chi'^3)$, the need for a bond dimension $\mathcal{O}(10)$ times larger significantly complicates the task of obtaining converged dynamics.  In practice this leads to reductions of an greater than an order of magnitude in the computational times required to obtain comparable accuracy using the multi-set compared to single-set methods for this model.  While it is possible that other representations of the wavefunction may potentially lead to improved performance of single-set approaches, the present results suggest that the multi-set method provides a powerful tool for treating the complicated vibronic coupling models arising in chemical systems.

In this section, we have presented two example applications of the pyTTN software package to the simulation of exciton dynamics in molecular systems. These examples are intended to demonstrate the ease with which models arising in chemical dynamics can be set up within this software package, and the ease with which different wavefunction ans\"atze can be compared.  In the remainder of this work, we present example applications using the open quantum system simulation tools present in the pyTTN package.

\section{Illustrative Applications: Open Quantum System Dynamics \label{sec:oqs_dynamics}}
In this section we consider a series of model open quantum system problems in order to demonstrate the use of some of the open quantum system simulation tools provided by the pyTTN software package.

\subsection{Dynamics of the Spin-Boson Model}
\begin{figure}[tp]
\includegraphics[width=0.9\columnwidth]{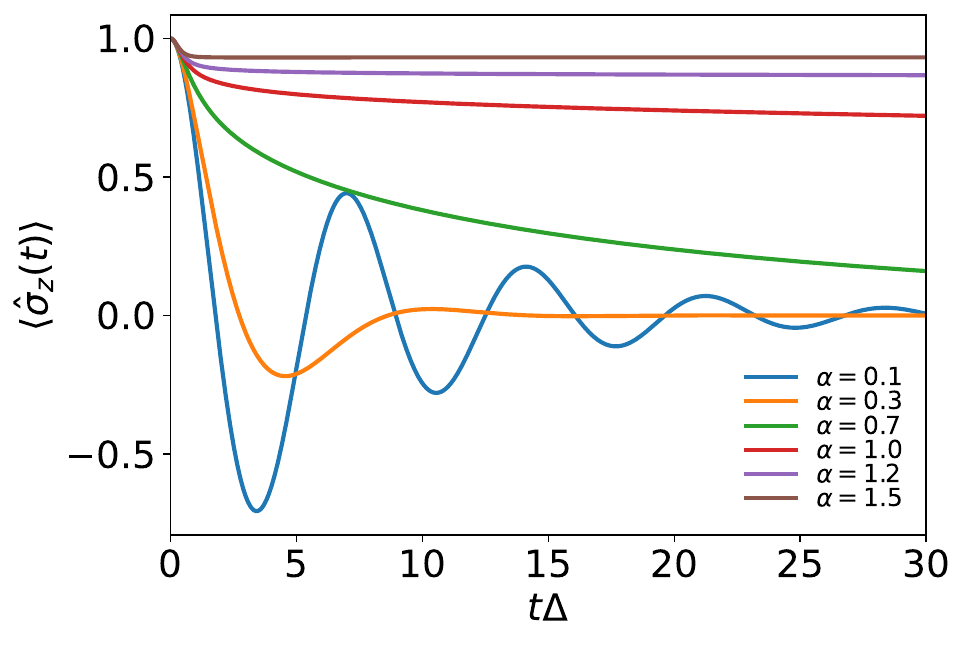} \vspace{-1em}
\caption{ \label{fig:sbm_zero_temp_alpha} The converged magnetisation dynamics, $\avg{\hat{\sigma}_z(t)}$, of the spin-boson model at $T=0$ with $\omega_c=5 \Delta$ and varying values for the Kondo parameter $\alpha$, obtained using the interaction picture chain Hamiltonian.  These results are obtained using a one-site algorithm with subspace expansion, with expansion tolerance parameters of $\epsilon_{\mathrm{S}} = 10^{-6}$ and a maximum allowed bond dimension of $\chi=32$.  A balanced binary tree representation is used for the bath.
The results presented here were generated using the scripts \protect\path{pyttn/examples/oqs/spin_boson/sbm_unitary.py}.}
\end{figure}

We start this section by considering the spin-boson model, a prototypical open quantum system model, in which a single two-level system (TLS) is linearly coupled to a harmonic bath, giving rise to a model in which the impact of the bath is entirely characterised by Gaussian fluctuations~\citep{FEYNMAN1963118, garg1985,Leggett1987,nitzan2006,Weiss2012}.
For problems involving a reservoir containing a macroscopic number of degrees of freedom couples to a small quantum system it is often sufficient to employ a Gaussian fluctuation model to capture the dynamics of the system~\cite{CALDEIRA1983374, Makri1999}. As such, in addition to arising naturally in quantum optics~\cite{PhysRevA.99.013807,PhysRevLett.111.243602}, models based on linear coupling to a harmonic bath have found application in the study of condensed phase chemical reactions~\cite{garg1985,CRAIG2007, 10.1063/1.5116800, LindoyMandalReichman, doi:10.1063/1.3611050}, decoherence of superconducting qubits due to interaction with their environment~\cite{PhysRevLett.66.810, PhysRevResearch.6.033215}, tunneling of defects in disordered metals~\cite{PhysRevB.54.4629}.
The spin-boson model Hamiltonian may be written in the form
\begin{equation}
\hat{{H}}_s=\frac{\varepsilon}{2}\hat{{\sigma}}_{z}+\frac{\Delta}{2}\hat{{\sigma}}_{x}+\hat{{\sigma}}_{z}\sum_{k=1}^Ng_{k}(\hat{{a}}_{k}^{\dagger}+\hat{{a}}_{k})+\sum_{k=1}^N\omega_{k}\hat{{a}}_{k}^{\dagger}\hat{{a}}, \label{eq:sbm_hamiltonian_2}
\end{equation}
 where $\hat{{a}}_{k}^{\dagger}$ and $\hat{{a}_{k}}$ are
the bosonic creation and annihilation operators associated with the
$k$-th bath mode. As in Sec.~\ref{sec:numerical_performance}, we consider a continuum Ohmic bath~\cite{Leggett1987} with an exponential cutoff \begin{equation}
J(\omega)= \frac{\pi}{2} \, \alpha \, \omega \, e^{-\omega/\omega_{c}}.
\end{equation}
Here we consider the spin-boson parameter regime considered in Ref.~\onlinecite{Strathearn2018}, corresponding to $\epsilon=0$, $\Delta=0.5$, $\omega_c=5$, and we consider ranges of $\alpha\in[0.1,1.5]$, which at zero temperature span the underdamped regime (at small $\alpha<0.5$),  the overdamped regime (at intermediate $\alpha$), and finally the localised regime (at large $\alpha$).  This is clearly seen in Fig.~\ref{fig:sbm_zero_temp_alpha}, where we present the zero-temperature dynamics of this model obtained using the interaction picture chain form for the bath contributions to the Hamiltonian.

In Fig.~\ref{fig:sbm_zero_temp_alpha} we consider the time-dependent polarisation dynamics of the expectation value of the two-level system's $\hat{\sigma}_z$ operator, starting from an initially uncorrelated configuration in which the system is in the $\ket{\uparrow}$ state and the bath in the vacuum state. In order to accurately capture the bath correlation function for this initial condition and up to times $t\Delta=30$, we discretise the bath using a maximum frequency cutoff of $\omega_{\mathrm{max}}=10 \, \omega_c$, and a total of $400$ bath modes obtained using the orthogonal polynomial discretisation scheme.  The dynamics we obtain agree well with the results presented in Ref.~\onlinecite{Strathearn2018} using the TEMPO approach.  In particular, we observe the expected underdamped oscillations for $\alpha < 0.5$, followed by a trend towards overdamped decays for $\alpha >0.5$, which as $\alpha$ increases further approaches the localised regime, where the magnetisation does not approach 0 at long times.

\subsubsection{Finite-Temperature Dynamics}
We next consider the dynamics of this model at finite-temperature through the use of a temperature-dependent spectral density (Eq.~\ref{eq:effective_spec_dens}).  Here we consider the interaction picture chain form of the system-bath Hamiltonian, and discretise the bath using the orthogonal polynomial based method with a total of $N=580$ bath modes. This is necessary to ensure accurate bath correlation functions up to $t\Delta=30$, owing to the wider bandwidth associated with the effective temperature-dependent spectral density.
\begin{figure}[tp]
\hspace{-2em}\includegraphics[width=0.85\columnwidth]{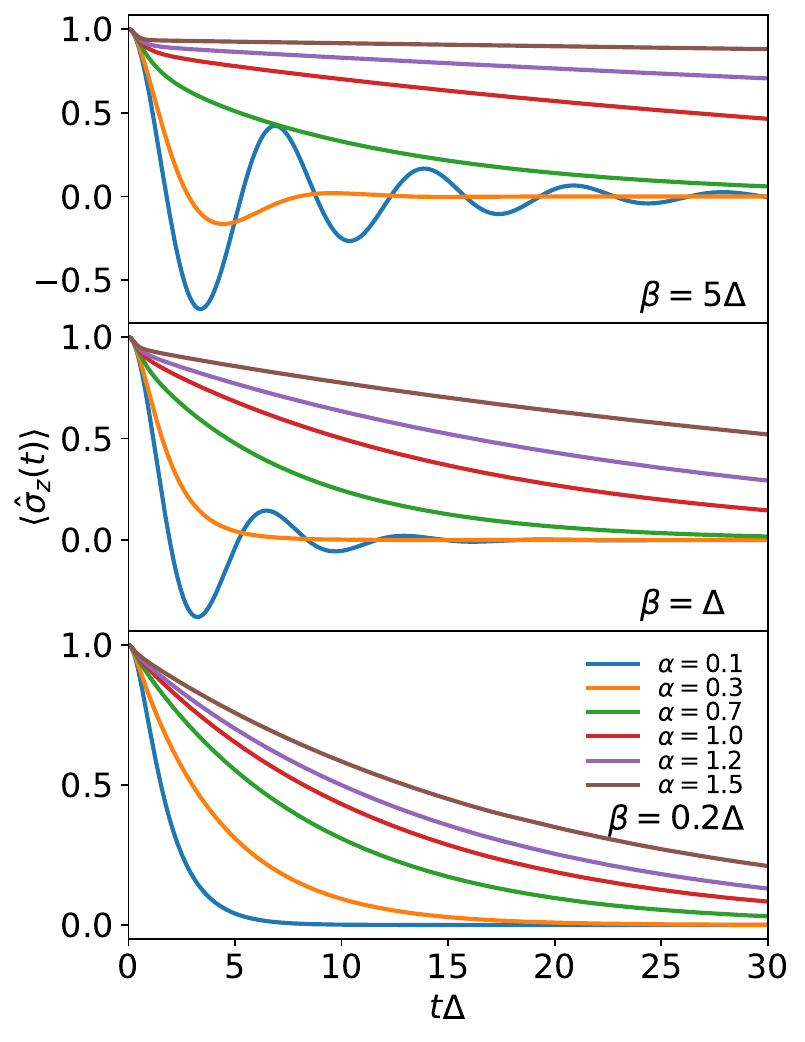}  \vspace{-1em}
\caption{ \label{fig:sbm_finite_temp_alpha} The dynamics of the magnetisation, $\avg{\hat{\sigma}_z(t)}$, of the spin-boson model obtained with the use of a thermal spectral density and the interaction picture chain Hamiltonian with $\omega_c=5\Delta$ at varying values of $\beta$ (Top: $\beta=5\Delta$, Middle: $\beta=1\Delta$, and Bottom: $\beta=0.2\Delta$). The results presented here are obtained using a balanced binary tree for handling the bath degrees of freedom; the subspace expansion integrator with a subspace tolerance of $\epsilon_S=10^{-6}$; and a maximum allowed bond dimension of $\chi=128$. The results presented in this figure were generated using the script \protect\path{pyttn/examples/oqs/spin_boson/sbm_unitary.py}.}
\end{figure}

In Fig.~\ref{fig:sbm_finite_temp_alpha} we present the polarisation dynamics obtained for a range of $\alpha$ parameters and at three different temperatures $\beta = 5\Delta, 1\Delta, $ and $0.2\Delta$.  At low temperature ($\beta=5\Delta$), the short-time dynamics are consistent with what observed at $T=0$; however, at longer times we observe the onset of the expected decay towards the equilibrium magnetisation $\langle\hat{\sigma}_z\rangle=0$, regardless of the coupling strength.  Increasing the temperature to $\beta=1\Delta$ we observe more rapid exponential decay at long times, noting a reduction in the magnitude of the oscillations in the magnetisation at weak coupling.  Further increasing the temperature to $\beta=0.2\Delta$, we observe exponential decay following a short-lived transient for all values of the Kondo parameter that have been considered.

\subsubsection{Impact of Bath Topology}
We now consider the impact of the bath topology---namely the star representation, the chain representation, and the interaction picture chain representation---on the convergence of the spin dynamics.  For ease of comparison, throughout this section we compare the accuracy of fixed-bond dimension single-site calculations.  Here we consider the case of an MPS representation for the bath degrees of freedom.

\begin{figure}[tp]
\hspace{-2em}\includegraphics[width=0.95\columnwidth]{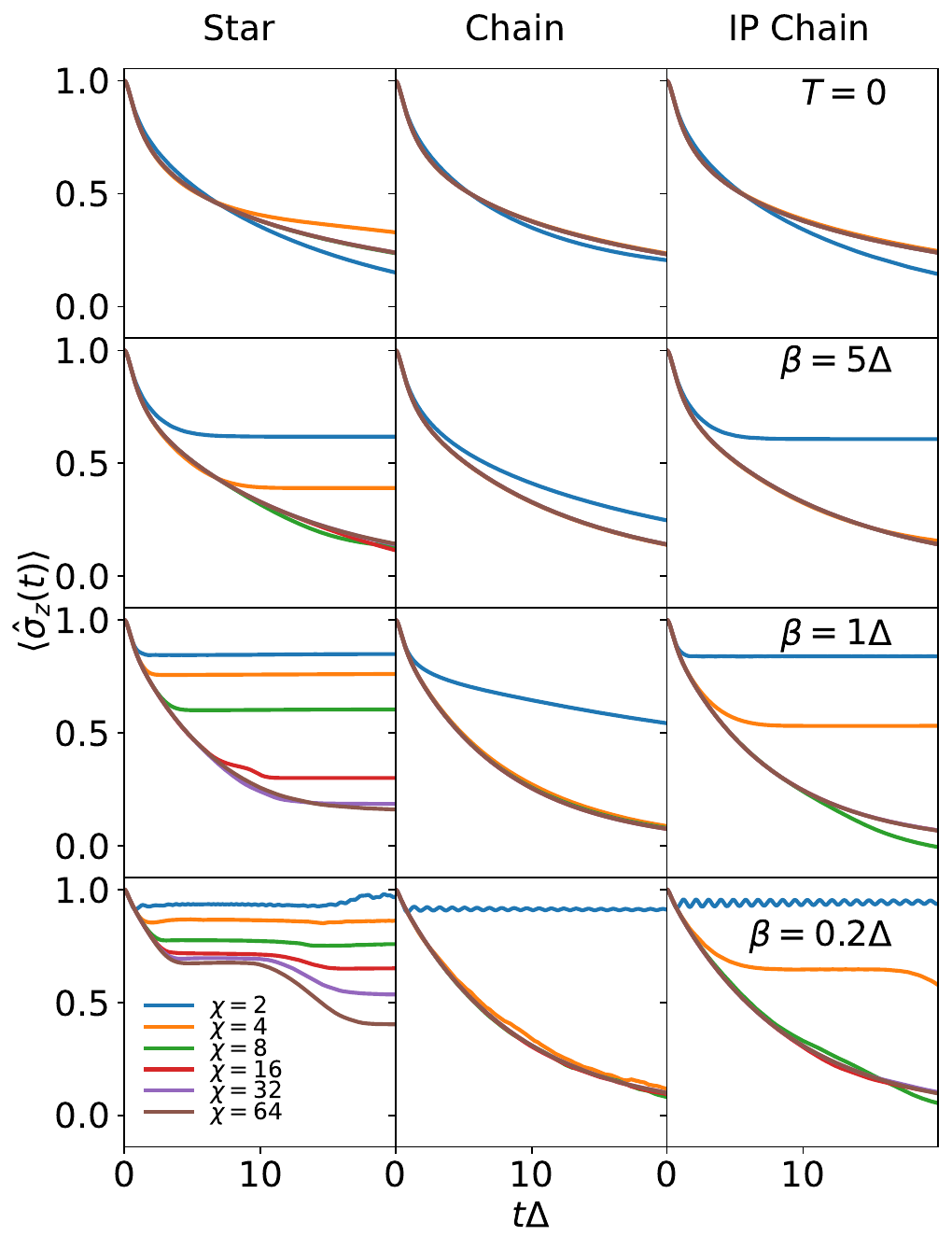}  \vspace{-1em}
\caption{ \label{fig:sbm_finite_temp_conv} The dynamics of the magnetisation, $\avg{\hat{\sigma}_z(t)}$, and its convergence with respect to bond dimension of the spin-boson model at finite temperature using the star (left), chain (middle), and interaction picture chain (right) Hamiltonians at various temperatures. From top to bottom, results are obtained at $T=0$, $\beta=5\Delta$, $\beta=1\Delta$, and $\beta=0.2\Delta$.  In all calculations an MPS topology with an optimised boson basis is used for the bath modes with fixed bond dimension, $\chi$, shown in the figures. The results presented in this figure were generated using the script \protect\path{pyttn/examples/oqs/spin_boson/sbm_unitary.py}.}
\end{figure}
In Fig.~\ref{fig:sbm_finite_temp_conv} we present the magnetisation dynamics of the spin-boson model with $\alpha=0.7$ obtained using the star, chain, and interaction picture chain representations at various fixed maximum bond dimensions and temperatures. In all representations we observe that convergence becomes more challenging upon increasing temperature, with small bond dimension sufficient for convergence for all topologies at $T=0$.  Upon increasing temperature, large bond dimensions are required for all methods.  However, we observe that the chain-topology converges most rapidly across all parameters considered, with convergence to within the thickness of the lines shown with $\chi=16$ for all temperatures considered.  The interaction picture chain results show larger deviations at a given bond dimension when compared to the chain representation, although with similar convergence characteristics.  In contrast, we observe significant convergence issues when employing the star representation at high temperatures.  In this representation we find significant deviations at high temperatures, with $\chi=64$ being insufficient to converge the long-time dynamics for $\beta\leq 1\Delta$.  For high temperature models we observe significant plateaus in the long time dynamics arising due to the use of an insufficient bond dimension.  For the case of $\beta=0.2\Delta$ these plateaus arise for $t\Delta \lesssim 4$ for all cases considered in Fig.~\ref{fig:sbm_finite_temp_conv}. The onset of this plateau shifts to longer times with increasing bond dimension; however, for this model significant increases in bond dimension only lead to minor improvements in the quality of the dynamics. Additionally, considerably larger local Hilbert space dimensions are required for the star-topology at high temperature, since the use of the thermal spectral density leads to strong coupling of low frequency boson modes to the spin dynamics.  In general, these two factors lead to simulations of the star-topology Hamiltonian being prohibitively expensive when attempting to access high-temperature dynamics of this Ohmic spin-boson model.

The details of this convergence depend on the specific tree topology used to represent the wavefunction.  For further details see Sec.~\ref{Sec:SI_sb_bath_topo} of the Supplementary Information, where analogous results are obtained using a balanced binary tree tensor network representation for the bath configurations.  When using the balanced binary tree ansatz, we find that the interaction picture chain representation exhibits the most rapid convergence with respect to bond dimension, with convergence typically observed with lower bond dimensions than those required when using the MPS bath topology.  In contrast, the convergence issues associated with the star-topology Hamiltonian persist regardless of the choice of an MPS or balanced binary tree tensor network for the bath.  The dynamics obtained using the chain representation also exhibit considerably poorer convergence when employing a balanced binary tree tensor network, with comparable performance to the star-topology.  These results demonstrate that unitary methods can represent an efficient approach for simulating the dynamics of zero- and finite-temperature open-quantum system dynamics; however, care must be taken when selecting the Hamiltonian representation and tensor network representation of the wavefunction.  Here we have considered this for a relatively simple Ohmic spin-boson model, but how well these conclusions generalise across model parameters and forms of the bath spectral density is left for future work.

\subsubsection{Non-unitary Methods: HEOM and the Pseudomode Method \label{sec:spin_boson_non-unitary}}
We now consider the application of the non-unitary HEOM and generalised pseudomode methods for simulating spin-boson models. We provide a set of examples demonstrating the use of pyTTN to simulate the HEOM and generalised pseudomode methods for the spin-boson model that are considered above using unitary dynamics schemes. In particular, we are considering intermediate-time dynamics of both zero- and finite-temperature dynamics of an Ohmic spin-boson model with an exponential cutoff. All results presented in this section were generated using the scripts \protect\path{pyttn/examples/oqs/spin_boson/sbm_heom.py} and \protect\path{pyttn/examples/oqs/spin_boson/sbm_pseudomode.py}.

The bath correlation function for the exponential cutoff form of the spectral density does not have a convenient analytic decomposition as a sum of exponential functions---in contrast to the case of rational function spectral densities. Here, following the discussion in Ref.~\onlinecite{10.1063/5.0209348}, we use the estimation of signal parameters via rotational invariant techniques (ESPRIT)~\cite{1457851, 9000636,10.1063/5.0209348,park2024quasilindbladpseudomodetheoryopen} to construct such representation, as discussed in Sec.~\ref{sec:non-unitary_dynamics}. In all calculations presented in this section we use $K=6$ exponential terms to decompose the bath correlation function using the ESPRIT method. Additionally, we employ a frequency-based cutoff scheme closely related to that presented in Ref.~\onlinecite{Lindoy2023} to determine the auxiliary density operator elements to be included within each calculation, and have ensured convergence with respect to the cutoff parameters. Within this scheme, we included all states up to an occupation $n_k$ for each bosonic mode $k$ satisfying
\begin{equation}
    n_k \leq \mathrm{max}\left(L_{\mathrm{min}}, \mathrm{min}\left(L_{\mathrm{max}}, \frac{\nu_{\mathrm{max}}}{\nu_k}\right)\right).
\end{equation}
For this system we obtain convergence using a maximum bosonic mode occupation of $L_{\mathrm{max}}=30$, a minimum value of $L_{\mathrm{min}}=6$, and a cutoff frequency of $\nu_{\mathrm{max}} = 10 \, \omega_c$.

\begin{figure}[tp]
\hspace{-2em}
\includegraphics[width=\columnwidth]{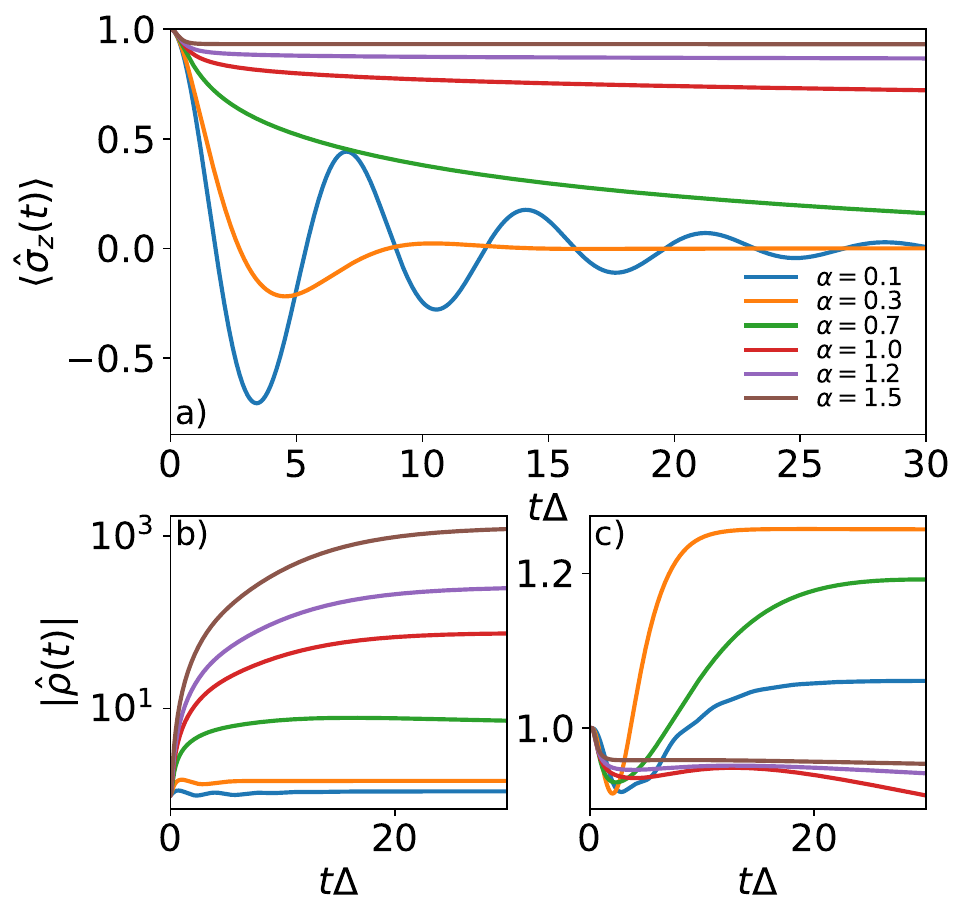}  \vspace{-1em}
\caption{ \label{fig:sbm_zero_temp_alpha_heom} The magnetisation dynamics, $\avg{\hat{\sigma}_z(t)}$, of the spin-boson model at $T=0$ with $\omega_c=5 \Delta$ and varying values for the Kondo parameter $\alpha$, obtained using the HEOM method (a). In the lower two panels we plot the norm of the auxiliary density operator vector $\left|\hat{\rho}(t)\right)$ obtained using HEOM (b) and the generalised pseudomode (c) methods.  These results are obtained using a one-site algorithm with subspace expansion, with expansion tolerance parameters of $10^{-7}$ and a maximum allowed bond dimension of $\chi=96$.  Here a balanced binary tree representation is used for the bath.
The results presented here were generated using the script \protect\path{pyttn/examples/oqs/spin_boson/sbm_nonunitary.py}.}
\end{figure}

In Fig.~\ref{fig:sbm_zero_temp_alpha_heom}a we consider the time-dependent polarisation dynamics for the expectation value of the two-level systems $\hat{\sigma}_z$ operator, obtained with a tree tensor network solver for HEOM (Eq.~\ref{eq:fpheom_op}), starting from a configuration in which the system is found in the $\ket{\uparrow}$ state and an initially uncorrelated zero-temperature bath state.  Here we observe excellent agreement with the expected dynamics (see Fig.~\ref{fig:sbm_zero_temp_alpha} for dynamics obtained using unitary methods), with only minor deviations ($\simeq \mathcal{O}(10^{-4})$) from the results obtained using unitary dynamics methods observed at later times for stronger system bath interactions.

Having confirmed the accuracy of the implementation, we can now consider various aspects of the numerical properties of each method. In the remaining panels of Fig.~\ref{fig:sbm_zero_temp_alpha_heom}, we consider the norm of the vector of auxiliary density matrix elements obtained from the HEOM calculations (b) and the equivalent quantity in the pseudomode calculations (c), the norm of the pseudomode density operator.  As is to be expected due to the non-unitary nature of the dynamics, neither method conserves this quantity; however, we observe significant differences in the behaviour of the two approaches.  In particular, we find rapid initial growth in the norm of the state vector in the HEOM calculations. In contrast, when applying the generalised pseudomode method, the norm of the state vector remains close to $1$ for all times considered here. While this difference does not significantly impact the results obtained here, the rapid growth in norm observed in the HEOM calculations has the potential to lead to numerical convergence issues when applying tensor network based solvers, in particular as the density operator is obtained from a projection onto the vacuum state of the effective boson modes.  As the norm of the vector grows, the system density operator provides a smaller relative proportion to the total norm of the vector. When applying the TDVP algorithm to evolve a tensor network representation of the ADO matrix, it is necessary to apply various linear algebra operations, such as singular values decompositions.  When the norm of vector is sufficiently large propagation of floating point arithmetic through these steps can lead to amplification of errors, and may significantly impact the accuracy of the system density matrix.

\begin{figure}[tp]
\hspace{-2em}\includegraphics[width=0.95\columnwidth]{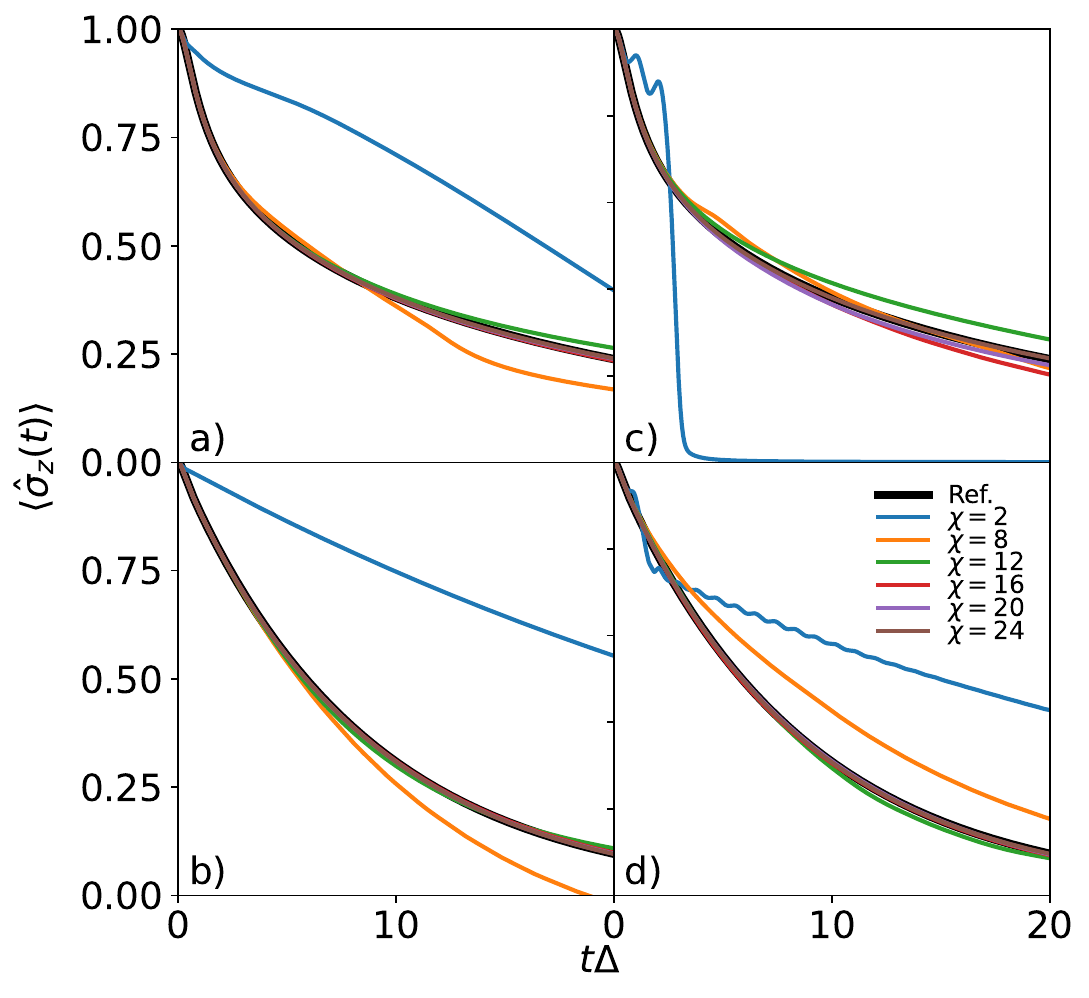}  \vspace{-1em}
\caption{ \label{fig:sbm_heom_chi_conv} Convergence with respect to bond dimension of the finite-temperature polarisation dynamics obtained using HEOM at zero temperature (a) and $\beta=0.2\Delta$ (b), and the generalised pseudomode method  at zero temperature (c) and $\beta=0.2 \Delta$ (d).   In all calculations a balanced binary tree topology is used for the bath modes with fixed bond dimension, $\chi$, shown in the figures.  Black lines labelled reference correspond to the converged results obtained from the unitary dynamics calculations presented in Fig.~\ref{fig:sbm_finite_temp_alpha}. The non-unitary dynamics results presented here were generated using the script \protect\path{pyttn/examples/oqs/spin_boson/sbm_nonunitary.py}.}
\end{figure}

Finally, we consider the impact of the choice of representation, namely the Lindblad form as compared to the normal-mode representation employed within HEOM (see Sec.~\ref{sec:supplementary_fpheom} of the Supplementary Information for further details), on the numerical performance of the tensor network based solver.  In Fig.~\ref{fig:sbm_heom_chi_conv} we present the convergence of the dynamics obtained from HEOM and pseudomode calculations for a spin-boson model with $\alpha=0.7$ at zero temperature and high temperature $\beta=0.2\Delta$ for several fixed bond dimension used for a balanced binary tree tensor network representation for the bath states.  Here we observe a weak dependence on the required bond dimension to obtain convergences, which is achieved to within the thickness of the line in all cases by $\chi\sim 24$.  At zero temperature we observe slightly more rapid convergence when using HEOM compared to the pseudomode approach, obtaining converged results for $\chi=16$ with HEOM, whereas significant deviations are still observed for $\chi=20$ with the pseudomode methods.

\begin{figure}[h]
\hspace{-2em}
\includegraphics[width=\columnwidth]{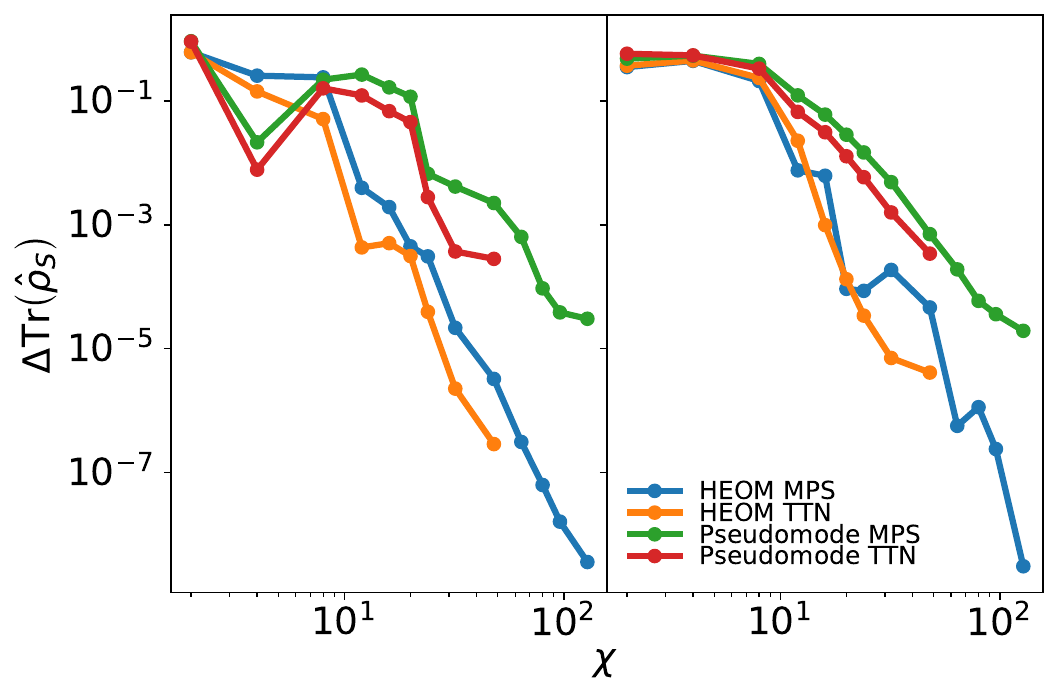}  \vspace{-1em}
\caption{ \label{fig:sbm_trace_conservation} Convergence of the trace of the density matrix, as measured by the time averaged absolute deviation of the trace (Eq.~\ref{eq:trace_error}), obtained using HEOM and pseudomode method calculations as a function of the bond dimension used.  Here we present results at zero temperature (left) and $\beta=0.2 \Delta$ (right) and consider both a MPS and balanced binary tree tensor network (labelled TTN) representation of the bath ADO hierarchy.  The results presented here were generated using the script \protect\path{pyttn/examples/oqs/spin_boson/sbm_nonunitary.py}.}
\end{figure}

In general, tensor network based solvers for the HEOM or generalised pseudomode methods are not guaranteed to conserve the trace of the reduced system density operator~\cite{10.1063/5.0027962}. These conservation rules have previously been enforced within HEOM calculations through the use of Lagrange multiplier constraints~\cite{10.1063/5.0027962}, with analogous approaches possible for the generalised pseudomode method; however, here we monitor trace convergence as an additional convergence parameter in the simulation.  In Fig.~\ref{fig:sbm_trace_conservation}, we consider a metric for trace conservation, defined by
\begin{equation}
   \Delta \mathrm{Tr}\left(\hat{\rho}_S\right) = \frac{1}{t_{\mathrm{max}}}\int_{0}^{t_{\mathrm{max}}} \|\mathrm{Tr}\left(\hat{\rho}_S(t)\right) - 1\| \mathrm{d}t, \label{eq:trace_error}
\end{equation}
where $t_{\mathrm{max}}\Delta =30$, as a function of the bond dimension used in HEOM and pseudomode calculations using either a MPS or balanced binary TTN representation for the bath degrees of freedom.  Here we consider results with $\alpha=0.7$ and at zero temperature and $\beta=0.2\Delta$, observing broadly consistent results at both temperatures.  As to be expected, as we increase the bond dimensions the TTN ansatz becomes more expressive and is better able to reflect the true dynamics of the system giving rise to a systematic reduction in the deviation of the trace.  In general, we find that the use of a balanced binary tree tensor network for the ADO degrees of freedom leads to a slight improvement in the trace conservation compared to the use of an MPS representation with the same bond dimension for both HEOM and pseudomode, albeit for additional memory and computation costs.  Additionally, we observe that the TDVP-based HEOM approach significantly outperforms the TDVP-based generalised pseudomode method.  This is illustrated by a $1-2$ order of magnitude reduction in the deviation in the system density operator trace for HEOM compared to the generalised pseudomode method at a given bond dimension for both choices of tensor network.  Failure to conserve this quantity significantly impacts the estimation of observables, and largely accounts for the long-time errors observed in the zero-temperature pseudomode calculations obtained with $\chi=16$ and 20. Hence, accurate conservation of the trace is important for large-scale tensor network based HEOM calculations. A potential avenue for future work is determining optimal schemes for enforcing trace conservation within tensor network-based HEOM and pseudomode calculations.  In particular, it is an open question as to whether Lagrange multiplier constraint based approaches~\cite{10.1063/5.0027962} are optimal, or whether it is possible to construct approaches analogous to the matrix product density operators (MPDO),~\cite{PhysRevLett.93.207204} which are commonly used in the tensor network communities for treating unitary, and Lindblad dynamics of density operator, suitable for use with the HEOM and pseudomode methods.

Finally, it is worth noting that both the HEOM and pseudomode method calculations required considerably less CPU time compared to the unitary dynamics based approaches,  with larger improvements observed in the strong coupling, high temperature regimes.  In particular, for $\alpha=1.5$ and $\beta=0.2\Delta$ the unitary dynamics calculations required $\mathcal{O}(10)$ CPU hours compared to the $~\sim 15$ minutes required for the HEOM calculations for the same spin-boson parameters. From a comparison of Figs.~\ref{fig:sbm_finite_temp_conv} and \ref{fig:sbm_heom_chi_conv}, we observe that similar bond dimensions are required to obtain accurate dynamics using both unitary and non-unitary approaches for this spin-boson model. However, due to the dramatic reduction in the number of bosonic modes needed to accurately approximate the bath correlation function over this time period using HEOM, combined with the ESPRIT method, compared to the unitary approach, we observe a dramatic reduction in CPU time.

In addition to the examples presented here, the example folder (\protect\path{pyttn/examples/spin_boson_model}) contains implementations of the Minimally Entangled Typical Thermal States (METTS) algorithm~\cite{white2009, Stoudenmire2010, binder2017} for the evaluation of the thermal (\path{pyttn/examples/oqs/spin_boson/sbm_metts.py}) and symmetrised \path{pyttn/examples/oqs/spin_boson/sbm_symmetrised_metts.py} correlation functions of the spin boson model, using the procedure as outlined in Ref.~\onlinecite{10.1063/5.0224880}. In contrast to the uncorrelated initial conditions consider within the more standard open quantum system treatments, in which the system and bath start to interact at time $t=0$, METTS allows for the study of the dynamics when the system and bath are initially at thermal equilibrium and a perturbation is applied at time $t=0$, a case commonly encountered in experiments.

\subsection{Dissipative XY Spin Models}
As a more challenging open quantum system application we consider the spread of a spin excitation in a set of anisotropic XY spin-$\frac{1}{2}$ models, where each spin is coupled to a local dissipative bosonic bath.  We consider a Hamiltonian of the form
\begin{equation}
\begin{split}
    \hat{H} &=\!\! \sum_{\langle i,j \rangle}\left( (1-\eta)\hat{S}_{xi}\hat{S}_{xj}\!+\!(1+\eta)\hat{S}_{yi}\hat{S}_{yj}\right) \\
    +\!&\sum_{i}\! \hat{S}_{zi}+\!\!\sum_{i} \sum_{k} \left(g_k(\hat{a}_{ik}^\dagger+a_{ik}) \hat{S}_{zi} + \omega_k \hat{a}_{ik}^\dagger \hat{a}_{ik}\right),
\end{split}
\end{equation}
where $\hat{S}_{\alpha i}$ is the $\alpha \in x,y,z$ spin-$\frac{1}{2}$ operator for spin $i$, and $\langle i, j \rangle$ denotes sets of pairs connected by the coupling topology. We consider two cases for the coupling topology. In the first case, we consider a 21-site anisotropic spin chain where each spin is locally coupled to a dissipative bath. This model has previously been treated using process tensor matrix product operator (PT-MPO) based methods~\cite{PhysRevResearch.5.033078}, and represents a more challenging problem than the spin-boson model due to the considerably large system Hilbert space dimension making direct HEOM or pseudomode calculation impractical. The second model we consider contains 22 spins with a degree 3 Cayley tree coupling graph. For each case we consider anisotropic spin interactions, with $\eta=0.04$, and consequently the total magnetisation of the system is not conserved throughout the dynamics.

\begin{figure}[tp]
\centering
\includegraphics[width=0.85\columnwidth]{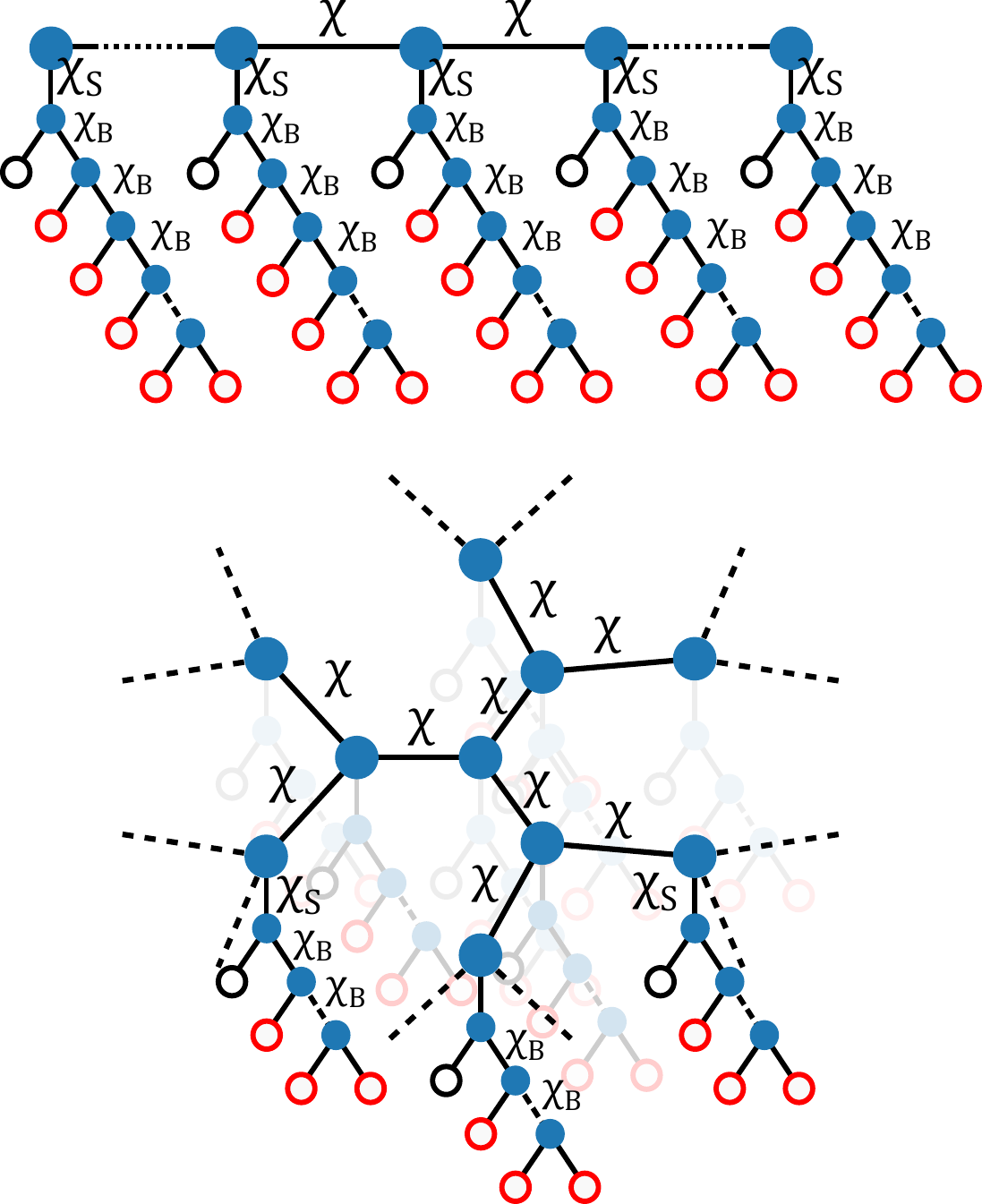}  \vspace{-1em}
\caption{ \label{fig:dissipative_xy_fmps} Schematic of the tensor network topologies used for the chain (top) and Cayley tree (bottom) dissipative XY models. For the dissipative spin chain, this topology is commonly referred to as a fork-MPS.  Black and red nodes correspond to the physical degrees of freedom of the system with the black nodes representing the spin degrees of freedom and the red nodes representing potentially composite bosonic bath modes.  The blue nodes are interior nodes and are responsible for handling correlations between the physical degrees of freedom. }
\end{figure}

Following Ref.~\onlinecite{PhysRevResearch.5.033078}, we consider an Ohmic bath spectral density with Gaussian cutoff, given by
\begin{equation}
    J(\omega) = \pi \sum_{k} g_k^2 \, \delta(\omega - \omega_k) = 2\pi \, \alpha \, \omega \, \exp\left(-\frac{\omega^2}{\omega_c^2}\right).
\end{equation}
For these problems, the number of degrees of freedom of the system is large enough that a full treatment of the system density operator is impractical.  Within equation of motion based approaches it is possible to address the dynamics of this system by using a tensor network ansatz for both the bosonic (pseudo-)bath modes and the system degrees of freedom.  For these calculations, we use a matrix product state representation for each spin and local bath, and use a network with the same topology as the interaction Hamiltonian for handling the coupling between each local site.  These topologies are illustrated in Fig.~\ref{fig:dissipative_xy_fmps}. In all calculations in this section we fix the maximum bond dimension that each bond can take, with $\chi$ defining the maximum bond dimension for bonds connecting different spin degrees of freedom, $\chi_S$ the maximum number of functions used to represent each spin and bath state, and $\chi_B$ the maximum bond dimension associated with each bath chain.

\subsubsection{XY Spin Chain}
\begin{figure}[h!]
\includegraphics[width=0.95\columnwidth]{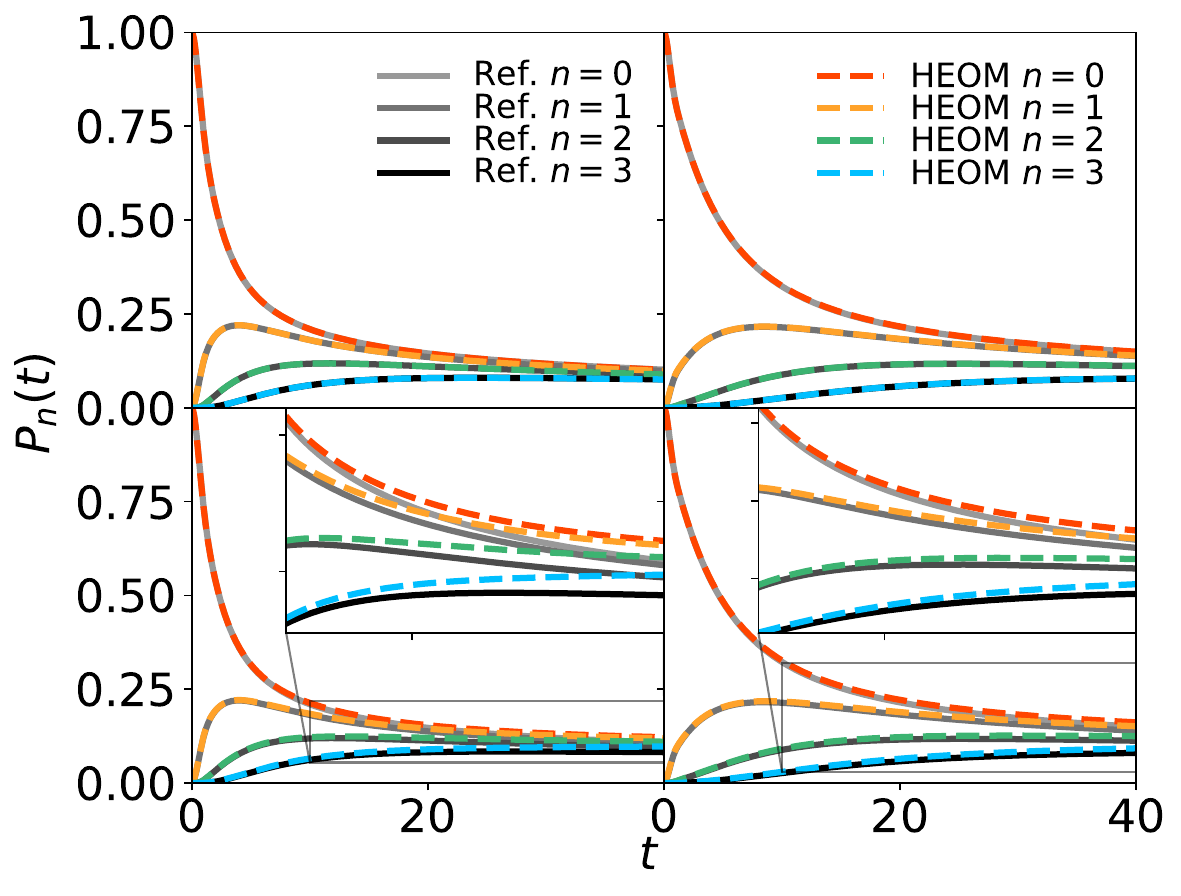} \vspace{-1em}
\caption{ \label{fig:dissipative_xy_dynamics} Population dynamics $P_n(t) = \frac{1}{2} - \left\langle S_{zn}(t)\right\rangle$ obtained for the four sites $n=0, 1, 2, 3$ for the isotropic ($\eta=0$ top) and anisotropic ($\eta=0.04$ bottom) cases obtained at $\beta = 0.625$ for system-bath coupling strengths $\alpha=0.16$ (left) and $\alpha=0.32$ (right).  For the HEOM calculations we have used the fork-MPS topology shown in Fig.~\ref{fig:dissipative_xy_fmps} with $\chi=128$, $\chi_S=96$, and $\chi_B=128$ .  In each of these panels we compare the HEOM results against the PT-MPO based results presented in Ref.~\onlinecite{PhysRevResearch.5.033078}.  The HEOM results presented in this figure were generated using the script \protect\path{pyttn/examples/dissipative_xy/dissipative_xy_nonunitary.py}.}
\end{figure}

We start by treating the model problems considered in Ref~\onlinecite{PhysRevResearch.5.033078}, corresponding to finite temperature with $\beta=0.625$, and with different couplings strengths $\alpha=0.16$ and $0.32$ and different values for the anisotropy, $\eta=0$ and $0.04$.  In order to simulate this regime we use the example script \protect\path{pyttn/examples/dissipative_xy/chain/dissipative_xy_nonunitary.py}, which provides an implementation of the Free-Pole HEOM method for the simulation of the dissipative XY chain model. We use the ESPRIT method to extract the decomposition of the bath correlation function into a sum of exponential terms and obtain accurate representations of the bath correlation functions with $K=6$ terms, which correspond to $12$ effective bosonic modes per spin.  We consider a system in which there is an initial spin excitation at spin $0$, with all other spins in the down state, and all baths initially in thermal equilibrium with respect to the bath only Hamiltonian. For all calculations, an integration timestep of $\delta t=0.025$ is used and the hierarchy of ADOs is truncated using the scheme discussed in Sec.\ref{sec:spin_boson_non-unitary}, with a maximum hierarchy depth of $L_{\mathrm{max}}=30$, a minimum depth of $L_{\mathrm{min}}=6$, and a cutoff frequency of $\nu_{\mathrm{max}} = 15 \, \omega_c$.  Details of the convergence of the HEOM calculations for this model are provided in Sec.~\ref{sec:si_dissipative_xy_chain} of the Supplementary Information.

\begin{figure}[h!]
\includegraphics[width=0.95\columnwidth]{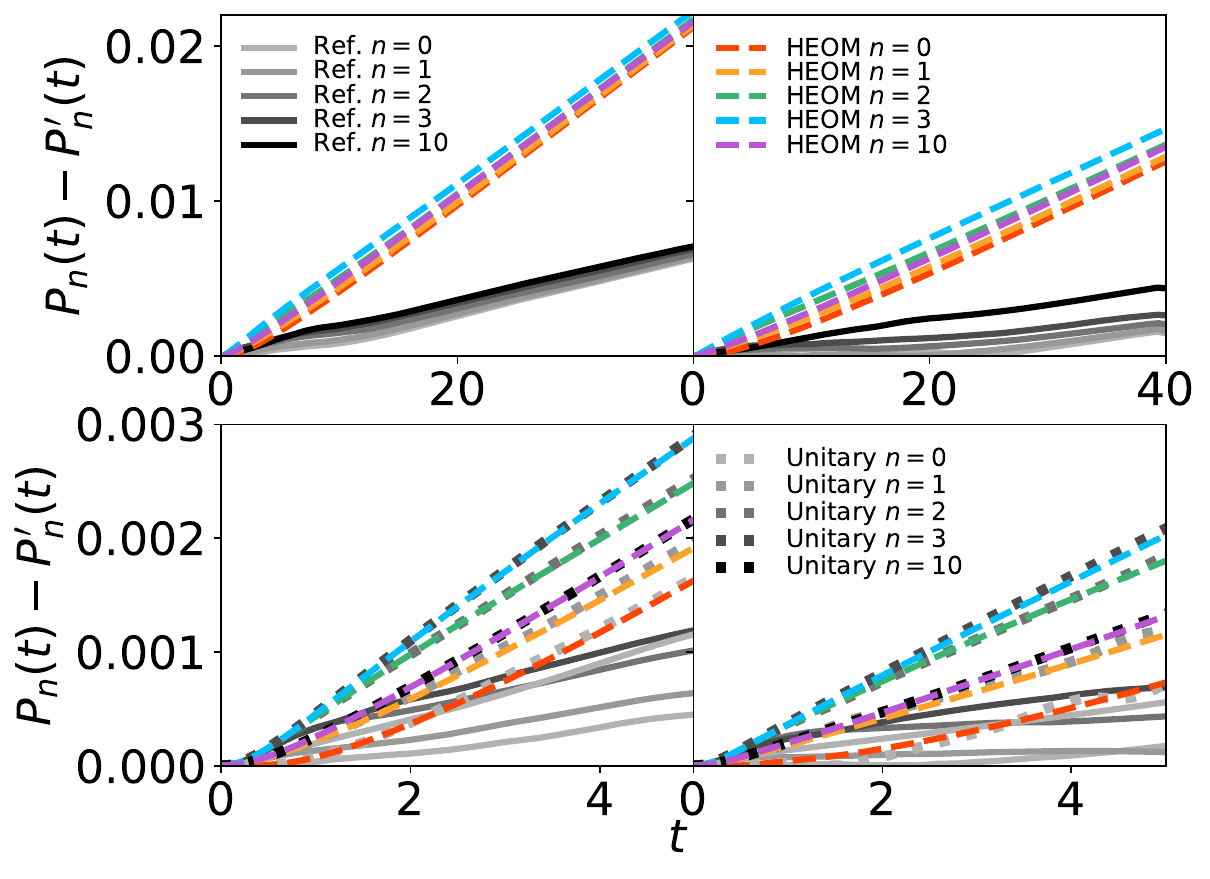} \vspace{-1em}
\caption{ \label{fig:dissipative_xy_dynamics_anisotropy} Difference in population dynamics obtained from isotropic ($\eta=0$) $P_n(t)$ and anisotropic ($\eta=0.04$) $P_n^\prime(t)$ for the four sites $n=0, 1, 2, 3$ and the site at the end of the chain $n=10$ at $\beta = 0.625$ for system bath coupling strengths $\alpha=0.16$ (left) and $\alpha=0.32$ (right).  In the top panels, we compare the HEOM results against the PT-MPO based results presented in Ref.~\onlinecite{PhysRevResearch.5.033078}.  The bottom panels highlight the short-time behaviour $t<5$ and additionally provide a comparison against results obtained with the unitary T-TEDOPA approach.  The HEOM results presented in this figure were generated using the script \protect\path{pyttn/examples/dissipative_xy/chain/dissipative_xy_nonunitary.py} while the unitary dynamics results were obtained using the script \protect\path{pyttn/examples/dissipative_xy/chain/dissipative_xy_unitary.py}.}
\end{figure}

In Fig.~\ref{fig:dissipative_xy_dynamics}, we compare the population dynamics, defined by $P_n(t) = \frac{1}{2} - \left\langle S_{zn}(t)\right\rangle$, for sites $n=0,1,2,3$ (the central site and the three sites to its right), obtained using the HEOM approach to the results presented in Ref.~\onlinecite{PhysRevResearch.5.033078}. For the isotropic case ($\eta=0$), we observe excellent agreement between the HEOM and the reference results for the entire time period considered.  In contrast, for the anisotropic case with $\eta=0.04$, significant deviations between the two approaches are observed, as highlighted in the insets.  In particular, we observe that the HEOM calculations predict larger populations for each spin site at long time, with this trend continuing for $n>3$ (not shown), and the deviation growing roughly linearly with time. This becomes more evident when considering the difference in the population dynamics between the isotropic and anisotropic cases, as shown in Fig.~\ref{fig:dissipative_xy_dynamics_anisotropy}.

In the upper panels of Fig.~\ref{fig:dissipative_xy_dynamics_anisotropy} we compare the difference in the population dynamics for the isotropic and anisotropic obtained using the HEOM calculations and the results presented in Ref.~\onlinecite{PhysRevResearch.5.033078}.  These results show clear differences in the dynamics obtained even at short times ($t<5$).  The bottom panels present a comparison of the HEOM dynamics and unitary dynamics obtained using the thermal spectral density and with a chain representation for the bath degrees of freedom (that is, the T-TEDOPA method), as implemented within the script \protect\path{pyttn/examples/dissipative_xy/chain/dissipative_xy_unitary.py}.  For the unitary dynamics calculations we have used baths containing $N=70$ bath modes, which is sufficient to converge the bath correlation functions over the time period considered. We note that in order to consider $t=40$ considerably larger baths are required.  These calculations used the tensor network topology presented in the top panel of Fig.~\ref{fig:dissipative_xy_fmps}, with maximum allowed bond dimensions, $\chi=160$, $\chi_S=120$, and $\chi_B=160$. Due to the considerably larger number of bath modes required and the larger bond dimensions needed to converge these finite-temperature results, it was not feasible to obtain converged dynamics up to $t=40$ using the unitary approach.  However, from these short-time results we observe excellent agreement between the unitary and HEOM calculations, and find with both approaches a consistently larger growth in the difference between the isotropic and anisotropic simulation than was observed in Ref.~\onlinecite{PhysRevResearch.5.033078}.

\begin{figure}[h!]
\includegraphics[width=0.95\columnwidth]{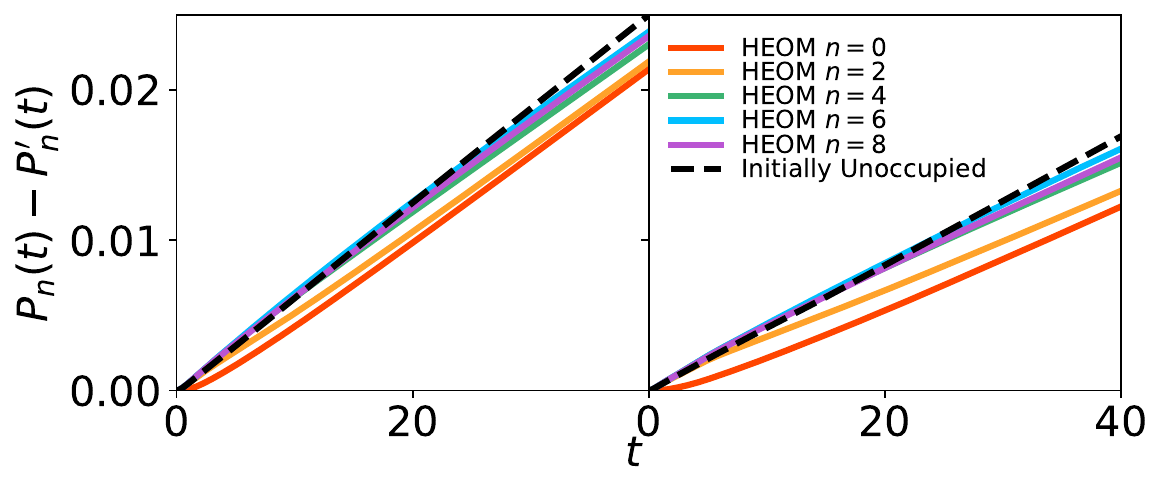} \vspace{-1em}
\caption{ \label{fig:dissipative_xy_dynamics_unoccupied} The difference in the population obtained dynamics obtained from isotropic ($\eta=0$) $P_n(t)$ and anisotropic ($\eta=0.04$) $P_n^\prime(t)$ for the sites $n=0, 2, 4, 6, 8$  at $\beta = 0.625$ following an initial excitation of site $n=0$ compared to the results obtained with no initial spin excitation.  Here we present results for system bath coupling strengths $\alpha=0.16$ (left) and $\alpha=0.32$ (right), and all results were generated using the script \protect\path{pyttn/examples/dissipative_xy/chain/dissipative_xy_nonunitary.py}.}
\end{figure}

The HEOM results demonstrate a clear linear deviation between the isotropic and anisotropic populations at long time.  At short times we observe that the difference in populations is largest for spins far from both the site of the initial excitation and the ends of the chain, with sites $n=0, 1$ and $10$ showing significantly slower growth in the difference in the population dynamics for the isotropic and anisotropic Hamiltonians.  This can be attributed to the anisotropic terms being of the form $\hat{S}_{+,i}\hat{S}_{+,i+1}+\hat{S}_{-,i}\hat{S}_{-,i+1}$, where the $\hat{S}_{\pm, i}$ are the ladder operators associated with spin $i$.  Consequently, the matrix element associated with the anisotropic terms between spins $i$ and $i+1$ is zero whenever these spins are not aligned.  For spins far from either the end of the chain or the initial excitation, we observe initial dynamics that closely resemble the dynamics observed for a chain with no initial spin excitation.  In Fig.~\ref{fig:dissipative_xy_dynamics_unoccupied}, we compare how the difference in population dynamics presented in Fig.~\ref{fig:dissipative_xy_dynamics_anisotropy} compare to the dynamics obtained for an initially unoccupied chain, here obtained using a chain of 5 spins with periodic boundary conditions.  For sites far from the initial excitation we observe excellent agreement with the unoccupied chain results up until the time at which the excitation reaches the site.  The unoccupied bath calculations require considerably fewer resource to converge the dynamics, and support the validity of the results we have obtained in the initially singly excited chain case.

\subsubsection{XY Spin Tree}
We now go beyond the simple chain geometry and consider the zero-temperature dynamics of a set of dissipative XY spin models on a degree 3 Cayley tree containing a total of 22 spins.  This corresponds to a central node (spin $0$, and labelled as layer $0$ below), and three layers of the tree structure labelled as layers $1$, $2$, and $3$, with these three layers containing the spins $S_{L}=\{1,2,3\}$, $\{4,5,6,7,8,9\}$, and $\{10, 11, 12, 13, 14, 15, 16, 17, 18, 19, 20\}$, respectively.

We consider the unitary dynamics of this system starting from an initial configuration in which the central spin is in the $\ket{\uparrow}$ configuration and all other spins are in the state $\ket{\downarrow}$, and where each of the local bosonic baths is in the vacuum state.  The results presented in this section were generated using the script \protect\path{pyttn/examples/dissipative_xy/cayley_tree/dissipative_xy_unitary.py}, which demonstrates some of the tools present within pyTTN to facilitate generation of more complicated tree structures.

\begin{figure}[h]
\includegraphics[width=0.95\columnwidth]{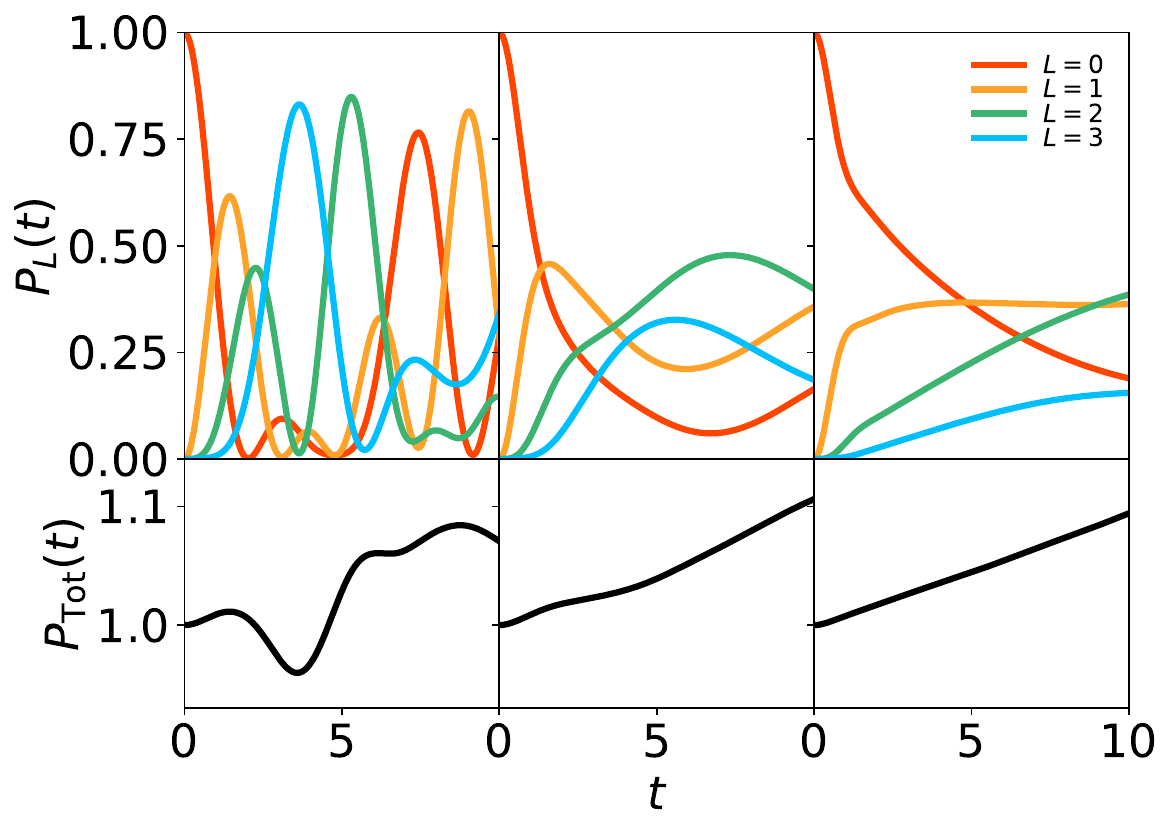} \vspace{-1em}
\caption{ \label{fig:dissipative_xy_tree_dynamics} Top) Total layer population dynamics, $P_L(t)$ obtained for the four different layers of the Cayley tree structure considered for the anisotropic ($\eta=0.04$) dissipative XY tree model obtained at $T = 0$ for system-bath coupling strengths of $\alpha=0$ (left), $\alpha=0.16$ (centre), and $\alpha=0.32$ (right).  Bottom) The total population dynamics across all spins as a function of time for the same models.  For the HEOM calculations we have used the tree structure topology shown in the bottom panel of Fig.~\ref{fig:dissipative_xy_fmps} with $\chi=48$, $\chi_S=64$, and $\chi_B=96$.   The results presented in this figure were generated using the script \protect\path{pyttn/examples/dissipative_xy/cayley_tree/dissipative_xy_unitary.py}.}
\end{figure}

In Fig.~\ref{fig:dissipative_xy_tree_dynamics} we present the dynamics of the total population of spin excitations associated with each layer of the Cayley tree
\begin{equation}
    P_{L}(t) = \sum_{n \in S_L} \left( \frac{1}{2} - \left\langle S_{zn}(t) \right\rangle\right),
\end{equation}
where $S_L$ corresponds to the set of sites that are a  distance $L$ from the central site of the Cayley tree. Throughout the dynamics, we would expect the excitation initially localised on the central site to spread layer-by-layer throughout the Cayley tree, in addition to extra spin excitations due to the aniotropic contributions.  In Fig.~\ref{fig:dissipative_xy_tree_dynamics}, we consider the case of an anisotropy of $\eta=0.04$ and system-bath coupling strengths of $\alpha=0$, $0.16$, and $0.32$. For $\alpha=0$ we observe rapid spread of the excitation throughout the layers of the tree observing transfer of population to the outer layer by $t=4$. Following this we observe recurrences with significant transfer of population back to the central site by $t=8$. For the cases of $\alpha>0.16$ and $0.32$, we observe significant slowdown in the transfer of spin excitations between layers.  For $\alpha=0.16$ we find signatures of the coherent transfer observed in the $\alpha=0$ case, although a significant broadening of peaks is observed.  For $\alpha=0.32$ we find slow population transfer, and the significant peaks in population are strongly suppressed.

In the bottom panels of Fig.~\ref{fig:dissipative_xy_tree_dynamics} we consider the total population of spin excitations.  Due to the anisotropic terms, the total population is not conserved. In all cases, the total population generally increases with time.  For $\alpha=0$, we observe a suppression of the population around $t=4$ corresponding to the time at which the excitation reaches the edge of the tree structure.  Introducing the system-bath coupling, these oscillations are suppressed and at $\alpha=0.32$ a monotonic increase in the total spin excitation population is observed. Due to the larger bond dimension required to obtain converged HEOM results, and the higher degree of connectivity of the spin nodes in tensor network representation of this system (Fig.~\ref{fig:dissipative_xy_fmps}), it was not practical to obtain converged results using the HEOM approaches for this model.

\subsection{Dynamics of a Single Impurity Anderson Model (SIAM) \label{sec:siam}}
Finally, we consider an application to a fermionic impurity model.  Here we consider a single-impurity Anderson model with two impurity spin-orbitals (with annihilation operators $\hat{d}_\uparrow$ and $\hat{d}_\downarrow$) that each independently couple to free Fermion baths ($\hat{c}_{k\sigma}$).  The Hamiltonian for this model is of the form
\begin{equation}
\begin{split}
    \hat{H} &= \sum_{\sigma} \epsilon \hat{d}^\dagger_\sigma\hat{d}_\sigma + U \hat{n}_{\uparrow} \hat{n}_\downarrow \\&+ \sum_{k\sigma} V_k \left(\hat{d}_\sigma^\dagger \hat{c}_{k\sigma} + \hat{c}_{k\sigma}^\dagger \hat{d}_{\sigma}\right) + \sum_{k\sigma} \epsilon_{k\sigma} \hat{c}^\dagger_{k\sigma}\hat{c}_{k\sigma}.
    \end{split}
\end{equation}
 We consider a flat-band hybridization function of width $W$, which sets the energy scales of the problem 
\begin{equation}
    V^2(x) = V^2 \,  \Theta(W-|\omega|),
\end{equation}
and explore how the impurity depth $-\epsilon$ and the strength of the hybridization interaction $V^2$ impact the dynamics of the impurity magnetisation. We start from an initial state where the impurity is in the state $\hat{c}_{\uparrow}^\dagger\ket{0}$ and the bath is uncorrelated from the system and at thermal equilibrium at a range of temperatures. Further, we consider the case where the interaction term $U=15 \, W$ is the largest energy scale in the problem, strongly suppressing the double-impurity occupancy state. This setup is similar to the unbiased lead case considered in Ref.~\onlinecite{PhysRevB.109.115101}, and is relevant to experimental studies of quantum dots.
\begin{figure}[tp]
     \centering
     \includegraphics[width=\linewidth]{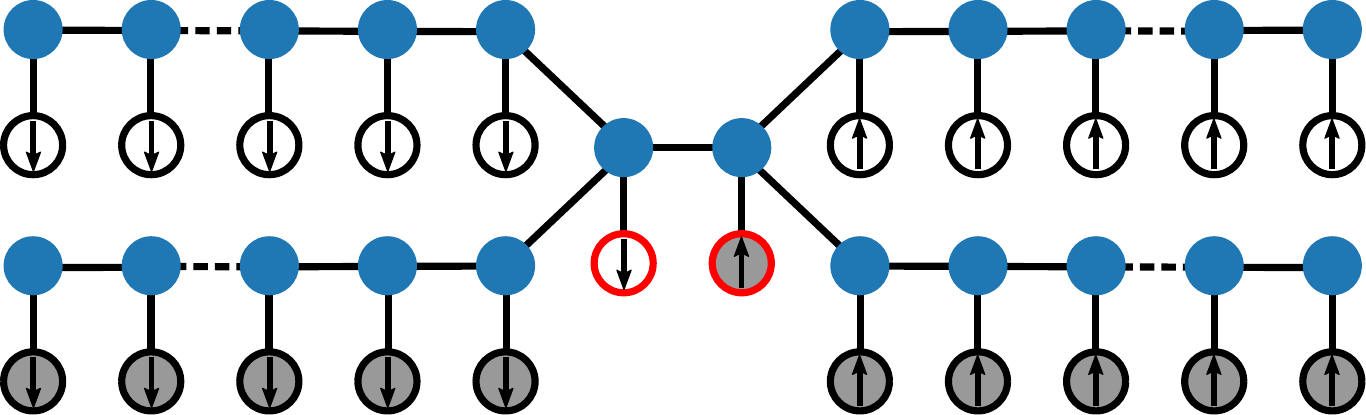}
     \caption{
     \label{fig:siam_ttn_schematic} Schematic of the Tree Tensor Network state used in the calculations, with the initial product state configuration shown by illustrating the filled (gray) and empty (white) orbitals. The left side of the structure handles down spin orbitals whereas the right half handles up spins, and the two impurity orbitals (red outlined circles) are contained at the centre of the tree structure. We used a thermal spectral density and have transformed that bath orbitals to a representation containing (temperature-dependent) initially filled and unfilled orbitals. }
 \end{figure}

As with the spin-boson model, it is possible to perform unitary transformations to the bath orbital to change the coupling topology.  In what follows, we employ a mapping introduced in Ref.~\onlinecite{PhysRevB.104.014303} that maps each bath onto two chains corresponding to the initially occupied and unoccupied bath orbitals.  Here we use unitary dynamics methods for simulating this problem, that is, we consider a discretised version of the continuous hybridisation, and we ensure that we have employed enough bath orbitals to accurately converge dynamics up to the timescale of interest.  Finite-temperature effects are included using a thermal spectral density approach~\cite{PhysRevB.104.014303}, in which we introduce a set of ancilla bath fermions ($\hat{c}_{2k\sigma}$) in addition to our original set of bath fermions, which we now label as $\hat{c}_{k\sigma}\rightarrow\hat{c}_{1k\sigma}$, and that have the same on-site energy as the original bath fermion of the same index $k$ and do not directly couple to the impurity degree of freedom ($V_{2k\sigma}=0$).  We consider an initial configuration in which both the bath and ancilla bath Fermions are initially in their filled non-interacting Fermi sea state, e.g. $\ket{\Psi} = \ket{\mathrm{FS}_1} \otimes \ket{\mathrm{FS}_2}$.  Defining a new set of fermionic operators by
 \begin{equation}
 \begin{pmatrix}
     \hat{f}_{1k} \\
     \hat{f}_{2k} \\
 \end{pmatrix} = \begin{pmatrix}
     \cos(\theta_k) & -\sin(\theta_k) \\
     \sin(\theta_k) & \cos(\theta_k) \\
 \end{pmatrix}
  \begin{pmatrix}
     \hat{c}_{1k} \\
     \hat{c}_{2k} \\
 \end{pmatrix} ,
 \end{equation}
and making the choice $\sin^2(\theta_k) = N(\epsilon_k; \epsilon_F, \beta)$, where
\begin{equation}
    N(\epsilon; \epsilon_F, \beta) = \begin{cases}
        \frac{1}{\exp(\beta (\epsilon-\epsilon_F)) + 1} & \beta < \infty\\
        1-\Theta(\epsilon-\epsilon_F) & \beta \rightarrow \infty
    \end{cases}
\end{equation}

is the Fermi function and $\epsilon_F$ is the Fermi energy, giving $\hat{f}_{1k} \ket{\Psi} = 0$ and $\hat{f}^\dagger_{2k} \ket{\Psi} = 0$. This means that our initial state is the vacuum of the $\hat{f}_{1k}$ fermions and the fully filled state of the $\hat{f}_{2k}$ fermions.  Additionally, we find that
\begin{equation}
    \bra{\Psi}\hat{c}_{1k}^\dagger \hat{c}_{1k} \ket{\Psi} = \sin^2(\theta_k),
\end{equation}
so this pure state acting on the space of original and ancilla fermions captures thermal expectation values of the states of the original fermions.

 \begin{figure}[tp]
     \hspace{-2em}
     \includegraphics[width=0.95\linewidth]{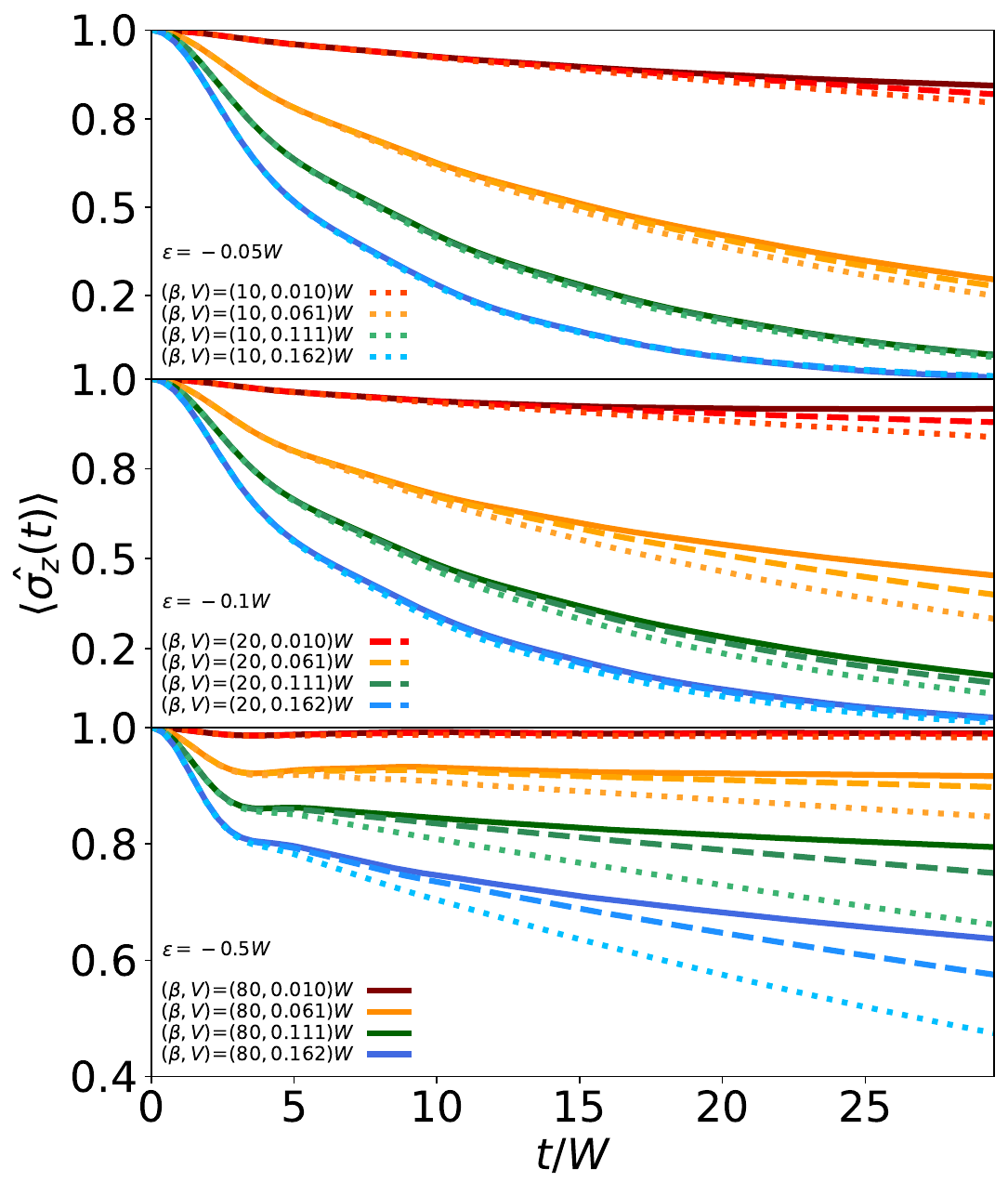} \vspace{-1em}
     \caption{
     \label{fig:siam_results}  Impurity magnetisation dynamics obtained at various temperatures ($\beta^{-1}$) and hybridisation strengths, $V$, (shown in captions) for three different impurity depths ($\epsilon=-0.05 \, W$ (top), $\epsilon=-0.1 \, W$ (middle) $\epsilon=-0.5 \, W$ (bottom).  In each panel results for all three temperatures and four hybridisation strengths considered are shown. These results were generated using the script \protect\path{pyttn/examples/anderson_impurity_model/siam.py}.}
 \end{figure}

In order to represent the state of this wavefunction, we use the tree tensor network ansatz illustrated in Fig.~\ref{fig:siam_ttn_schematic}.  Within this representation, we have split the system into spin-up and spin-down degrees of freedom.  For each spin sector, we split the bath into filled and empty orbitals and employ an energy ordering of the bath orbitals, a representation expected to give well-controlled growth of entanglement in time for the zero-temperature bath regime~\cite{PhysRevB.96.085107}. All calculations are performed using the TDVP algorithm with adaptive subspace expansion using a tolerance of $\epsilon_S=10^{-6}$.  We restrict the maximum bond dimension of the bond separating the two spin sectors to $\chi_S = 64$, and use a constant maximum allowed bond dimension throughout the bath chains of $\chi_B = 180$. We find that the bond dimension required to obtain accurate long-time dynamics increases significantly with increasing temperature.  This is consistent with Ref.~\onlinecite{PhysRevB.96.085107}, as an increase in temperature leads to an increase in the overlap of the effective hybridisation functions used within the thermofield mapping onto filled and unfilled chains, consequently increasing the bond dimension required to capture the entanglement within the system.  In all calculations, we have converged the dynamics with respect to the number of bath orbitals, and in all calculations presented below we have used $N=200$ bath orbitals for each of the two baths corresponding to the different spin states.  These simulations were performed using the example script \protect\path{pyttn/examples/anderson_impurity_model/siam.py}.

 Fig.~\ref{fig:siam_results} shows the impurity magnetisation dynamics for different values of the impurity depth $-\epsilon$,  temperature $\beta^{-1}$, and hybridisation strength $|V|$.   Following a temperature independent transient, we observe a long-time exponential decay towards the equilibrium magnetisation.  For $\epsilon=-0.05 \, W$, this long-time decay shows a comparably weak temperature dependence, with the significance of temperature reducing as the hybridisation strength increases.  In this regime, in which the impurity is near the Fermi energy, the impurity is freely able to hybridise with the unoccupied bath degrees of freedom, allowing for rapid population transfer between impurity and bath orbitals.  As we increase the depth of the impurity below $\epsilon_F$, the rate of decay of magnetisation reduces significantly, and additionally temperature has a significant larger impact on the resultant decay timescales. For $\epsilon =-0.1 \, W$, we observe a significant temperature dependence of the long-time exponential decay of the magnetisation for $V < 0.1 \, W$, however, for larger $V$ in which the interaction energy between the impurity and bath is larger than the impurity depth we again observe a roughly temperature independent decay of the magnetisation.  For $\epsilon=-0.5 \, W$, the temperature-dependent exponential decay is observed for all hybridisation couplings considered.


\section{Conclusion\label{sec:conclusions}}
In this paper we have introduced the pyTTN software package, a code that implements the projector splitting algorithm based implementation of the multilayer multiconfiguration time-dependent Hartree method.  This software package is designed to be user friendly, and includes several features to this aim, such as efficient, symbolic generation of Hamiltonian operators and adaptive subspace based expansion of the bond dimensions in the network, which enable a relatively low entry barrier to setting up simulations for new systems.

We have demonstrated the use of these tools for efficiently constructing and simulating the dynamics of both open and closed quantum systems, employing a range of different representations of the problem.  We have first considered several applications to molecular systems.  In particular, we have applied pyTTN to the simulation of the unitary dynamics of a widely used benchmark model for the photo-excitation dynamics of a 24 mode pyrazine model, in which we were able to reach convergence of the auto-correlation function and associated spectra as a function of the number of single-particle functions ($\chi$).  Following this, we have considered a model for exciton dynamics at the interface of a $n$-oligothiophene donor-C$_{60}$ fullerene acceptor system, in which we demonstrated the utility of the multi-set ansatz implemented within pyTTN for obtaining long time dynamics of excitonic system that are not routinely accessible with ML-MCTDH or MPS based calculations for this system due to the large bond dimensions required for converged dynamics.  Importantly, we observe systematic improvement of the ML-MCTDH and MPS based calculations towards the multi-set results, suggesting that previously reported discrepancies between the methods can likely be attributed to convergence issues. These results demonstrate the utility of multi-set tensor network methods for the treatment of exciton transfer dynamics beyond the idealised Holstein models considered in Ref.~~\onlinecite{PhysRevLett.123.126601}.

We have next considered applications to open quantum system dynamics.  We first considered the spin-boson model in which a spin couples to a bosonic bath.  We have applied both unitary dynamics based approaches and non-unitary HEOM and pseudomode based methods for the spin-boson case.  Within the unitary dynamics framework, we have provided examples showing the use of different bath topologies for treating the dynamics of the system, and have explored the impact of this choice on the numerical performance at both zero and non-zero temperatures.  We find that both chain and interaction picture chain representations of the Hamiltonian provide compact representations of the wavefunction suitable for capturing dynamics of the spin-boson model to long times at both zero and finite temperature, in stark contrast to the use of the star topology which requires considerably enhanced.  However, we observe that the interaction picture chain form exhibits less strong dependence of the performance on the choice of tensor network topology.
For the non-unitary dynamics simulations, we observe considerably improved performance, however, find that in general the pseudomode method, when coupled with a TDVP based tensor network solver, demonstrates significantly larger violations of trace conservation. 

We next considered a series of more challenging dissipative spin models, namely spin chains and trees with each spin coupled to a local dissipative bath. Here we have also employed both unitary and non-unitary approaches.  The methods discussed throughout these examples are relevant to simulations of environment-induced noise in coupled superconducting qubits~\cite{PhysRevResearch.6.033215}, and accurate modeling of such situations including contributions from additional two-level system fluctuations~\cite{Agarwal_2024} is a potential avenue for future work. This section concluded with  treatment of a fermionic impurity model, the Anderson impurity model, in which a single impurity orbital couples to a free electron bath.  Here we presented a series of zero- and finite-temperature calculations obtained using unitary dynamics methods to explore spin decoherence of an initially spin-polarised impurity due to interactions with electronic reservoirs.

This software package is under continuous development, for an up to date status of the features of the software please visit \url{https://gitlab.npl.co.uk/qsm/pyttn}.
Future releases will aim to extend the open quantum system interface further to provide a more black-box approach for treating general open quantum systems.  Additionally, the inclusion of the fermionic HEOM~\cite{10.1063/1.2938087,10.1063/1.2713104,10.1063/1.5034776,PhysRevB.107.195429} and pseudofermion methods~\cite{PhysRevResearch.5.033011,park2024quasilindbladpseudomodetheoryopen} is planned for future work.
One key aspect that limits black-box usage of the code is the need to specify tree structures.  To help mitigate the impact of this, pyTTN provides several helper functions to enable easy construction of common tree structures and modification of the tree structures.  In recent years, several approaches have been developed for automatic construction of tree topologies~\cite{PhysRevResearch.5.013031,Mendive-Tapia2023,10.1093/ptep/ptad018}, the implementation of which is an area for future work.  Additionally, the inclusion of particle number conservation and other abelian symmetries, as are commonly used within Matrix Product State based tensor network codes~\cite{10.21468/SciPostPhysCodeb.4,10.1063/5.0180424, 10.21468/SciPostPhysCodeb.41, renormalizer}, which both improve reliability and considerably improved efficiency when working with symmetric problems, is planned for future releases.

Finally, we reiterate that in addition to the examples discussed in the paper, we have provided a set of tutorials outlining the use of the core features of the library:
\begin{itemize}[leftmargin=*,topsep=0pt]
    \item The construction of tree tensor networks with arbitrary topologies (\protect\path{pyttn/tutorials/tree_topologies.ipynb}).
    \item The generation of arbitrary operators using both predefined (supporting a standard set of spin, bosonic, and fermionic operators) and user-defined single-site operators, with support for both sparse and dense formats for matrix representations of operators (\protect\path{pyttn/tutorials/operator_generation.ipynb}).
    \item The use of some additional features of the pyTTN package that enable efficient simulation of open quantum system dynamics (including both bosonic and fermionic baths) with wavefunction and pseudomode/hierarchical equations of motion (HEOM) based approaches (found in \protect\path{pyttn/tutorials/oqs/}).
\end{itemize}
With these features, we expect that this software package provides a simple-to-use and useful tool for both the simulation of the quantum dynamics of molecular systems as well as for large-scale open quantum system dynamics simulations. Additionally, we expect that the features provided by pyTTN for the use of tensor network methods for the simulation of non-markovian open quantum system dynamics, and the implementation of several different approaches within the same package, may help facilitate benchmarking efforts to better understand the performance of various open quantum system approaches in different parameter regimes.

    {\noindent}{\bf Code availability}: The pyTTN software package is available on Gitlab at \url{https://gitlab.npl.co.uk/qsm/pyttn}.  Example scripts are provided as part of the repository that can be used to generate all results presented in this manuscript.

    {\noindent}{\bf Acknowledgements}:
    L.P.L. acknowledges helpful discussions with Nathan Ng.
    This work was supported by the Engineering and Physical Sciences Research Council [grant EP/Y005090/1], Innovate UK - UKRI through grant 1468/10074167, and the UK Government Department of Science, Innovation and Technology through the UK National Quantum Technologies Programme.


    {\noindent}{\bf Conflict of Interest}:
    The authors have no conflicts to disclose.

    {\noindent}{\bf Data Availability}:
    The data that support the findings of this study are available upon reasonable request.  

\bibliographystyle{aipnum4-2}
\bibliography{references}
\clearpage
\newpage
\onecolumngrid
\title{Supplementary Information}
\widetext
\begin{center}
\textbf{\large Supplementary Information: pyTTN: An Open Source Toolbox for Quantum Dynamics Simulations
Using Tree Tensor Network States}
\end{center}
\setcounter{equation}{0}
\setcounter{figure}{0}
\setcounter{table}{0}
\setcounter{page}{1}
\setcounter{section}{0}
\makeatletter
\renewcommand{\theequation}{S\arabic{equation}}
\renewcommand{\thefigure}{S\arabic{figure}}
\makeatother

\section{Comparison of hSOP and SOP representations for Quartic Hamiltonians \label{sec:si_hsop_vs_sop}}
\begin{figure}[h]
\includegraphics[width=0.6\columnwidth]{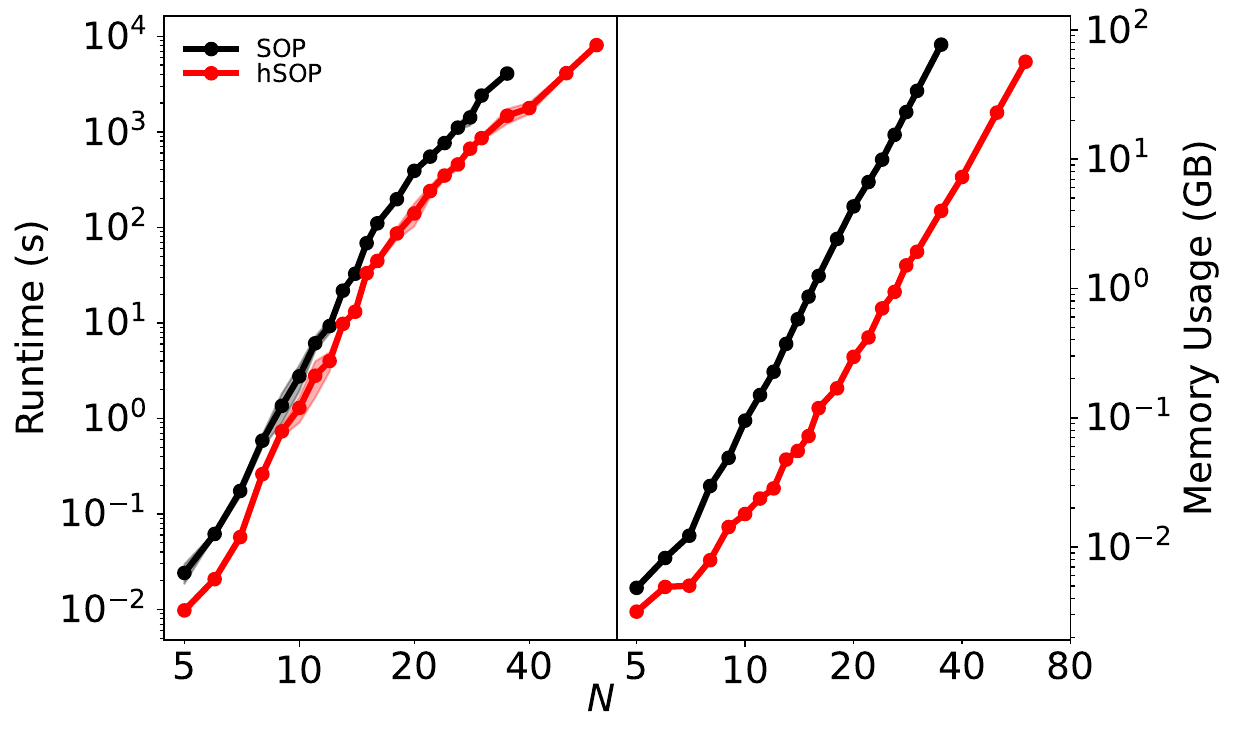} 
\caption{\label{fig:si_es_scaling} Left) The average walltime cost (over 5 steps) of a one-site TDVP step for the electronic structure Hamiltonian as a function of the number of electronic spin-orbitals included.  Right) The peak memory used by pyTTN to perform the integration step with this Hamiltonian.  In both panels we compare results obtained using the SOP and hSOP representations, obtained for a balanced binary tree tensor network representation of the wavefunction. For all calculations we have used a constant bond dimension of $\chi=20$, and have considered random coefficients for the kinetic energy and potential matrix elements.
The results presented here were generated using the script \protect\path{pyttn/examples/scaling/electronic_structure_hamiltonian.py}.} 
\end{figure}

 In Fig.~\ref{fig:si_es_scaling}, we compare the performance of the SOP and hSOP representations for TDVP simulations of the electronic structure Hamiltonian,
 \begin{equation}
     \hat{H} = \sum_{ij}^N h_{ij} \hat{c}^\dagger_i \hat{c}_j + \sum_{ijkl}^N U_{ijkl} \hat{c}^\dagger_i \hat{c}^\dagger_j \hat{c}_l \hat{c}_k,
 \end{equation}
 with varying numbers of orbitals, $N$.
 Here we make use of a balanced binary tree tensor network representation of the wavefunction with $\chi=20$ throughout, and without the use of mode combination.  We  observe an approximately constants speed up in performance when using the hSOP representation compared to the SOP representation, however, in contrast to the main text this does not lead to an asymptotic improvement in the runtime scaling with the number of spin orbitals. As discussed in the main text, here the dominant numerical cost arises from Hamiltonian evaluations across bonds in the tensor network that are close to partitioning the system into approximately equal halves.  The cost of this scales as $\mathcal{O}(N^4)$, regardless of whether the SOP or hSOP representation is used.  The hSOP representation can provide improvements when evaluating interaction terms towards the leaves of the tree, but the overall scaling remains the same as the SOP representation as it does not impact the dominant costs significant.  However, the used of the hSOP representation does reduce the total number of SPF and SHF matrices that need to be stored in the evaluation of the a single TDVP/DMRG sweep.  As shown in the right panel of Fig.~\ref{fig:si_es_scaling} this does significantly reduce the memory requirements needed to perform calculations, enabling signficantly larger system sizes to be treated.

The use of approximate techniques may allow for a reduction of the $\mathcal{O}(N^4)$ scaling observed here. The use of approximate tensor network compression techniques may reduce the bond dimension of the constructed hSOP objects. Alternatively, recent work has explored the use of the canonical polyadic decomposition-based compression schemes~\cite{10.1063/5.0192012} to construct more compact sum-of-product operator representations for the full Hamiltonian.  As an alternative, it may be possible to reduce the number of terms in the Hamiltonian through the use of strategies that screen long-range 2 electron integrals through the use of localised bases, reducing the scaling to $\mathcal{O}(N^2)$~\cite{10.1063/1.2345196,doi:10.1021/acs.jctc.8b00780,10.1063/1.5047207,PhysRevB.107.205119,10.1063/5.0025055}.  A comparison of such approximate schemes is left for future work.  
\pagebreak
\section{Equivalence of Free-Pole HEOM and the Generalised Pseudomode Method up to a Similarity Transformation \label{sec:supplementary_fpheom}}

The Hierarchical Equations of Motion (HEOM) and pseudomode methods are two approaches for obtaining numerically exact quantum dynamics of systems linearly coupled to Gaussian baths.  In this section we show that the generalised Quasi-Lindblad pseudomode method can be transformed into the Free Pole HEOM method through the use of a normal mode transformation in the auxiliary boson space.  The strategy for doing this is very closely related to the work presented in Ref.~\onlinecite{xu2023universalframeworkquantumdissipationminimally} for transforming from FP-HEOM form to a more common pseudomode representation.

\subsection{Free Pole Hierarchical Equations of Motion}
The Hierarchical Equations of Motion (HEOM) method is a numerically exact method for the dynamics of a quantum subsystem bi-linearly coupled to a Gaussian bath~\cite{TanimuraJPSJ1989,IshizakiJPSJ2005,TanimuraJCP2020}.The HEOM method makes use of the Vernon-Feynman influence functional representation of the reduced system dynamics in terms of the bath correlation functions.  By employing a discretisation of the bath correlation function in terms of a sum of decaying exponential 
\begin{equation}
    C(t) = \sum_{k=1}^K \alpha_k e^{-\nu_k t}, \label{si:eq:corr_func_expansion} 
\end{equation}
it becomes possible to construct a linear set of equations of motion for the reduced system density operator and an infinite set of auxiliary density operator (ADO), that fully captures the dynamics of the system.  A general form for the FP-HEOM is:
\begin{equation}
\begin{split}
    \frac{\partial}{\partial t}\hat{\rho}_{\boldsymbol{m},\boldsymbol{n}}(t) = &- \left(i\mathcal{L}_s(t) + \sum_{k=1}^K (\nu_k m_k + \nu_k^* n_k) \right)\hat{\rho}_{\boldsymbol{m}, \boldsymbol{n}}(t)\\&- i\sum_{k=1}^K \left[\sqrt{m_k \alpha_k}\hat{S} \hat{\rho}_{\boldsymbol{m}_k^-, \boldsymbol{n}}(t)  - \sqrt{n_k\alpha_k^*}\hat{\rho}_{\boldsymbol{m}, \boldsymbol{n}_k^-}(t)\hat{S}\right] \\ & - i\sum_{k=1}^K \left[\sqrt{(m_k +1)\alpha_k} \left[\hat{S}, \hat{\rho}_{\boldsymbol{m}_k^+, \boldsymbol{n}}(t)\right] + \sqrt{(n_k +1)\alpha_k^{*}} \left[\hat{S}, \hat{\rho}_{\boldsymbol{m}, \boldsymbol{n}_k^+}(t)\right]\right], 
\end{split}\label{eq:fpheom_SM_rescaled}
\end{equation}
where $\boldsymbol{m}_k^\pm$ ($\boldsymbol{n}_k^\pm$) corresponds to the set of indices $\boldsymbol{m}$ ($\boldsymbol{n}$) but with the $k$-th element incremented (+) or decremented (-) by one.  Here there are a total of $2K$ bath modes for the $K$ terms in the bath correlation function.  Truncation of this dynamics at some fixed depth of the hierarchy provides an efficient scheme for generating open-quantum system dynamics. Now, as introduced in Ref.~\onlinecite{10.1063/1.3125003}, it is possible to perform an arbitrary rescaling of the ADOs,
\begin{equation}
\hat{\rho}_{{\boldsymbol{m},\boldsymbol{n}}} \rightarrow \left(\prod_{k=1}^K c_k^{-m_k}c_k^{*-n_k} \right) \hat{\rho}_{{\boldsymbol{m},\boldsymbol{n}}}
\end{equation}
in order to rewrite the HEOMs as  
\begin{equation}
\begin{split}
    \frac{\partial}{\partial t}\hat{\rho}_{\boldsymbol{m},\boldsymbol{n}}(t) = &- \left(i\mathcal{L}_s(t) + \sum_{k=1}^K (\nu_k m_k + \nu_k^* n_k) \right)\hat{\rho}_{\boldsymbol{m}, \boldsymbol{n}}(t)\\&- i\sum_{k=1}^K \left[c_k^{-1} \sqrt{m_k \alpha_k}\hat{S} \hat{\rho}_{\boldsymbol{m}_k^-, \boldsymbol{n}}(t)  - c_k^{*-1}\sqrt{n_k\alpha_k^*}\hat{\rho}_{\boldsymbol{m}, \boldsymbol{n}_k^-}(t)\hat{S}\right] \\ & - i\sum_{k=1}^K \left[c_k\sqrt{(m_k +1)\alpha_k} \left[\hat{S}, \hat{\rho}_{\boldsymbol{m}_k^+, \boldsymbol{n}}(t)\right] + c_k^{*}\sqrt{(n_k +1)\alpha_k^{*}} \left[\hat{S}, \hat{\rho}_{\boldsymbol{m}, \boldsymbol{n}_k^+}(t)\right]\right], 
\end{split}\label{eq:fpheom_SM}
\end{equation}
\subsection{Quasi-Lindblad Pseudomode Methods}
Within the generalised Quasi-Lindblad pseudomode method~\cite{park2024quasilindbladpseudomodetheoryopen}, we once again consider a discretisation of the bath correlation function in terms of a sum of decaying exponentials
\begin{equation}
    C(t) = \sum_{k=1}^K (V_k-iM_k) e^{-(\gamma_k + i E_k) t}(V_k+iM_k)^*,\label{si:eq:corr_func_expansion} 
\end{equation}
where $E_k, \gamma_k \in \mathbb{R}$.  
Within this approach, the reduced system dynamics is obtained by evolving a generalised Quasi-Lindblad equation of motion, in which the system couples to a set of $K$ auxiliary pseudo-boson modes.  This Quasi-Lindblad equation can be written as~\cite{park2024quasilindbladpseudomodetheoryopen}
\begin{equation}
\begin{split}
    \frac{\partial}{\partial t}\hat{\rho}(t) =& -i \left(\mathcal{L}_s(t)\hat{\rho} + \sum_{k=1}^K (E_k \hat{a}_k^\dagger\hat{a}_k\hat{\rho} - E_k \hat{\rho} \hat{a}_k^\dagger\hat{a}_k) \right) -i \sum_{k=1}^K 2i\gamma_k \left(\hat{a}_k \hat{\rho} \hat{a}_k^\dagger -\frac{1}{2}\left[\hat{a}_k^\dagger\hat{a}_k\hat{\rho}+ \hat{\rho}\hat{a}_k^\dagger\hat{a}_k\right]\right) \\
    & -i \sum_{k=1}^K 2iM_k \left[ \hat{a}_k\hat{\rho} \hat{S}- \frac{1}{2} \left(\hat{S}\hat{a}_k \hat{\rho}  + \hat{\rho}\hat{S}\hat{a}_k\right)\right] -i \sum_{k=1}^K 2iM_k^* \left[ \hat{S}\hat{\rho}\hat{a}_k^\dagger  - \frac{1}{2} \left(\hat{S}\hat{a}_k^\dagger \hat{\rho}  + \hat{\rho}\hat{S}\hat{a}_k^\dagger\right)\right]\\
    &- i\sum_{k=1}^K\left(\hat{S}(V_k^* \hat{a}_k^\dagger + V_k\hat{a}_k)  \hat{\rho}- \hat{\rho}\hat{S}(V_k^*\hat{a}_k^\dagger + V_k\hat{a}_k) \right).
\end{split}\label{eq:pseudomode_SM}
\end{equation} 
We now move to a Liouville space representation for the bosonic degrees of freedom, using an orthonomal ket-basis of the form $\ket{\boldsymbol{n}, \tilde{\boldsymbol{m}}} $, to represent the combined ket-bra state $\ket{\boldsymbol{n}}\bra{\boldsymbol{m}}$, where $\ket{\boldsymbol{n}}$ is the commonly used occupation number basis of the pseudomode bosons.  In moving to this representation we define \emph{tilde} creation $\hat{\tilde{a}}_k^\dagger$ and annihilation operators $\hat{\tilde{a}}_k$ by
\begin{equation}
    \begin{split}
        &\tilde{a}_k \ket{\boldsymbol{n}, \tilde{\boldsymbol{m}}} = \ket{\boldsymbol{n}}\bra{\boldsymbol{m}} \hat{a}_k^\dagger \\
        &\tilde{a}_k^\dagger \ket{\boldsymbol{n}, \tilde{\boldsymbol{m}}} = \ket{\boldsymbol{n}}\bra{\boldsymbol{m}} \hat{a}_k.
    \end{split}
\end{equation}
With this definition it is straight to demonstrate that the tilde and non-tilde operators satisfy the standard bosonic commutation relations, and therefore can be treated as representing independent bosonic degrees of freedom.  Additionally we introduce the system superoperators $\hat{S}^L$ and $\hat{S}^R$ defined by 
\begin{equation}
    \begin{split}
        &\hat{S}^L\hat{O} = \hat{S} \hat{O} \\
        &\hat{S}^R\hat{O} = \hat{O} \hat{S} \\
    \end{split}
\end{equation}

Using these definitions, we may rewrite the Quasi-Lindblad pseudomode equation as 
\begin{equation}
\begin{split}
    \frac{\partial}{\partial t}\hat{\rho}(t) =& -i \mathcal{L}_s(t)\hat{\rho} \! - i\sum_{k=1}^K \left(2i\gamma_k \hat{a}_k\hat{\tilde{a}}_k \!
    -\!i\left[\nu_k \hat{a}_k^{\dagger}\hat{a}_k\! +\!\nu_k^* \hat{\tilde{a}}_k^{\dagger}\hat{\tilde{a}}_k\right]\right)\hat{\rho} -i \sum_{k=1}^K2i \left[  M_k ^*\hat{S}^L\hat{\tilde{a}}_k + M_k \hat{S}^R\hat{a}_k \right] \hat{\rho}\\
    &- i\sum_{k=1}^K \left( \left[V_k^* - i M_k^*\right] \hat{S} ^L\hat{a}_k^{\dagger} \!  -\!\left[V_k + i M_k\right] \hat{S}^R\hat{\tilde{a}}_k^{\dagger} \right)\hat{\rho}
    - i\sum_{k=1}^K \left(\left[V_k - i M_k \right]\hat{S} ^L \hat{a}_k \!-\! \left[V_k^* + i M_k^* \right] \hat{S}^R\hat{\tilde{a}}_k   \right)\hat{\rho},
\end{split}\label{eq:pseudomode_superop}
\end{equation}
where $\nu_k = \gamma_k + i E_k$.  If we now only consider terms that act on the bath degrees of freedom, we obtain an harmonic, non-hermitian bath operator of the form 
\begin{equation}
\begin{split}
    \hat{M}_B = \sum_{k=1}^K \left(2i\gamma_k \hat{a}_k\hat{\tilde{a}}_k \!
    -\!i\left[\nu_k \hat{a}_k^{\dagger}\hat{a}_k\! +\!\nu_k^* \hat{\tilde{a}}_k^{\dagger}\hat{\tilde{a}}_k\right]\right).
\end{split}\label{eq:pseudomode_bath_ham}
\end{equation}
This bath Hamiltonian has a block diagonal coupling matrix structure with no coupling terms between Liouville space bosons associated with different $k$ values.  For each $k$ the diagonal block has the form 
\begin{equation}
\begin{split}
    \hat{M}_{B, k} &= \! \frac{1}{2}\!\!\begin{pmatrix}
        \hat{a}_k^{\dagger}\! \\  \hat{\tilde{a}}_k^{\dagger} \!\\  \hat{a}_k\! \\ \hat{\tilde{a}}_k\!
    \end{pmatrix}^T\!\!\!\!\!\!
    \begin{pmatrix}
        -i\nu_k &  0 & 0 & 0\\
        0 & -i \nu_k^* & 0 & 0 \\
        0 & \!i(\nu_k\!\!+\!\!\nu_k^*)\! & -i \nu_k & 0 \\
        \!i(\nu_k\!\!+\!\!\nu_k^*)\! & 0 & 0 & -i\nu_k^*
    \end{pmatrix}\!\!\!
    \begin{pmatrix}
        \hat{a}_k \!\\  \hat{\tilde{a}}_k\! \\  \hat{a}_k^{\dagger}\! \\ \hat{\tilde{a}}_k^{\dagger}\!
    \end{pmatrix}  + \frac{i}{2}(\nu_k\!\! +\!\! \nu_k^*)\\
   & =    \frac{1}{2}\boldsymbol{a^\dagger}_k\!\,^T 
    \boldsymbol{H}
    \boldsymbol{a}_k + \frac{i}{2}(\nu_k\!\! +\!\! \nu_k^*),
\end{split}
\end{equation}
where we have used that $\gamma_k = \mathrm{Re}(\nu_k) = \frac{\nu_k + \nu_k^*}{2}$.
The off-diagonal, bath-bath coupling term in this expression gives rise to terms in the equations of motion that are not present in HEOM formulations.  This expression corresponds to a non-hermitian harmonic Hamiltonian, and so following the Ref.~\onlinecite{PhysRevA.96.062130}, we seek to find a diagonal representation of the form
\begin{equation}
    \begin{split}
    \hat{H}_{B, k} &= \! \frac{1}{2}\!\!\begin{pmatrix}
       \hat{\overline{b}}_{1k}^{\dagger}\! \\ \hat{\overline{b}}_{2k}^{\dagger}\! \\ \hat{b}_{1k} \!\\  \hat{b}_{2k}\! 
    \end{pmatrix}^T\!\!\!\!\!\!
    \begin{pmatrix}
        -i\nu_k &  0 & 0 & 0\\
        0 & -i \nu_k^* & 0 & 0 \\
        0 & 0 & -i \nu_k & 0 \\
        0 & 0 & 0 & -i\nu_k^*
    \end{pmatrix}\!\!\!
    \begin{pmatrix}
        \hat{b}_{1k} \!\\  \hat{b}_{2k}\! \\  \hat{\overline{b}}_{1k}^{\dagger}\! \\ \hat{\overline{b}}_{2k}^{\dagger}\!
    \end{pmatrix}  + \frac{i}{2}(\nu_k\!\! +\!\! \nu_k^*)\\ &=  -i \nu_k \hat{\overline{b}}^{\dagger} _{1k} \hat{b}_{1k} -i \nu_k^* \hat{\overline{b}}_{2k}^\dagger \hat{b}_{2k}, 
    \end{split}
\end{equation}
where we have new set of complex normal mode bosonic operators, that are related to the original by a generalised Bogoliubov transformation.  In general, we have that $\hat{\overline{b}}^{\dagger}_{ik} \neq \hat{b}^{\dagger}_{ik}$, however these operators satisfy the bosonic commutation relations
\begin{equation}
    [\hat{b}_{ik}, \hat{\overline{b}}^{\dagger}_{jk}] =  \delta_{ij}, [\hat{b}_{ik}, \hat{\overline{b}}_{jk}] =  [\hat{b}_{ik}^\dagger, \hat{\overline{b}}_{jk}^\dagger]=0
\end{equation}
These operators can be found by considering the eigendecomposition of the matrix $\boldsymbol{M}\boldsymbol{H}$~\cite{PhysRevA.96.062130}, where 
\begin{equation}
    \boldsymbol{M} = \begin{pmatrix}
        I & 0 \\ 0 & -I \\
    \end{pmatrix},
\end{equation}
and $\boldsymbol{M}\boldsymbol{H}$ determines the commutators of the bath Hamiltonian with the original boson operators.

For the case of the Quasi-Lindblad pseudomode bath Hamiltonian considered above,  $\boldsymbol{MH}$ is diagonalisable with eigendecomposition matrix
\begin{equation}
    \begin{split}
        &\boldsymbol{MH} = \boldsymbol{W} \boldsymbol{\Lambda} \boldsymbol{W}^{-1}\\
        &= \begin{pmatrix}
    1 & 0 & 0 & 0 \\ 
    0 & 1 & 0 & 0 \\
    0 & 1 & 1 & 0 \\
    1 & 0 & 0 & 1 \\
    \end{pmatrix}
    \begin{pmatrix}
       \!-i \nu_k & 0 & 0 & 0 \\
       0 & \!\!-i\nu_k^* & 0 & 0 \\ 
       0 & 0 & \!i\nu_k & 0 \\
       0 & 0 & 0 &\!i\nu_k^*
    \end{pmatrix}
    \begin{pmatrix}
    1k & 0 & 0 & 0 \\ 
    0 & 1 & 0 & 0 \\
    0 & \!\!-1 & 1 & 0 \\
    \!-1 & 0 & 0 & 1 \\
    \end{pmatrix} 
    \end{split}.
\end{equation}

Giving the new bosonic operators
\begin{equation}
    \begin{pmatrix}
    \hat{b}_{1k} \\ \hat{b}_{2k} \\ \hat{\overline{b}}^{\dagger}_{1k} \\ \hat{\overline{b}}^{\dagger}_{2k} 
    \end{pmatrix}= \boldsymbol{W}^{-1} \boldsymbol{a}_k = 
    \begin{pmatrix} \hat{a}_k \\ \hat{\tilde{a}}_k \\ \hat{a}^{\dagger}_k - \hat{\tilde{a}}_k \\\hat{\tilde{a}}^{\dagger}_k  - \hat{a}_k
    \end{pmatrix},
\end{equation}
where we note that the new creation operators are now a linear combination of creation and annihilation operators acting on bra and ket states. This transformation can equivalently be expressed using the operators 
\begin{equation}
    \hat{W}_k^\pm = e^{\pm\hat{a}_k\hat{\tilde{a}}_k}
\end{equation}
\noindent as
\begin{equation}
    \begin{pmatrix}
    \hat{b}_{1k} \\ \hat{b}_{2k} \\ \hat{\overline{b}}^{\dagger}_{1k} \\ \hat{\overline{b}}^{\dagger}_{2k} 
    \end{pmatrix}= 
    \begin{pmatrix} \hat{W}_k^-\hat{a}_k\hat{W}_k^+ \\ \hat{W}_k^-\hat{\tilde{a}}_k\hat{W}_k^+ \\ \hat{W}_k^-\hat{a}^{\dagger}_k\hat{W}_k^+ \\ \hat{W}_k^-\hat{\tilde{a}}^{\dagger}_k \hat{W}_k^+
    \end{pmatrix},
\end{equation}
Now transforming Eq.~\ref{eq:pseudomode_superop} to this new representation, gives
\begin{equation}
\begin{split}
    \frac{\partial}{\partial t}\hat{\rho}(t) =& -i \mathcal{L}_s(t)\hat{\rho} \!  -i \sum_{k=1}^K (-i\nu_k \hat{\overline{b}}_{1k}^{\dagger}\hat{b}_{1k} -i\nu_k^* \hat{\overline{b}}_{2k}^\dagger \hat{b}_{2k}^{\dagger})\hat{\tilde{\rho}} -i \sum_{k=1}^K2i \left[  M_k ^*\hat{S}^L\hat{b}_{2k} + M_k \hat{S}^R\hat{b}_{1k} \right] \hat{\rho}\\
    - i\sum_{k=1}^K & \left(\!\left[V_k^* \!-\! i M_k^*\right] \hat{S} ^L(\hat{\overline{b}}_{1k}^{\dagger} \!+\! \hat{b}_{2k}) \!  -\!\left[V_k \!+\! i M_k\right] \hat{S}^R(\hat{\overline{b}}_{2k}^{\dagger} \!+\! \hat{b}_{1k}) \!\right)\!\hat{\rho}
    - i\sum_{k=1}^K \left(\!\left[V_k \!-\! i M_k \right]\hat{S} ^L \hat{b}_{1k} \!-\! \left[V_k^* \!+\! i M_k^* \right] \hat{S}^R\hat{b}_{2k}  \! \right)\!\hat{\rho},
\end{split}\label{eq:pseudomode_superop}
\end{equation}
which upon rearranging  and considering the general gauge choice from Ref.~\onlinecite{park2024quasilindbladpseudomodetheoryopen}, 
\begin{align}
V_k - iM_k &= \kappa_k \sqrt{\alpha_k} \\
V_k + iM_k &= \kappa_k^{*-1}\sqrt{\alpha_k^*}
\end{align}
with $\kappa_k\in\mathbb{C}$, 
gives
\begin{equation}
\begin{split}
    \frac{\partial}{\partial t}\hat{\tilde{\rho}}(t) &= -i \mathcal{L}_s(t)\hat{\tilde{\rho}}   -i \sum_{k=1}^K (-i\nu_k \hat{\overline{b}}_{1k}^{\dagger}\hat{b}_{1k} -i\nu_k^* \hat{\overline{b}}_{2k}^\dagger \hat{b}_{2k}^{\dagger})\hat{\tilde{\rho}} \\
    &- i\sum_{k=1}^K \left(\kappa_k^{-1}\sqrt{\alpha_k} \hat{S} ^L\hat{\overline{b}}_{1k}^{\dagger}  -\kappa_k^{*-1}\sqrt{\alpha_k^*} \hat{S}^R\hat{\overline{b}}_{2k}^{\dagger} \right)\hat{\tilde{\rho}}
- i\sum_{k=1}^K \left(\hat{S} ^L\! -\!\hat{S}^R\right) \!\!\left(\kappa_k\sqrt{\alpha_k} \hat{b}_{1k} + \kappa_k^{*}\sqrt{\alpha_k^*} \hat{b}_{2k} \right)  \hat{\tilde{\rho}}.
\end{split}\label{eq:pseudomode_superop_general}
\end{equation}
As a final step, we now consider expanding this equation in a basis for the bosonic degrees of freedom. 

In order to do this we define the states
\begin{equation}
\begin{split}
    &\ket{\boldsymbol{n}_{b}} = \left(\prod_{ik} \frac{(\hat{\overline{b}}_{ik}^\dagger)^{n_{ik}}}{\sqrt{n_{ik}!}} \right) \ket{0_b} \\
    &    \ket{\boldsymbol{m}_{\overline{b}}} = \left(\prod_{ik} \frac{(\hat{b}_{ik}^\dagger)^{m_{ik}}}{\sqrt{m_{ik}!}} \right) \ket{0_{\overline{b}}}, 
\end{split}
\end{equation}
which, form a biorthogonal set~\cite{PhysRevA.96.062130}
\begin{equation}
    \braket{\boldsymbol{m}_{\overline{b} }}{\boldsymbol{n}_{b} }= \prod_{ik} \delta_{m_{ik} n_{ik}}.
\end{equation}
Within this representation we obtain
\begin{equation}
\begin{split}
    \frac{\partial}{\partial t}\braket{\boldsymbol{m}_{\overline{b}}}{\hat{\tilde{\rho}}(t) }= &- i\left(\mathcal{L}_s(t) - i\sum_{k=1}^K (\nu_k m_{1k} + \nu_k^* m_{2k}) \right)\braket{\boldsymbol{m}_{\overline{b}}}{\hat{\tilde{\rho}}(t)}  \\&- i\sum_{k=1}^K \left[\kappa_k^{-1}\sqrt{m_k \alpha_k}\hat{S} \braket{\boldsymbol{m}_{\overline{b}} \!-\! \boldsymbol{1}_{1k}}{ \hat{\tilde{\rho}}(t)}    -\kappa_k^{*-1}\sqrt{n_k\alpha_k^*}\braket{\boldsymbol{m}_{\overline{b}} \!-\! \boldsymbol{1}_{2k}}{ \hat{\tilde{\rho}}(t)}\hat{S}\right] \\ & - i\sum_{k=1}^K \left[\kappa_k\sqrt{(m_k +1)\alpha_k} \left[\hat{S}, \braket{\boldsymbol{m}_{\overline{b}} \!+\! \boldsymbol{1}_{1k}}{ \hat{\tilde{\rho}}(t)}\right]  + \kappa_k^{*}\sqrt{(n_k +1)\alpha_k^*} \left[\hat{S}, \braket{\boldsymbol{m}_{\overline{b}} \!+\! \boldsymbol{1}_{2k}}{ \hat{\tilde{\rho}}(t)}\right]\right], 
\end{split}\label{eq:pseudo_mode_fpheom}
\end{equation}
where $\boldsymbol{1}_{ik}$ is the vector that is zero everywhere except at index $ik$ where it takes value $1$.  Inspection of these equations shows that they are exactly of the form of the FP-HEOM equations given in Eq.~\ref{eq:fpheom_SM_rescaled}.

Within the FP-HEOM approach the system density matrix is obtained from the total state vector by projecting onto the vacuum state, that is $
    \rho_S(t) = \big\langle \boldsymbol{0}_{\overline{b}}\big|\hat{\tilde{\rho}}(t) \big\rangle$.
Transforming the density operator back to the Quasi-Lindblad pseudomode representation gives
\begin{equation}
\begin{split}
    \rho_S(t) &= \big\langle \boldsymbol{0}\big| \prod_k \hat{W}_k^+ \big|\hat{\tilde{\rho}}(t) \big\rangle \\ 
    &= \big\langle\boldsymbol{I}\big|\hat{\tilde{\rho}}(t)\big\rangle.
\end{split}
\end{equation}
That is, the reduced density matrix is obtained by taking the trace over bosonic modes of the pseudomode density operator, as expected.
\pagebreak

\section{Spin-Boson Model: Impact of Bath Topology for Binary Tree Bath Representation\label{sec:si_spin_boson}}

\begin{figure}[h]
\hspace{-2em}\includegraphics[width=0.45\columnwidth]{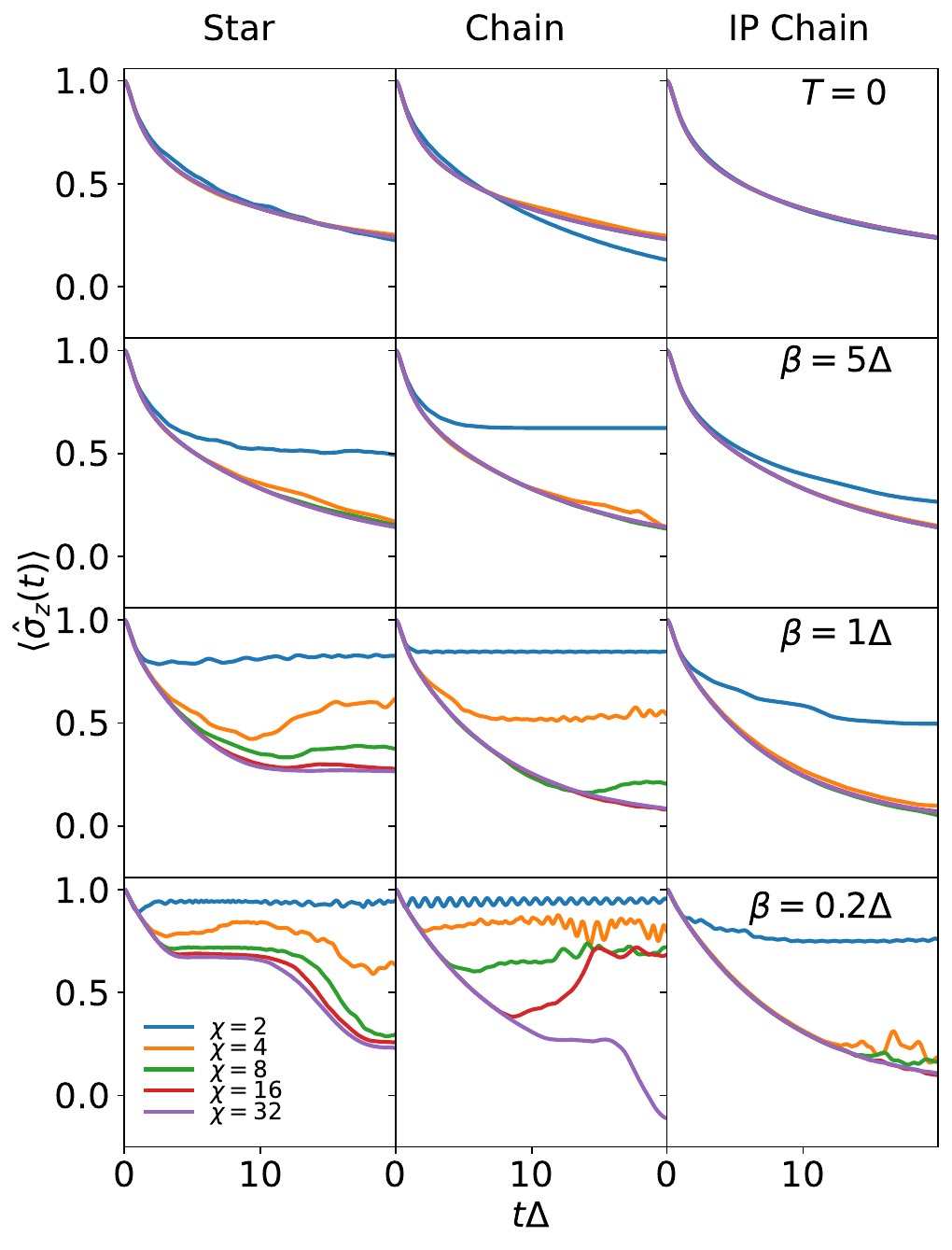} 
\caption{ \label{fig:sbm_finite_temp_conv_binary} Convergence with respect to bond dimension of the finite temperature polarisation dynamics obtained using the star (left), chain (middle) and interaction picture chain (right) Hamiltonians at various temperatures.  From top to bottom results were obtained at $T=0$, $\beta=5$, $\beta=1$, and $\beta=0.2$.  In all calculations a balanced binary tree topology (with an optimised boson basis) is used for the bath modes with fixed bond dimension, $\chi$, shown in the figures.  All results presented here were generated using the scripts \protect\path{pyttn/examples/oqs/spin_boson/sbm_unitary.py}}
\end{figure}

In Fig.~\ref{fig:sbm_finite_temp_conv} of the main text we present a comparison of the convergence with respect to bond dimension of polarisation dynamics of a spin-boson model with $\alpha=0.7$, corresponding to the overdamped, delocalised regime at $T=0$, using an MPS representation for the wavefunction.  In this section we consider the convergence of the same dynamics, but with the use of a balanced binary tree representation for the bath degrees of freedom. These results are shown in Fig.~\ref{fig:sbm_finite_temp_conv_binary}, in which we compare the dynamics obtained using the star, chain, and interaction picture chain topologies at four different temperatures.  

At $T=0$, we observe results comparable with those presented in the main text for the case of an MPS representation of the wavefunction.  We observe the rapid convergence of the dynamics with respect to bond dimension regardless of topology, with all three sets of results showing convergence to within the thickness of the line by $\chi=8$. For the star and interaction picture chain topologies, we continue to observe results consistent with those presented in the main text as the temperature is increased.  For the star topology increasing temperature leads to the need for larger bond dimension to accurately capture dynamics regardless of topology, and at high-temperatures significant plateaus are observed in the dynamics.  These plateaus are moved to later times with increasing bond dimension, but large changes in bond dimension lead to only minor increases in the time over which the dynamics is accurate.  For the interaction picture chain representation, we observe small bond dimensions remain sufficient to capture the finite-temperature dynamics.

For the chain geometry, significant differences are observed when using a balanced binary tree representation for the bath wavefunction compared to the MPS results presented in the main text.  Here we observe a significant increase in the required bond dimension upon increasing temperature, and similar to the case for the star geometry observe significant deviations following a time that depends strongly on the bond dimension used.  This difference can likely be attributed to the MPS topology better representing the structure of entanglement generated through dynamics under the chain Hamiltonian.

\section{Dissipative XY Chain: Convergence of Dynamic with Respect to Maximum Bond Dimension \label{sec:si_dissipative_xy_chain}}
\begin{figure}[h!] 
\includegraphics[width=0.65\textwidth]{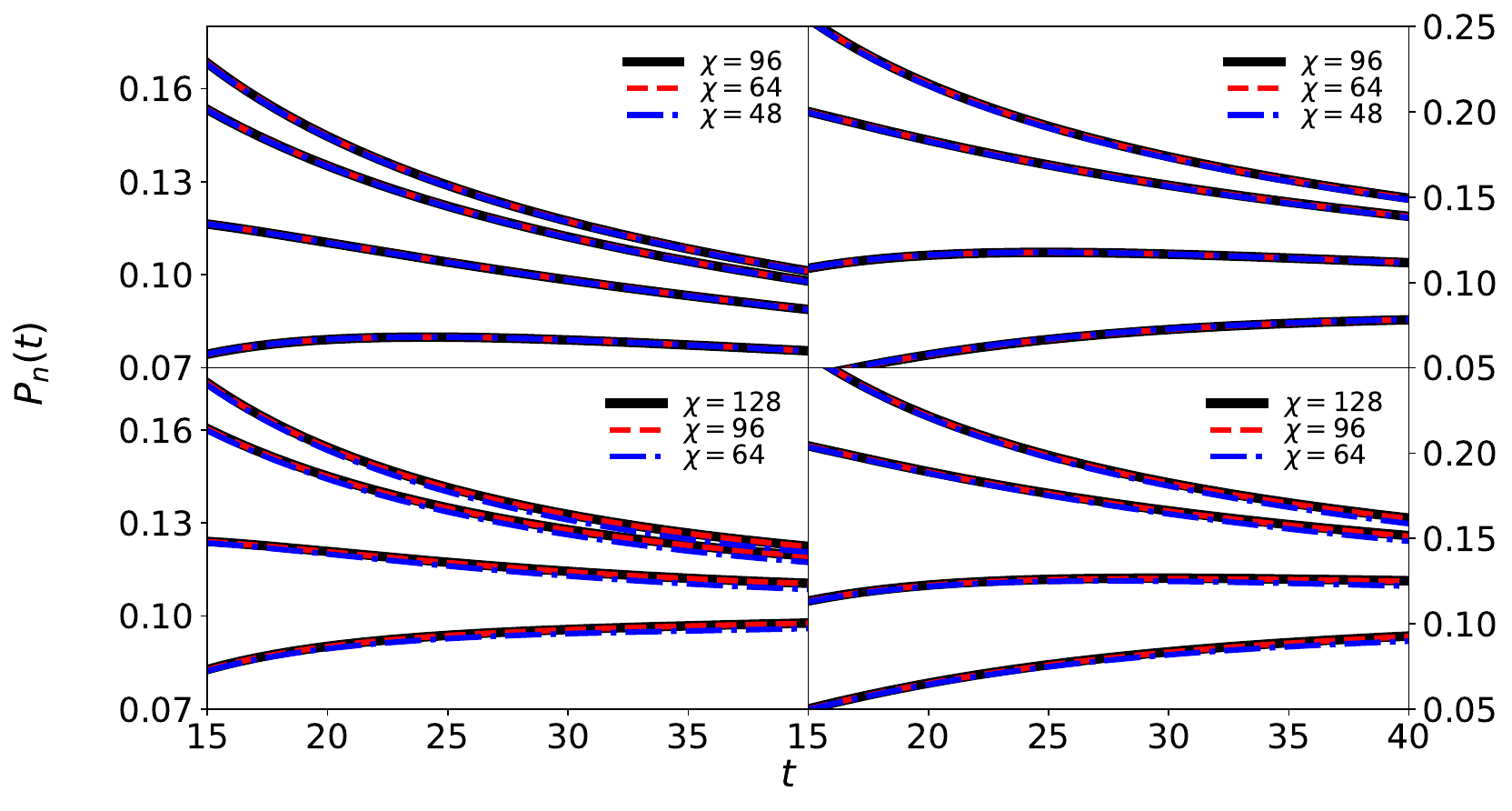}
\caption{
\label{fig:dissipative_xy_dynamics_chi_conv} Convergence of the dissipative XY spin chain population dynamics $P_n(t) = \frac{1}{2} - \left\langle S_{zn}(t)\right\rangle$ with respect to bond dimension.  Here we present the dynamics obtained for the four sites $n=0, 1, 2, 3$ for the isotropic case ($\eta=0.0$ top) and anisotropic case ($\eta=0.04$ bottom) obtained at $\beta = 0.625$ for system-bath coupling strengths $\alpha=0.16$ (left) and $\alpha=0.32$ (right).  In the captions we present the maximum allowed bond dimension $\chi$ in the tensor network presented in the top panel of Fig.~\ref{fig:dissipative_xy_fmps} and have used $\chi_S = \frac{3}{4}\chi$ and $\chi_B=\chi$.}
\end{figure}

In Fig.~\ref{fig:dissipative_xy_dynamics_chi_conv}, we present results showing the convergence of the HEOM dynamics for the dissipative XY spin chain dynamics with respect to bond dimension.  Here we observe systematic convergence of the dynamics with increasing bond dimension.  For the isotropic case $\eta=0$, we find that $\chi=48$ is sufficient to converge the dynamics on the timescales of interest.  For the anisotropic case, $\eta=0.04$, $\chi=96$ is required to see a similar degree of convergence (to within the thickness of lines presented here).  For the case considered here, this is not surprising as the anisotropic term does not conserve the total magnetisation. As a result, the dynamics in the anisotropic case explores a larger set of spin configurations, requiring a larger bond dimension to accurately reproduce the dynamics.  

\end{document}